\RequirePackage{fix-cm}
\documentclass[twocolumn]{svjour3}          
\smartqed  
\usepackage{graphicx}
\usepackage{url}
\usepackage{microtype}
\usepackage{hyperref}
\usepackage{booktabs}
\usepackage{listings}

\usepackage{xcolor}
\usepackage{textcomp}
\begin{document}

%

\title{
GeantV
}
\subtitle{
Results from the prototype of concurrent vector particle transport simulation in HEP
}

\titlerunning{GeantV} 

\author{
G.~Amadio$^{a}$,
A.~Ananya$^{a}$,
J.~Apostolakis$^{a}$,
M.~Bandieramonte$^{{a},{b}}$,
S.~Banerjee$^{c}$,
A.~Bhattacharyya$^{d}$,
C.~Bianchini$^{{e},{f}}$,
G.~Bitzes$^{{a}}$,
P.~Canal$^{c}$, 
F.~Carminati$^{a}$,
O.~Chaparro-Amaro$^{g}$,
G.~Cosmo$^{a}$,
J.~C~De~Fine~Licht$^{a}$,
V.~Drogan{$^{{a},{h}}$}
L.~Duhem$^{j}$,
D.~Elvira$^{c}$,
J.~Fuentes$^{k}$
A.~Gheata$^{a}$,
M.~Gheata$^{{a},{l}}$,
M.~Gravey$^{m}$,
I.~Goulas$^{a}$,
F.~Hariri$^{a}$,
S.~Y.~Jun$^{c}$,
D.~Konstantinov$^{a,u}$,
H.~Kumawat$^{d}$,
J.~G.~Lima$^{c}$,
A.~Maldonado-Romo$^{g}$,
J.~Mart\'inez-Castro$^{g}$,
P.~Mato$^{a}$,
T.~Nikitina$^{{a},{p}}$,
S.~Novaes$^{e}$,
M.~Novak$^{a}$,
K.~Pedro$^{c}$,
W.~Pokorski$^{a}$,
A.~Ribon$^{a}$,
R.~Schmitz$^{s}$,
R.~Seghal$^{d}$,
O.~Shadura$^{{a},{r}}$,
E.~Tcherniaev$^{a}$,
S.~Vallecorsa$^{{a},{p}}$,
S.~Wenzel$^{a}$,
Y.~Zhang$^{{a},{t}}$
}

\authorrunning{G.~Amadio, A.~Ananya, J.~Apostolakis et al.} 

\institute{
$^{a}$CERN, EP Department, Geneva, Switzerland \\
$^{b}$University of Pittsburgh, Pittsburgh, PA, 15260, USA \\
$^{c}$Fermi National Accelerator Laboratory, Batavia, IL, 60510, USA \\
$^{d}$Bhabha Atomic Research Centre (BARC), Mumbai, India \\
$^{e}$S\~{a}o Paulo State University (UNESP), S\~{a}o Paulo, Brazil \\
$^{f}$Mackenzie Presbyterian University, S\~{a}o Paulo, Brazil \\
$^{g}$Centro de Investigaci\'on en Computaci\'on. Instituto Polit\'ecnico Nacional. Mexico City. Mexico \\
$^{h}$Taras Shevchenko National University of Kyiv, Kyiv, Ukraine \\
$^{j}$Intel Corporation, Santa Clara, CA, 95052, USA \\
$^{k}$University of Bio Bio, Dept. of Computer Science and Information Technologies, Bio Bio, Chile \\
$^{l}$Institute of Space Sciences, Bucharest-Magurele, Romania \\
$^{m}$Univ. of Lausanne, Institute of Earth Surface Dynamics, Lausanne, Switzerland \\
$^{p}$Gangneung-Wonju National Univ., Gangneung, South Korea \\
$^{r}$National Technical University of Ukraine, Kiev Polytechnic Institute, Kiev, Ukraine \\
$^{s}$Univ. of California, Santa Barbara, California, USA \\
$^{t}$KIT, Karlsruhe, Germany \\
$^{u}$NRC `Kurchatov Institute' IHEP, Protvino, Russia
}

\date{Received: date / Accepted: date}

\maketitle
\begin{abstract}
\begin{sloppypar}Full detector simulation was among the largest CPU consumer in all CERN experiment software stacks for the first two runs of the Large Hadron Collider (LHC). In the early 2010s, it was projected that simulation demands would scale linearly with increasing luminosity, with only partial compensation from increasing computing resources. The extension of fast simulation approaches to cover more use cases that represent a larger fraction of the simulation budget is only part of the solution, because of intrinsic precision limitations. The remainder corresponds to speeding up the simulation software by several factors, which is not achievable by just applying simple optimizations to the current code base. In this context, the GeantV R\&D project was launched, aiming to redesign the legacy particle transport code in order to benefit from features of fine-grained parallelism, including vectorization and  increased locality of both instruction and data. This paper provides an extensive presentation of the results and achievements of this R\&D project, as well as the conclusions and lessons learned from the beta version prototype.\end{sloppypar}
\end{abstract}

\section{Introduction}
\label{sec:introduction}
With ever-increasing data acquisition rates and detector complexity, the experimental particle physics program is reaching the exascale in terms of the data produced. The high-luminosity phase of the Large Hadron Collider (LHC) will produce about 150 times more data than its first run~\cite{Albrecht2019}; hence, a proportional increase in computing requirements is expected. All steps in the data processing chain are expected to cope with the increased throughput, under the assumption of a flat computing budget.

Particle transport simulation is an essential component in all phases of a particle physics experiment, from detector design to data analysis. Its main role is trying to predict the detector response to the traversal of particles, which is a very complex task involving a large number of models. Among the most used particle transport libraries in high energy physics (HEP) are Geant4~\cite{Agostinelli:2002hh}, Fluka~\cite{Ferrari:2005zk}, and Geant3~\cite{Brun:118715}. Simulation is one of the most computationally demanding applications in HEP, utilizing more than half of the distributed computing resources of the LHC. The increasing demand for simulated data samples can be satisfied in part with approximate (so-called) fast simulation techniques, but accelerating the detailed simulation process remains essential for increasing simulation throughput.

The ambitious experiment upgrades are occurring in a context where computing technology is rapidly evolving. Since the historical approach to improve CPUs, increasing the clock speed and shrinking the transistors, is now limited by quantum leakage, industry is exploring alternative solutions for the next technological breakthrough. The main hardware manufacturers now favor parallel (or vector) processing units as well as heterogeneous hardware solutions with accelerators such as GPUs, FPGAs, and ASICs, facilitating a performance boost for many domain-specific applications. Most HEP applications are not optimized for Single Instruction Multiple Data (SIMD) parallelism or coprocessors and therefore do not make efficient use of these new resources.

The SIMD model utilizes specialized CPU vector registers to execute the same sequence of instructions in parallel for multiple data. The Single Instruction Multiple Threads (SIMT) model has the same concept as SIMD but the common code (kernel) is executed by multiple synchronous threads. The main practical difference between the two models is the length of the data vector: short in the case of SIMD, usually found on CPUs, and much longer in the case of SIMT, usually found on GPUs. Also, SIMD requires the strict alignment in a single register of all the data to be processed, while each thread in SIMT processes data in its own, separate register. Vectorized applications are easier to port to coprocessors that implement the SIMT model.

The benefits of SIMD and SIMT have been demonstrated for applications featuring massive data parallelism, such as linear algebra and graphics. However, bringing these vectorization techniques to complex code with significant branching presents a different type of challenge. Particle transport simulation has many features hostile to SIMD, including sparse memory access into large data structures, deep conditional branching, and long algorithmic chains and deep function call stacks per data unit (a track, representing a particle state) with poor code locality.

The GeantV simulation R\&D project~\cite{geantv:2018} aimed to exploit modern CPU vector units by re-engineering the simulation workflow implemented in Geant4~\cite{Agostinelli:2002hh} and the associated data structures. The goal was to enhance instruction locality by regrouping data (tracks) according to the tasks to be executed, rather than executing a sequence of tasks for the same track. The advantage of such an approach, besides the temporal locality, is that it enables new forms of data parallelism that were inaccessible before, such as SIMD and SIMT. Other computational workflows in HEP, such as reconstruction or physics analysis, could benefit from the same optimizations and it is expected that the lessons learned from the GeantV R\&D can be applied to these areas.

The target of the GeantV prototype was to speed up particle transport simulation applications by a factor of 2--5 on modern CPUs~\cite{geantv:2018}, compared to Geant4 in similar conditions. Gains from SIMD and better instruction cache locality were foreseen, along with code and algorithm refactoring. To support multi- and many-core platforms, thread parallelism was supported starting with the very early versions of the prototype. Another design requirement of this study was to ensure portability to various hardware architectures. This entails keeping the same code and preserving the ability to migrate the data model representation in a device-friendly format.

\section{Concepts and architecture}
\label{sec:architect}

Particle transport simulation is peculiar in terms of workflow and data access patterns. In most HEP event processing applications, the data lifetime is rather short: data is filtered and processed to produce results or derived quantities that are consumed by subsequent tasks. It is common for the same data to be used as constant input by several algorithms, but it is less common for that data to be recursively changed while being processed. The latter is the case for simulation, which follows the life cycle of a \emph{track}, representing a particle traveling through the detector. The track is the central data object used by most of the transport algorithms: geometry computations, propagation in electric or magnetic fields, or physics processes affecting the associated particle. From a computational perspective, the track represents a state taken as input and modified subsequently by a sequence of tasks, collaborating to perform a \emph{step} that moves it from one point to another. There is a design choice in the ordering of individual steps. In the traditional design, simulation engines perform consecutive steps on a single track until it completes its transportation.
To enhance code locality, one can chose an alternative approach, grouping tracks undergoing
similar stepping tasks (e.g. the same physics model actions). This requires deep changes
in the track handling and step ordering compared to the classical approach, which is
the basic direction taken by GeantV.

An important feature of simulation that drives the application design is unpredictability: particle physics is stochastic by nature, implying that the next physics process affecting a particle has to be chosen according to probability distribution functions. One cannot generate a sequence of processes in advance, because their probabilities are dependent on the material properties of the geometry location and the kinematic properties of the current track.
Hence, the scalar (per track) data flow consists in a sequence of tasks which,
depending on the previous one, cannot be known a priori.

The most convenient concept for handling the multitude of alternative algorithms is run-time polymorphism or virtual inheritance. Moreover, the large diversity and complexity of physics and geometry algorithms typically generate deep simulation call stacks and expensive branching logic, with a corresponding loss of computational efficiency.

The main GeantV concept is to change the focus from being data-centric to being algorithm-centric, making simulation SIMD- and SIMT-friendly. Instead of following a workflow from a track's perspective, static processing stages are defined that handle track populations being processed by each stage. This change of viewpoint helps to enhance spatial and temporal instruction locality, at the price of using more memory and likely worse data caching. Bundling more work together also enables more fine-grained parallelism and favors deployment on heterogeneous computing resources.

Another important exploration in the context of simulation is parallelism. Multi-threading parallelism is an important lever for making use of the full processing power of modern CPUs. Even if most HEP workflows are embarrassingly parallelizable on input data (such as individual LHC collision events), most of our applications are memory-bound and simulation is not an exception. Event-level parallelism has already been used in production for several years in Geant4, with very good overall scaling performance in multi-threaded mode and rather small memory overhead coming from each additional thread. The only problem is that, while multi-threading allows effective use of many-core CPUs, it does not produce any increase in the throughput per thread.

Vectorization is one of the throughput-increasing acceleration techniques and becomes beneficial when the code produces a large percentage of SIMD instructions. Although compiler authors are striving to provide solutions for automatic vectorization, in practice there are only a few kinds of problems for which auto-vectorization works out of the box. Auto-vectorization is more likely to be successful within confined data loops with reduced branching complexity and without any dependence on the input data. Since, in simulation, relatively few algorithms have natural internal loops, there are only limited benefits from auto-vectorization. GeantV explores  percolating track data into low-level algorithms, aiming to loop over this data internally. This approach requires being able to schedule reasonable data populations for each vectorized algorithm.

In this approach, data first needs to be accumulated into per-algorithm containers (``baskets'' in GeantV jargon), before being processed. The algorithms need to expose a new interface to handle an input basket and provide implementations that handle the basket data in a vectorizable manner. Note that the tracks coming from a single event may not suffice to fill baskets efficiently, given the complex branching of simulation code and the sheer variety of physics processes needed. One framework prerequisite is, therefore, to be able to mix tracks belonging to many concurrent events in the same processing unit.

\begin{figure}[ht]
    \centering
    \includegraphics[width=\columnwidth]{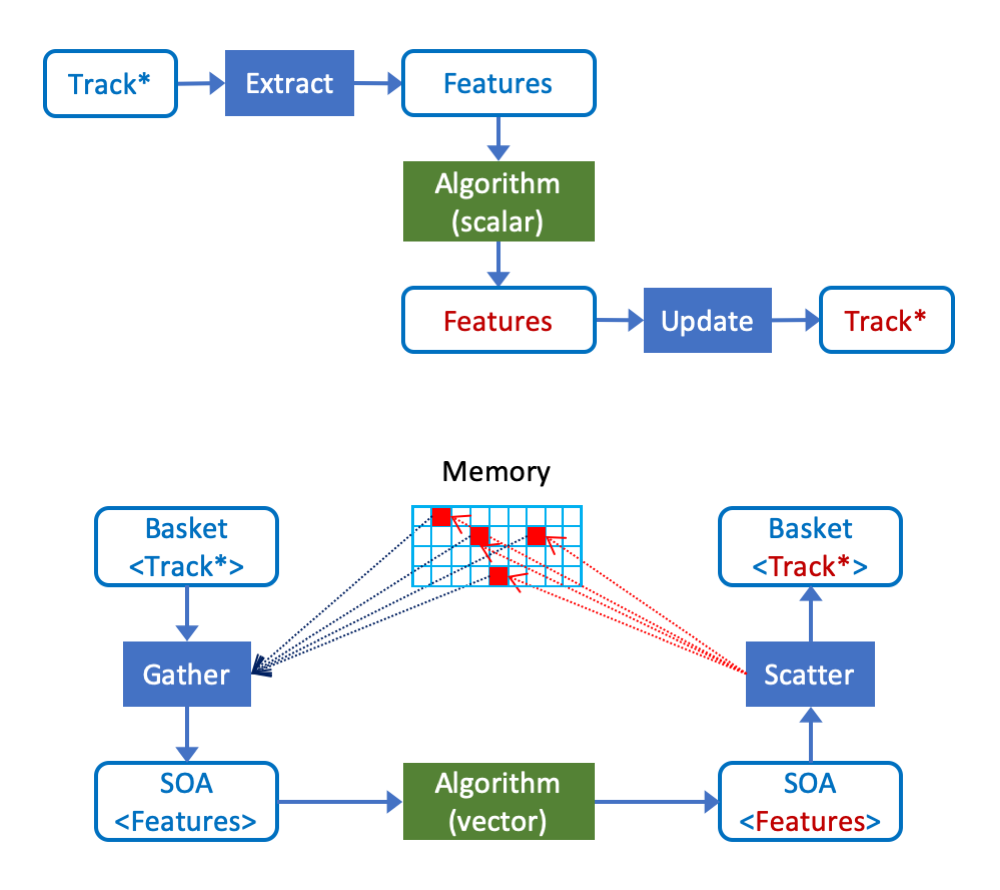}
    \caption{An algorithm-centric view of the operations performed for updating the track
      state during a single step for the scalar and vector cases. Tracks are handled by
      pointer (Track*) rather than by value due to the need to reshuffle per algorithm.}
    \label{fig:gather_scatter}
\end{figure}

Moving one level below, the requested track data has to be gathered and copied into the vector registers. For this to happen, the data are copied into arrays, each entry corresponding to the data of one track. In this scenario, the algorithm can be expressed as an easily-vectorized loop over C-like arrays. Scattering the algorithm output data to the original tracks completes the procedure and allows the processed tracks to be dispatched to subsequent algorithms. This schema requires a data transformation layer on top of each algorithm as shown in Fig.~\ref{fig:gather_scatter}.

\begin{sloppypar}
During this study, available vectorization techniques were thoroughly investigated in terms of programmability, performance, and portability. The techniques evaluated include auto-vectorization, compiler pragmas, SIMD libraries, and compiler intrinsics. The conclusion was that the higher the control over vectorization performance, the lower the portability and programmability. Assembly code or intrinsics are both difficult to write and maintain. On the other hand, auto-vectorization and compiler pragmas do not guarantee vectorization as an outcome, and this is an effect that worsens with increasing algorithm complexity.
Our preferred choice was to use SIMD libraries offering a high-level approach to vectorization via SIMD types and higher-level constructs, while keeping the complexity at a reasonable level and leveraging the portability of the library. It was decided to decouple as much as possible the implementation of algorithms from the concrete SIMD libraries, leading to the creation of VecCore~\cite{veccore:acat2017}, an abstraction layer on top of SIMD types and interfaces, supporting both scalar and vector backends (such as Vc~\cite{ref:vclib}, and UME::SIMD~\cite{umesimd}). The scalar backend supports SIMT as well.\end{sloppypar}

\subsection{Software design}
GeantV transforms the scalar workflow into a vector one. Instead of handling one track at a time, algorithms can operate on \textit{baskets} of tracks. Once a basket is injected in the algorithm, the vectorization problem is reduced to transforming all scalar operations on track data into vector operations on basket data. To generate efficient SIMD instructions and to quickly load data into SIMD registers, the basket data needs to be transformed from an array of track structures (AOS) to a structure of arrays of track data (SOA). This copying operation is only necessary for the part of the track data needed by the algorithm.

The workflow is orchestrated by a central run manager. This coordinates the work of several components, among which there are the event generator, the geometry and physics managers, and the user application. The main event loop can be controlled by either the GeantV application or the user framework. Primary tracks, defining the original input collision event, are either generated internally or injected by the user, buffered by an event server. The track-stepping loop is re-entrant, executed concurrently in several threads. Each thread takes and processes tracks from the event server. Once all tracks from a given event are transported, another event is generated/imported. The scheduler respects the constraint not to exceed the maximum number of events in flight set by the user.

\subsubsection{GeantV scheduler}

The scheduler's main task is to gather data efficiently in baskets for all the components, in order to improve vectorization. Also, the multi-threaded approach needs to have good scaling to make efficient use of all available cores. During the study, several different approaches to achieve both of these goals were tested, resulting in several versions of the scheduler.

\begin{figure}[h]
    \centering
    \includegraphics[width=\columnwidth]{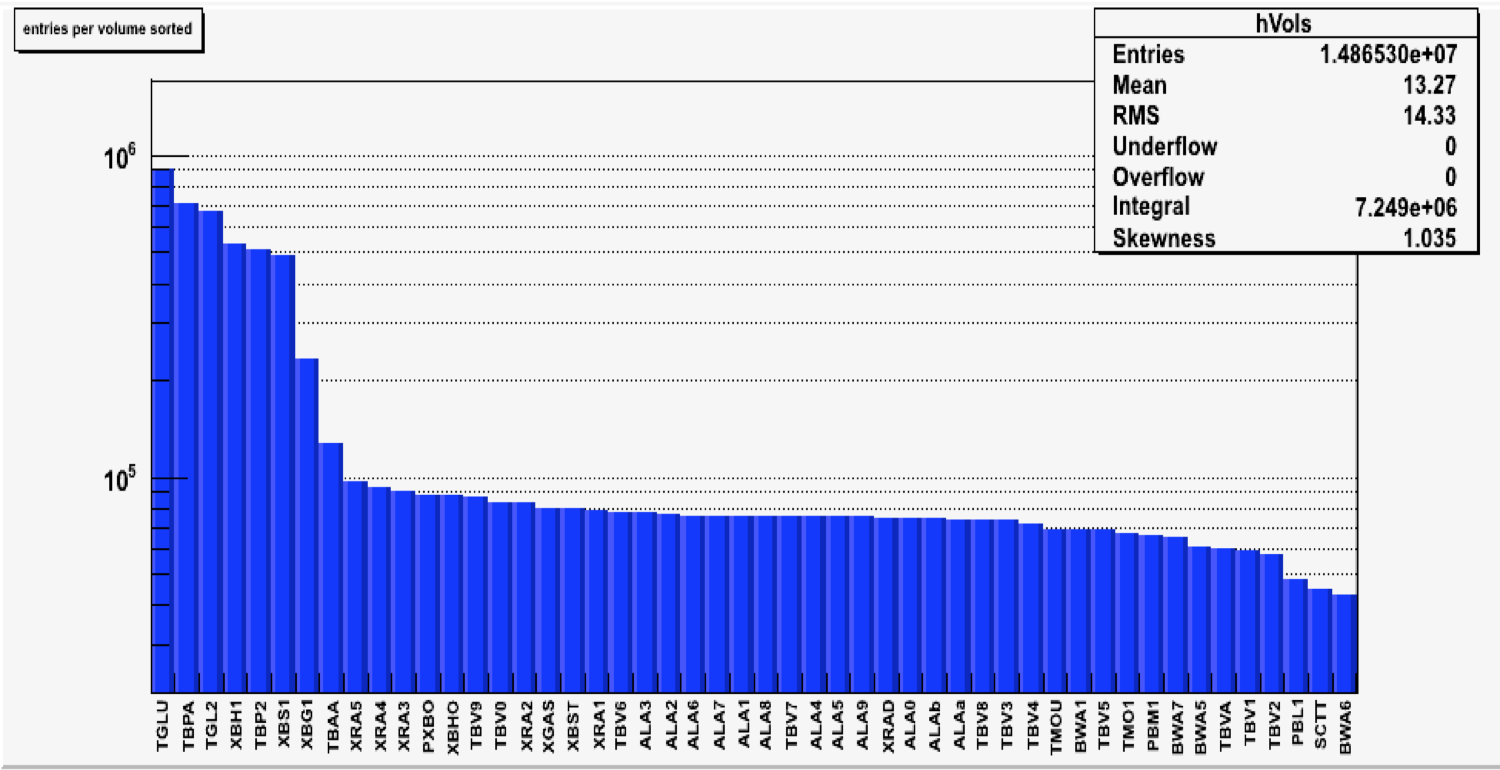}
    \caption{Geometry volume locality is observed in most detector simulations. This example illustrates the sorted number of simulated steps per volume in an ATLAS simulation from 2011. For this particular simulation, 50\% of the steps are executed in 50 out of 7100 logical volumes.}
    \label{fig:steps_per_volume}
\end{figure}

The first version of scheduling was mostly geometry-centric. It tried to benefit from the observation, illustrated in Fig.~\ref{fig:steps_per_volume}, that many track steps are done in a smaller number of important detector volumes/materials (volume locality). At the least, geometry calculations could be vectorized for such baskets. The model had a central work queue that handled baskets containing tracks located in the same geometry volume. Dedicated transport threads concurrently picked baskets from the queue and transported them to the next boundary. Whenever a track entered a new volume, it was copied into a pending basket for that volume. The worker thread that managed to fill a given basket beyond a threshold was then responsible for dispatching it to the work queue and replacing it with a recycled empty basket. A garbage collector thread was responsible for pushing partially filled baskets to the work queue whenever the queue started to be depleted. Merging produced hits and storing them to the output file was managed by a special I/O thread.

This first approach
focused on demonstrating track-level parallelism based on geometry locality, although vectorized algorithms for baskets were not available at the start of the project. This was an extremely useful step for understanding the differences and peculiarities of the basket-based track workflow compared to the single-track approach. However, the model had scaling issues due to high contention on specific baskets and frequent flushes done by the garbage collector during the event tails.

A second version of the scheduler
introduced support for explicit SIMD vectorization. The basket contained a track SOA with aligned arrays ready to be copied into the vector registers. Track data was copied in and out of the SOA, as tracks were passing from one basket to another. A simplified tabulated physics model was available in this version and, since it was not vectorized, the scheduler was still dealing only with geometry-local baskets. The prototype complexity increased and several tunable parameters were introduced in an attempt to implement an adaptive behavior, optimizing the performance of different setups and in different simulation regimes. Gathering and scattering data into the SOA baskets introduced new overheads due to extra memory operations, plus extra bottlenecks in the concurrent approach. To minimize the cost of memory operations, awareness of non-uniform memory access (NUMA) was introduced for handling basket data, leading to improvements of up to $10\%$ of the simulation time.

\begin{figure*}[h]
    \centering
    \includegraphics[width=\textwidth]{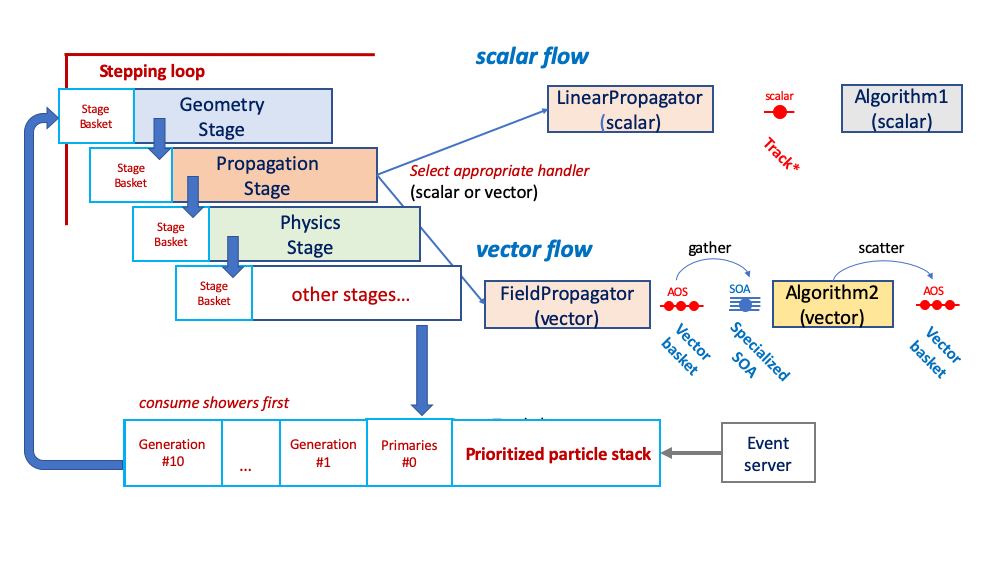}
    \caption{The final version of GeantV scheduler, accommodating both scalar and vector processing flows.}
    \label{fig:scheduler_v3}
\end{figure*}

The final version of the GeantV scheduler is shown in Fig.~\ref{fig:scheduler_v3}. 
Track data is described as a POD structure and pre-allocated in contiguous memory blocks.
Each thread takes pointers to primary tracks from an event server, storing them in an input
buffer (having the role of particle stack).  The stepping loop is implemented as a sequence of stages, each implementing a specific part of the processing required to make a single step for a population of tracks. Pointers to tracks tagged to execute a given stage are accumulated in the input \textit{stage basket}, processed by the stage algorithms, then dispatched to the input stage basket of the next stage. This implements a stepping pipeline for track populations. The scheduler takes bunches of track pointers (last generations first) and copies them in the \textit{input basket} of the first stage, triggering the pipeline execution. The stage basket is dispatched internally to specific handlers of specific processing tasks. For example, the propagation stage dispatches all neutral tracks to a linear propagator and the charged ones to a field propagator. The handlers of vectorized algorithms first accumulate (basketize) enough tracks to make the algorithm execution efficient. Subsequently, only the members of the track structure needed by the algorithm are gathered in an SOA before being processed, and then the results are scattered back to the original track pointers. Scalar algorithms make use directly of the stage basket track pointers, without having to gather/scatter data, so scalar and vector workflows can coexist. The last stage in the stepping pipeline implements the final stepping actions and calls the user application for scoring (tallying hits in sensitive detectors), before completing the cycle by copying the surviving tracks back to the prioritized particle stack. The scheduler has the role to push tracks in the stepping pipeline until exhausting the initial track population, then refilling it from the event server. Globally, the scheduler has also to balance the workload among concurrent threads and enforce policies to optimize the global workflow.
In addition to fixing many of the issues identified in the previous versions, such as contention in multi-threaded mode and memory behavior, this version introduced a generic model for basketizing, corresponding to the availability of more vectorized algorithms, in addition to geometry ones. The new framework significantly improved the basketizing efficiency, while also accommodating scalar and vector processing flows, switching from one to another depending on the workflow conditions.

\subsubsection{Scalar and vector workflows}
\label{sec:vec_workflow}
To support both scalar and vector workflows in the same framework, a common interface class called \textit{handler} was introduced to wrap all simulation algorithms in a common tasking system. The algorithm needs to implement the appropriate scalar and vector interfaces taking as input either a single track pointer or a vector (basket) of tracks. The vector method acts as a dispatcher for the SIMD version of the algorithm. It has to first gather the needed data from the container of tracks and copy it into a custom SIMD data structure. For example, geometry navigation requires only the track position and direction, while magnetic field propagation needs also the charge, momentum, and energy. The SIMD structure is then passed to the vectorized algorithm. The newly produced track state variables are then scattered to the original track pointers. To feed such \textit{handlers} in a workflow, tracks executing the same algorithm need to be gathered in SIMD baskets before being handed to the vector interface. In the case vectorization of a given algorithm is not implemented or inefficient, the scalar interface can be directly invoked, using a scalar pipeline for this algorithm.

\begin{figure}[h]
    \centering
    \includegraphics[width=\columnwidth]{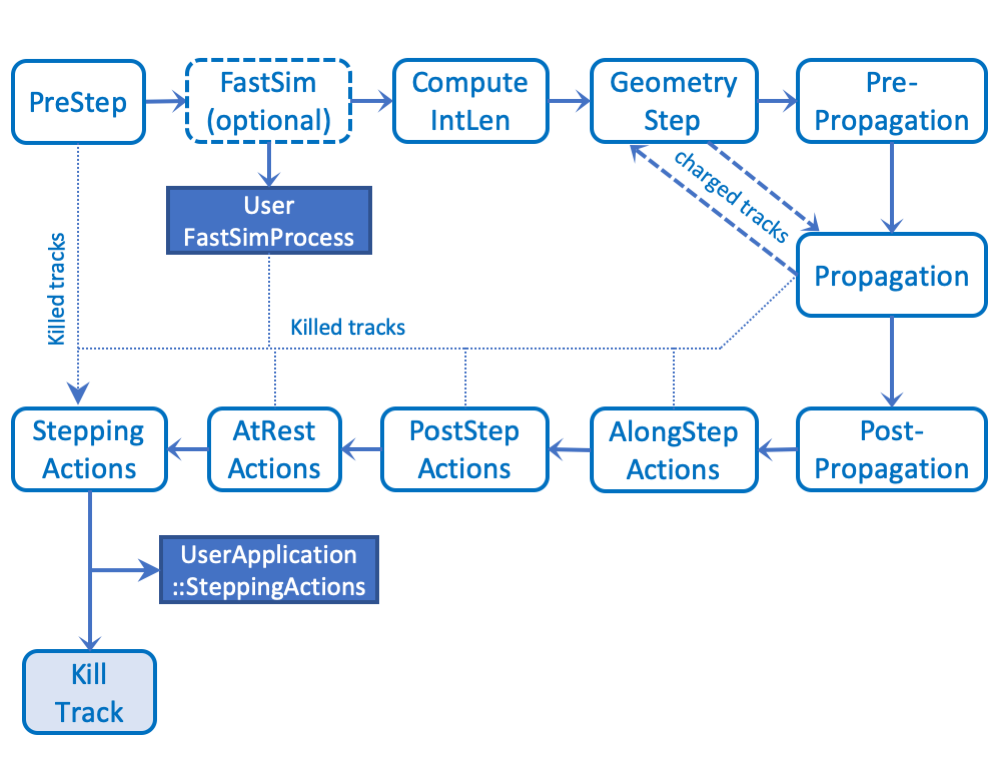}
    \caption{The sequence of stepping stages for baskets of particles in the GeantV prototype. Stages are connected in a predefined sequence similar to the stepping approach in Geant4, but there are also shortcuts or back connections that allow dumping stopped particles or repeating some stages. Each stage provides scalar and, in most cases, vector handlers for the stage algorithms.}
    \label{fig:simulation_stages}
\end{figure}

Algorithms of the same type are grouped into \textit{simulation stages}. The simulation stages refer to specific operations that have to be executed in a pipelined manner to perform a single step that moves a particle from one position to the next. The sequence of stages executed per step by baskets of tracks can be followed in Fig.~\ref{fig:simulation_stages}. At the beginning of the step, a \textit{PreStep} stage initializes the track flags and separates killed tracks, handling them to a final {\textit{SteppingActions}} to be accounted and scored. The remaining tracks enter the stage \textit{ComputeIntLen}, which samples physics processes' cross-sections and proposes an interaction length. Subsequently, a \textit{GeomQuery} stage computes the geometry stepping limits in the current volume and a \textit{PrePropagation} stage uses the actual step to determine in advance if multiple scattering will affect the current step. The actual track propagation is performed during a \textit{PropagationStage}, having one handler for neutral and one for charged particles. The multiple scattering deflection is added after the propagation in a \textit{PostPropagation} stage, and any continuous processes are subsequently applied by the \textit{AlongStepAction} stage. For steps limited by physics processes, a \textit{PostStepActions} stage is executed, and then the final \textit{SteppingActions} stage that accounts for stopped tracks and executes user actions. Every stage has an input basket per thread, used to execute the stage either in scalar mode, by looping over the contained tracks, or in vector mode, by passing the full basket to the interface.

The workflow is executed in the following manner. Each thread collects a set of primary tracks in a special buffer, called \textit{StackBuffer}, which emulates the functionality of a typical track stack (also used in Geant4). Secondary tracks of a higher generation are also pushed into this buffer and prioritized compared to their ancestors. The workload manager only copies the highest generation tracks into the basket of the first stage, then executes it. Once processed, the tracks are copied to the input basket of the second stage, and so on. Each stage has one or more follow-ups, so most particles get pushed along the stepping pipeline, but some particles may loop between stages before being able to execute the complete step. As an example, charged particle propagation requires repeated queries to the geometry before finally crossing the volume boundary. The stepping loop just pushes the input buffers executing the stages one after another, multiple times, until the baskets are empty. It then takes a new bunch of tracks from the \textit{StackBuffer}. During this loop, some tracks typically end up in unscheduled SIMD baskets, but a subsequent loop can fill these SIMD baskets and flush them back into the pipeline.

\subsubsection{Concurrency model}\label{sec:concmod}
The GeantV prototype implements parallelism at the track level. It supports an internal mode where the workload is parallelized among threads managed by the GeantV scheduler. It also supports an external mode implemented as a call to a re-entrant task transporting an event set, where the parallelism is controlled by the framework that makes the call.

\begin{figure}[h]
    \centering
    \includegraphics[width=\columnwidth]{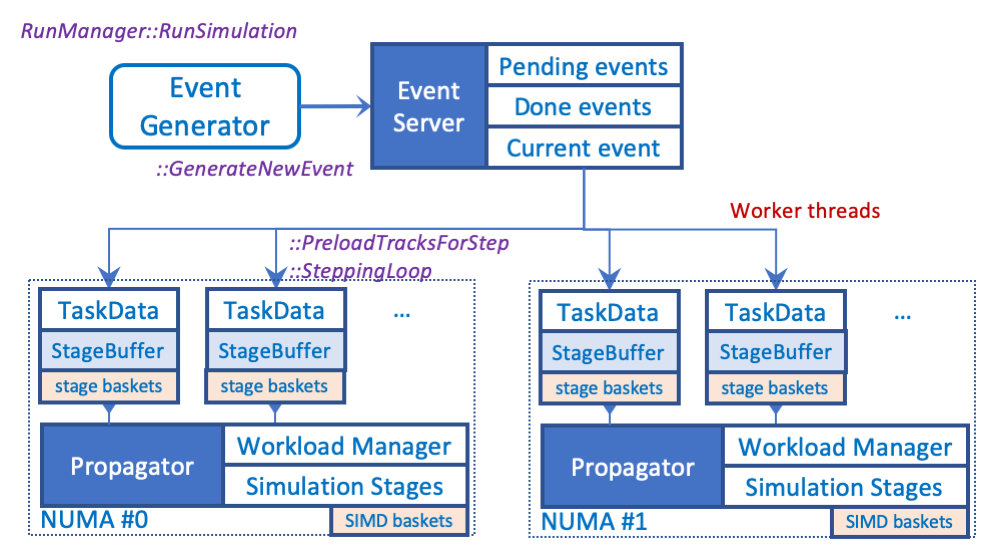}
    \caption{Schematic view of parallelism for the GeantV prototype. Multiple transportation tasks, handling their task data, share a set of propagator objects, one per NUMA domain. Each propagator shares a set of simulation stages with their SIMD baskets, but each transport task will handle its stack buffer of tracks. The system threads are scheduled by the run manager, and they will pick-up a free task data object from a concurrent queue. Each transport tasks will preload tracks from a concurrent event server, executing the stepping loop.}
    \label{fig:concurrency}
\end{figure}

Primary tracks produced by an event generator are stored in a concurrent event server and delivered to worker threads in bunches of customized size. The track data storage itself is pre-allocated to avoid dynamic memory management, partitioned per NUMA domain, and only pointers to tracks are delivered via the event server interfaces, as shown in Fig.~\ref{fig:concurrency}. Once a thread picks up a set of primary tracks, it becomes the only user of each track in the set for the given step. Due to this design, there is no synchronization needed when changing the state of a track. Threads handle tracks in their buffers; however, they share a single set of SIMD baskets per NUMA domain, so a thread may steal the tracks accumulated in these baskets by other threads. Even in scalar mode, when the SIMD baskets are empty, there is a mechanism allowing threads to steal tracks from each other as a mechanism of work balancing during the processing tail at the end of events.  

The concurrency model was designed to minimize the synchronization needs and to reduce contention in the concurrent services, while sharing track data to increase basket populations.
The thread-specific state data needed by the different methods cooperating for track
propagation is aggregated in specific objects (called \textit{TaskData}), different for every thread.
A \textit{TaskData} object is passed as argument to the stepping loop method executed by a given
thread, becoming visible to all the callees requiring it. This approach avoids the need of
syncronizing concurrent write operations on state data.

\begin{sloppypar}To maximize the basket population, vectorized handlers have a common SOA basket shared between threads. This was a requirement for enhancing the vector population, but it has a large cost of increased contention and loss of data locality. To improve this, thread-local copies of the SIMD basket are created for the handlers with the largest population of tracks, such as those for field propagation and multiple scattering. For these, the track population in a single thread is enough to fill them, without workflow perturbation or basket population loss. This allowed a large reduction in contention in the multi-threaded basket mode.\end{sloppypar}

An important feature for fine-grained workflows is load balancing. The GeantV workflow is naturally balanced by the event server, which acts as a concurrent queue. The main problem that occurs is the depletion of the stack buffers of each thread when most of the remaining particles reside in SIMD baskets that do not have a large enough population to execute efficiently in vectorized mode. Such a regime becomes blocking when the number of events in flight has already reached the maximum specified by the user, so the scheduler enters the so-called \textit{flush mode}. All SIMD baskets are simply flushed and the scalar \textit{DoIt} methods are executed by the first thread triggering this mode. Flushed particles are gathered in the stage baskets of this thread, which feeds the thread but depletes, even more, the other threads that were already starving. This unbalancing mechanism is compensated by a round-robin track sharing mechanism, which allows threads to feed not only from the event server, but also from the shared buffers of other threads. To preempt the depletion regime, threads always share a small fraction of their own track populations, but will consume those themselves if no other client has. This mechanism of weak sharing allows the reduction of contention in the normal regime. Sharing is dominant during event tails and is also more important when running with many threads.

The externally-driven concurrency mode is the so-called \textit{external loop} mode. In this mode, no internal threads are launched. The run manager provides an entry point that is called by a user-defined thread and that takes a set of events coming from the user framework. This will subsequently book a GeantV worker to perform the stepping loop, and will notify the calling framework via a callback. An example is provided, performing a simplified simulation of the CMS detector steered by a toy CMSSW~\cite{CMSSW} framework, mimicking the features of the full multi-threaded software framework of an LHC experiment. The GeantV simulation can be wrapped in a TBB (Threading Building Blocks, Intel{\textregistered})~\cite{IntelTBB} task and executed in a complex workflow, as described in Section~\ref{sec:exp-fwk-int}.

\section{Implementation}
\label{sec:implementation}
This section describes the core components of GeantV libraries and modules: VecCore, VecGeom, VecMath, propagation in a magnetic field and electromagnetic (EM) physics.
Auxiliary modules, such as I/O and user interfaces, are briefly summarized as well.

\subsection{Vector libraries: VecCore}
\label{sec:VecCore}

\begin{figure}[h]
  \centering
  \includegraphics[width=0.8\columnwidth]{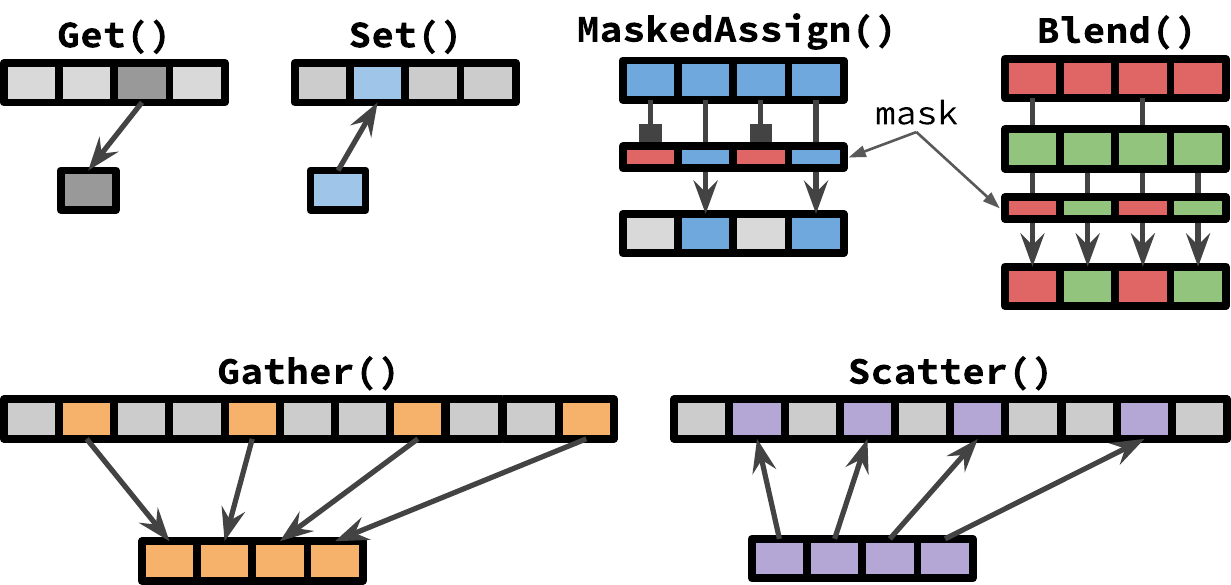}
  \caption{Illustration of VecCore API operations.}
  \label{fig:veccore-api}
\end{figure}

\begin{sloppypar}Portable and efficient vectorization is a significant challenge in large-scale software projects such as GeantV. The VecCore library~\cite{veccore:acat2017} was created to address the problem of lack of portability of SIMD code and unreliable performance when relying solely on auto-vectorization by the compiler. VecCore allows developers to write generic computational kernels and algorithms using abstract types that can be dispatched to different backend implementations, such as the Vc~\cite{ref:vclib} and UME::SIMD~\cite{umesimd} libraries, CUDA, and scalar. VecCore provides an architecture-agnostic API, illustrated in Fig~\ref{fig:veccore-api}, that covers the essential parts of the SIMD instruction set. These include performing arithmetic in vector mode, computing basic mathematical functions, operating on elements of a SIMD vector, and performing gather and scatter, load and store, and masking operations. Code written using VecCore can be annotated for running on GPUs with CUDA, and can be portable across ARM{\textregistered}, PowerPC{\textregistered}, and Intel{\textregistered} architectures, if not relying on features specific to a particular backend (e.g. using CUDA-specific variables such as thread and block indices, or calling external library functions that may be available only on the CPU).\end{sloppypar}

VecCore is used to implement vectorized geometry primitives in VecGeom (described in Section~\ref{sec:VecGeom}), and vectorized physics models in GeantV. A brief discussion of VecCore with code samples can be found in Ref.~\cite{veccore:acat2017}, and examples of VecCore usage within VecGeom and GeantV appear in the following sections as well.

\subsection{Geometry description: VecGeom}
\label{sec:VecGeom}
\subsubsection{Introduction}

Detector simulation relies on the availability of methods to describe and construct the detector layout in terms of elementary geometry primitives, 
as well as interfaces that allow the determination of positions and distances with respect to the constructed layout.
Well-known examples of such geometry modelers are the Geant4 geometry module and the ROOT TGeo library~\cite{Brun:1997pa}. 
Both enable users to build detectors out of hierarchical descriptions of (constructive) solids and their containment within each other.

The vectorized geometry package, named {\emph{VecGeom}}, was chosen as one of the first areas in which to study the optimal 
usage of SIMD and SIMT paradigms for passing vector data between algorithms, which is one of the main targets of GeantV. 
From this point of view, the primary development focus was implementing algorithms capable of operating on elements of baskets in parallel.  
This entails geometry primitives, such as a simple box, that offer kernels to calculate distances for a
group of tracks in one function call, in addition to the normal case where only one track is handled.

Below as an example the signature for a typical geometry primitive is followed
by the corresponding signature of the new vector/basket interface:
\begin{small}
\begin{verbatim}
class Box {
  // interface for a single track
  double DistanceToIn(Track) const;

  // interface for multiple tracks
  VectorOfdoubles DistanceToIn(VectorOfTracks) const;
};
\end{verbatim}
\end{small}

Moreover, data structures and algorithms in VecGeom are laid out to enable efficient operation in heavily multi-threaded frameworks. For instance, a clear separation of state and services enables frequent track or context switches in the navigation module. This module is responsible for predicting where a track will go in the geometry hierarchy along its straight-line path.
Multi-platform usage was targeted since the beginning: the same code base is intended to compile and run on CPUs as well as GPU accelerators.

Besides these primary goals, the development of VecGeom was guided and influenced by other requirements and circumstances.
The first is to continue offering traditional interfaces operating on the single-particle (scalar) level. 
This ensures backward compatibility with the Geant4 or TGeo systems and is, in any case, needed to treat particles that have not been put into a basket.
Secondly, another geometry project called USolids~\cite{Gayer:2012}, funded by the EU AIDA project, was already in place, aiming to review and modernize the algorithms of Geant4 and TGeo and to unify the geometry code base. 
The VecGeom project joined forces with the USolids project for better use of available resources. As a consequence, VecGeom was factored out into a standalone repository and with the potential to evolve independently of GeantV. 
Therefore, VecGeom serves GeantV, with a basketized treatment and GPU support, as well as making the modernized code available to clients in traditional scenarios using the single-particle interface.

The multitude of use cases and APIs to support (scalar, basketized, CPU, GPU) poses the risk of code duplication. In order to reduce this, VecGeom started with an approach, adopted by most GeantV modules, in which standalone (static) templated algorithmic kernels are instantiated multiple times with different types and specializations behind the public interfaces.
\begin{figure}[b]
    \centering
    \includegraphics[width=\columnwidth]{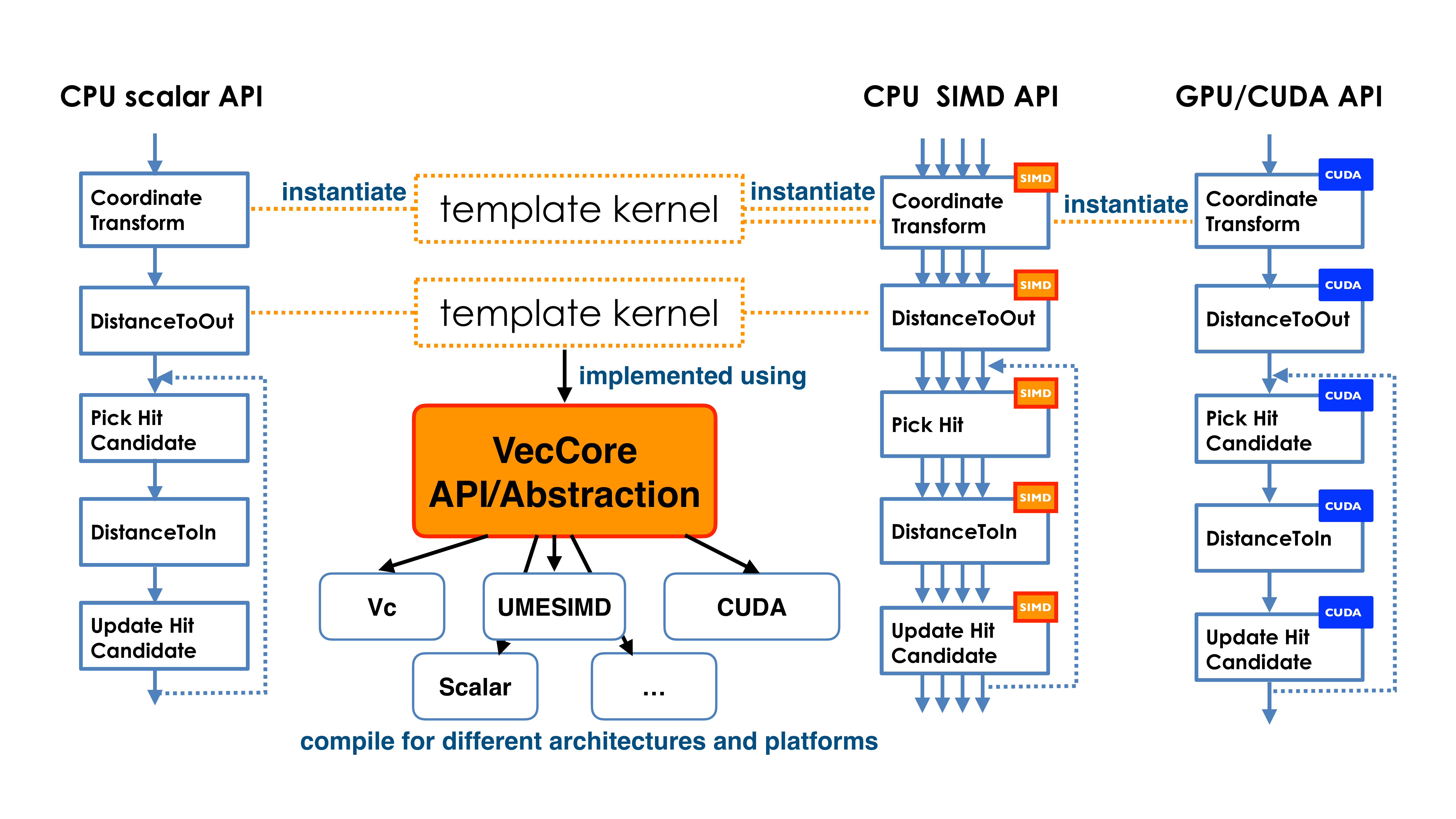}
    \caption{The code organization of VecGeom that motivates VecCore.}
    \label{fig:VecGeom_templatekernels}
\end{figure}
This development architecture is visualized in Fig.~\ref{fig:VecGeom_templatekernels}, where the typical use cases are depicted 
as functional chains of algorithms (scalar, vector, GPU), all implemented in terms of the same kernel templates. 
In order to make this happen, the kernels are written in such a way that they can be instantiated with 
native C++ types as well as with SIMD vector types (as offered by vectorization libraries such as Vc). Furthermore, all constructs used 
have the proper annotation to compile on the GPU (using CUDA). VecCore, prototyped within the VecGeom effort,
provides the abstractions needed to write these generic kernels.

\begin{sloppypar}Using this development approach, VecGeom has evolved into a geometry library that offers similar features to the classical Geant4 geometry or TGeo for transport simulation for 
single particle queries. On top of this, these algorithms are also made available for basket queries or for execution on CUDA GPUs.
In particular, all major geometry primitives have been implemented, and hierarchical detectors can be constructed from them via composition and placement.
To solve the complex geometry tasks typically needed in particle detector simulation, such as determining the minimum distance of particles to any other material boundary or computing
the intersection points with the next object along a particle's straight-line path, VecGeom offers navigator classes that operate on top of these primitives.\end{sloppypar}

\begin{sloppypar}Today VecGeom's objective is to be a high-performance library for these tasks {\emph{in general}}. A lesson learned in the development was that it is worth taking a more loosely defined approach
to achieve good performance and to benefit from SIMD instructions. In particular, VecGeom targets both basketized (or horizontal) vectorization as well as inner-loop (vertical) vectorization, depending
on the complexity of the algorithm. A simple box primitive is an example of the former, and a complicated tessellated shape is an example of the latter.
The best SIMD performance for a box is obtained with the use of baskets, yet a SIMD speedup for the tessellated solid is available even in scalar/single-particle mode and does not require basket input. However, processing baskets can still be beneficial due to positive cache effects.\end{sloppypar}

\begin{sloppypar}VecGeom has been discussed and presented in various publications~\cite{Apostolakis:2015:ACAT,vecgeom-SIMD-multifaceted,vecgeom-SIMD-navigation}. The following sections briefly review a few important results for specific aspects of VecGeom.\end{sloppypar}

\subsubsection{The performance of geometry primitives}
\begin{figure}[t]
\includegraphics[width=\columnwidth]{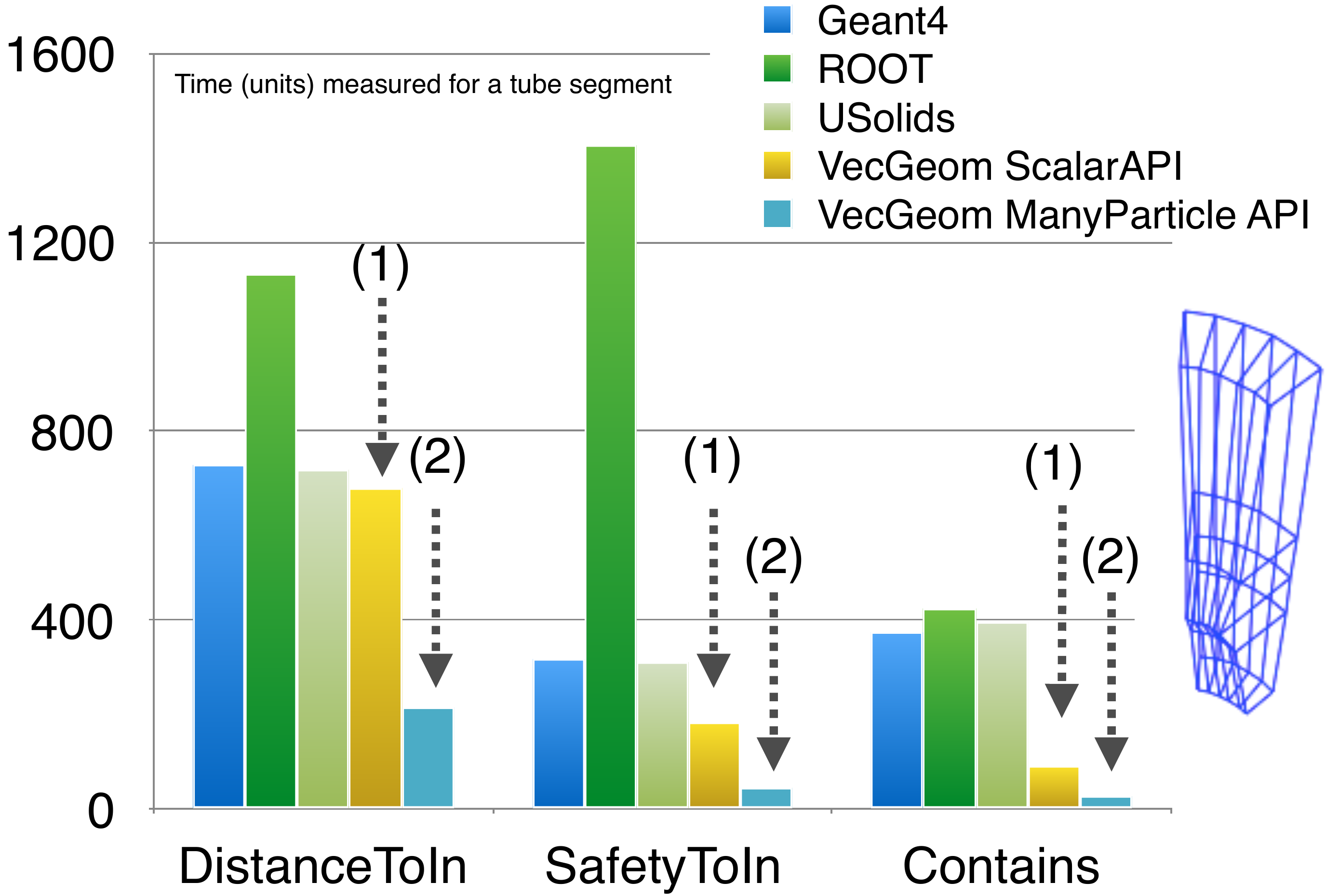}
\caption{\label{fig:shapebench1}Performance examples of VecGeom on the shape level for the case of a more elementary solid primitive, a tube segment. This demonstrates the performance
improvement of important functions for the one-particle interface (1) (better algorithms) as well as an additional SIMD acceleration
for the basket interface (2), automatically obtained by instantiating the same underlying kernel with Vc vector types.}
\end{figure}

\begin{figure}[t]
\includegraphics[width=\columnwidth]{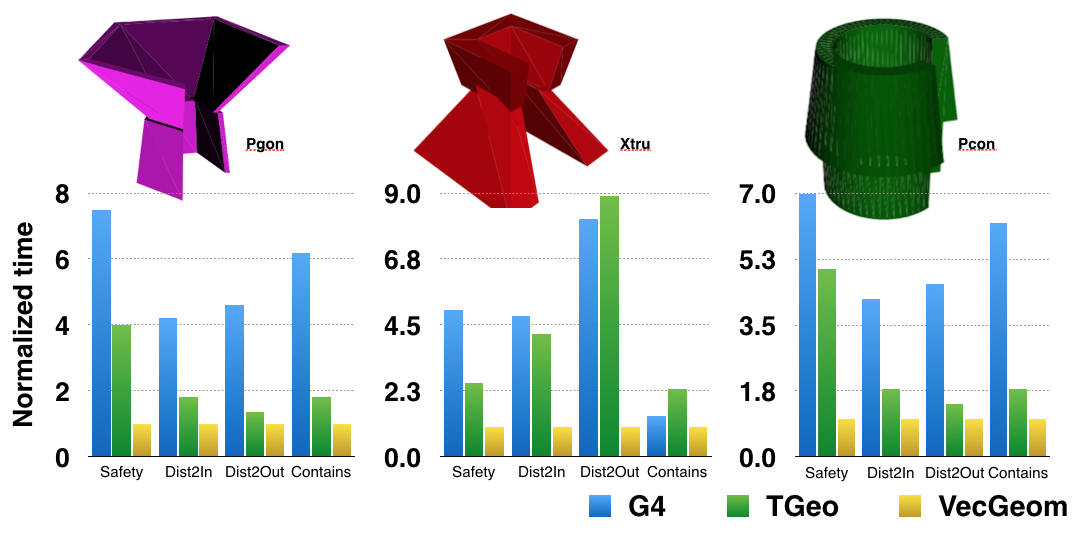}
\caption{\label{fig:shapebench2}
Performance speedups of more complex primitives (polycone, extruded solid, polyhedron) 
for scalar interfaces, compared with Geant4 and TGeo. Speedups are averages over all such solids found in the ALICE detector. In these complex cases, no additional basket SIMD acceleration is feasible.
}
\end{figure}
Geometry primitives (or solids) are, in addition to affine transformations, the basic building blocks of complex detectors.
The range goes from simple structures such as boxes, tubes, and cones, to more complex entities such as polycones, polyhedrons, and tessellated solids (see, e.g., the GDML reference manual~\cite{GDML} for a description).
In general, VecGeom offers improved performance of the solid algorithms 
with respect to previous implementations in Geant4 and TGeo and even with respect to USolids~\cite{Gayer:2012}. 
In most cases, the improvement is due to better algorithms, often as a natural consequence of the effort to restructure towards SIMD-friendly code.
In the case of simpler geometry primitives, the implementations provide real SIMD acceleration for basketized usage.
Figure~\ref{fig:shapebench1} exemplifies this for a tube segment, where the SIMD acceleration was found to be a factor of $2$ or better with the AVX instruction set (maximum of $4$ vector lanes) via the use of VecCore and Vc.

For the more complex solids, some performance improvements for the scalar interface are shown in Fig.~\ref{fig:shapebench2}.
In these cases, an additional SIMD acceleration for the basket interface is not feasible due to divergent code paths taken by different particles in a basket. However, as mentioned, the vector units can often still be utilized by vectorizing inner loops or inner computations.
This technique is used heavily in the tessellated solid, polyhedra, and multi-union cases~\cite{vecgeom-SIMD-multifaceted}, and it contributed to the excellent performance
gain compared to previous implementations. 

\subsubsection{The performance of navigation algorithms}
\label{sec:navigation_performance}
\begin{table*}[t]
\centering
\caption {\label{table:vectornav} The time (in seconds) to navigate a batch of test particles
in selected volumes of the CMS detector, and speedup factors for selected methods.
The time is dominated by the {\tt ComputeStep} navigation
interface of VecGeom in both the single-particle and basket (vector) mode.  The test is done using
the generic (normal) implementation of the navigator algorithm, as well as with specialized, generated code
that is tailored to the specific volume in question.
There is a small SIMD acceleration observed for baskets in simple volumes. This SIMD benefit can be
enhanced with specialized navigators. In case of complex volumes (with many containing other volumes), there is no
SIMD acceleration from the basket treatment. 
}
\begin{tabular}{l l c r c r c r c}
\hline\noalign{\smallskip}
Volume & Type & Normal Scalar & \multicolumn{2}{c}{Normal Vector} & \multicolumn{2}{c}{Specialized Scalar} & \multicolumn{2}{c}{Specialized Vector} \\

       &      &   time (s)         &  time (s)  & speedup & time (s) & speedup & time (s) & speedup \\

\noalign{\smallskip}\hline\noalign{\smallskip}
HVQX & simple & $12.6$ & $10.6$  & $1.2$ & $6.4$ & $2.0$ & $4.7$ & $2.7$ \\
ZDC\_EMFiber & simple & $10.1$ & $8.8$ & $1.2$ & $5.9$ & $1.7$ & $2.6$ & $3.9$ \\
ZDC\_EMLayer & complex & $27.0$ & $27.0$ & $1.0$ & $19.7$ & $1.4$ & $19.3$ & $1.4$ \\
\noalign{\smallskip}\hline
\end{tabular}
\end{table*}
\begin{table*}[t]
\centering
\caption{\label{table:simdbvh} The time (in seconds) to process all
  test rays for a list of complex detector volumes, and speedup factors with respect to Geant4. The worst time value for each volume is shown
in red, while the best is in blue. VecGeom's SIMD-enabled navigation performs consistently better than any existing solutions.}
\begin{tabular}{l l c r c r c r c}
\hline\noalign{\smallskip}
Volume & \# daughters & Geant4 & \multicolumn{2}{c}{TGeo}& \multicolumn{2}{c}{VG (SSE4.2)} & \multicolumn{2}{c}{VG (AVX2)} \\
 & & time (s) & time (s) & speedup & time (s) & speedup & time  (s) & speedup \\
\noalign{\smallskip}\hline\noalign{\smallskip}
ALIC (ALICE) & $65$ & $0.74$ & {\color{red} $1.07$} & 0.69 & $0.30$ & 2.47 & {\color{blue} $0.23$} & 3.22  \\
TPC\_Drift (ALICE) & $641$ & {\color{red} $14$} & $2.2$ & 6.3 & $1.2$ & 11.7 & {\color{blue} $0.9$} & 15.6  \\
MBWheel\_1N (CMS) & $789$ & $0.84$ & {\color{red} $1.09$} & 0.77 & $0.49$ & 1.71& {\color{blue} $0.35$} & 2.40  \\ 
\noalign{\smallskip}\hline
\end{tabular}
\end{table*}

Apart from solid primitives, VecGeom offers navigation algorithms for solving geometry problems such as distance calculations between particles and geometry boundaries in composite geometry scenes made up of many primitives. These navigation algorithms are the primary point of contact 
or interface between the geometry and the simulation engine. The algorithmic chain in Fig.~\ref{fig:VecGeom_templatekernels} is a simplified example of a typical navigation algorithm flow. This chain contains transformations of global particle coordinates to the frame of reference of the volume in which the particle is currently situated, and performs distance queries to the solids embedded in this volume.

Just as with the geometry primitives, complexity defines the performance scenario for SIMD acceleration.
\begin{enumerate}
\item \textbf{Simple geometry limit:} In this case the current volume contains only a few (simple) solids,
  e.g. in simple showering modules in calorimeters.

  In this limit, SIMD-accelerated geometry navigation of a basket is feasible for GeantV because
  most of the algorithmic chain can process baskets efficiently. Table~\ref{table:vectornav}
  gives a few benchmark performance values that show the gain from using baskets and SIMD for simple volumes. 
  In the generic case, the gain is rather modest because the SIMD throughput is limited by
  some non-vectorizable parts. The typical non-vectorizing operation is particle relocation after crossing boundaries, which becomes more expensive with increasing complexity, due to divergence of the location at the end of the step and the need to access a larger amount of non-local transformation matrix and 3D solid data. However, a process~\cite{vecgeom-SIMD-navigation} was developed that can auto-generate code implementing specialized navigators that take into account the specific properties of the geometry.
  This generated code can reduce the non-vectorizable parts significantly. 
  This increases the gain from baskets and SIMD, but is also beneficial in its own right to improve the performance
  of the scalar interface. The drawback of specialized navigators is that they require a generation workflow to be run
  before simulation, and so far they have not been extensively tested within GeantV. This explains, in part, why the overall gain from baskets in the current version of GeantV did not materialize for geometry navigation. Navigation specialization requires the analysis of all possible geometry state transitions for tracks crossing any placement of a given volume (possibly replicated) to its neighbours. Cached in the form of compiled code, the method performs global to local conversions between states. If the number of transitions is small (e.g. in the case of well-packed touching neighbour volumes), the number of crossing candidates to check can be much smaller than in the generic case, so the search can be accelerated. For complex structures, the number of combinations can lead to very large libraries with inefficient instruction caching.
  The navigation specialization approach is nevertheless very promising and will continue to be optimized in the context of future VecGeom developments. 

\item \textbf{Complex geometry limit:} In this case, the current volume contains many solids, which typically occurs for container volumes inside which many other modules are placed.

  In this limit, due to the large number of geometry objects to test, acceleration structures are typically used to reduce
  the complexity from $O(N)$ to $O(log(N))$, in the case of hit-detection or ray-tracing, where $N$ is the number of geometry objects present in this volume.
  This, in turn, makes it difficult to achieve a coherent instruction flow for all particles in a basket and to avoid
  branching. However, as for the tessellated solid, the navigation algorithms can benefit from SIMD acceleration
  via internal vectorization. In Ref.~\cite{vecgeom-SIMD-navigation}, a particular \emph{regular} tree data structure, based on bounding boxes, was proposed,
  which can be traversed with a SIMD speedup. The VecGeom implementations for tessellated solids and for navigation are based on the same data structure.
  Table~\ref{table:simdbvh} shows a comparison of the performance of the navigation algorithms in given complex volumes using Geant4, TGeo, or VecGeom. The benefit of the SIMD speedup is highlighted by the additional gain when switching from SSE4 to AVX2 instructions on the x86\_64 architecture.
  This benefit is also available in non-basketized modes via internal vectorization.

  There are many possible layouts of acceleration structures with SIMD support. This gives room
  for further improvement by selectively choosing the best possible acceleration structures for any given geometry volume.
  In this respect, VecGeom is ready to interface with kernels available from industrial ray-tracing libraries, such as Intel{\textregistered} Embree~\cite{Embree,Wald-Embree-2014}, which
  has SIMD support.
\end{enumerate}

\subsection{VecMath}
VecMath is a library that collects general-purpose mathematical utilities with SIMD and SIMT (GPUs) support based on VecCore. Templated fast math operations, pseudorandom number generators, and specific types (such as Lorentz vectors) were initially extracted from GeantV, then developed and extended within VecMath. The library is being extended to support vector operations for 2D and 3D vectors, and general-purpose vectorized algorithms. VecMath is intended as a core mathematical library, free of external dependencies other than VecCore and usable by vector-aware software stacks. 

\subsubsection{Fast Math}
The \texttt{Math.h} header in the VecMath library contains templated implementations for {\textit{FastSinCos}}, \textit{FastLog}, \textit{FastExp}, and \textit{FastPow} functions. The functions can take either scalar or SIMD types as arguments. While the scalar specializations redirect to the corresponding \textit{Vdt}~\cite{ref:vdt} implementations, the SIMD specialization is currently implemented based on Vc types. 

\subsubsection{Pseudorandom number generation}
\begin{sloppypar}The VecRNG class of VecMath provides parallel pRNGs (pseudorandom number generators) implementations for both SIMD and SIMT (GPU)
workflows via architecture-independent common kernels, using backends provided by VecCore.
Several state-of-the-art RNG algorithms are implemented as kernels supporting parallel generation of random numbers in scalar, vector, and CUDA workflows.
For the first phase of implementation, the following representative
generators from major classes of pRNG were selected: MRG32k3a~\cite{ref:LEcuyer2002}, Random123~\cite{ref:random123}, and MIXMAX~\cite{ref:MIXMAX}. These generators meet strict quality requirements, belonging to families of generators that have been examined in 
depth~\cite{ref:LEcuyer1996} or that have evidence from ergodic theory of 
exceptional decorrelation properties~\cite{ref:MIXMAX}. All pass major 
crush-resistant tests such as DIEHARD~\cite{ref:DIEHARD} and BigCrush of 
TestU01~\cite{ref:TestU01}.  In addition, 
constraints in the size of the state and the performance were considered: 1) a very long period {\small ($\geq 2^{200}$)}, obtained from a small state (in memory), 2) fast 
implementations and repeatability of the sequence on the same hardware 
configuration, and 3) efficient ways of splitting the sequence into long disjoint 
streams.\end{sloppypar}

The design choice for the class hierarchy was the exclusive use of static polymorphism, motivated by performance considerations.  
Every concrete generator inherits through the CRTP (curiously recurring
template pattern) from the VecRNG base class, which defines mandatory methods 
and common interfaces.  VecRNG is exclusively implemented in header files, and provides a minimal set of member methods.  This approach allows more flexibility in the higher-level interfaces for specific computing applications, but minimizes the overhead in compilation time. 

\begin{sloppypar}The essential methods of VecRNG interfaces are 
{\small{\textit{Uniform$<$Backend$>$()}}} and {\small{\textit{Uniform$<$Backend$>$(State\_t{\rm \&} s)}}},
which generate the backend type of double-precision u.i.i.d (uniformly
independent and identically distributed) numbers in [0,1), and update the internal {\it fState} and the given state $s$, respectively. The {\it State\_t} is defined 
in each concrete generator and provided to the base class through {\textit{RNG\_traits}}.
One of the associated requirements for each generator in VecRNG is to provide
an efficient skip-ahead algorithm, $s_{n+p} = f_p(s_n)$ (advance a
state $s_n$ by $p$ sequences, where $p$ is the unit of the stream length or an
arbitrary number) in order to assign disjointed multiple streams for different
tasks.
For example, the mandatory method, {\textit{Initialize(long n)}}, moves
the random state at the beginning of the given $n^{th}$ stream.
Each generator supports both scalar and vector backends with a common kernel.
Random123 has an extremely efficient stream assignment without any additional
cost since the key serves as the stream index, while MRG32k3a uses transition
matrices ($A$), which recursively evaluate $(A^{s}~{\rm mod}~m)$ using the
binary decomposition of $s$. The vector backend uses N (SIMD length)
consecutive substreams and also supports the scalar return type, which
corresponds to the first lane of the vector return type. Besides the Uniform method, some commonly used random probability distribution functions are also provided.\end{sloppypar}

\subsection{GeantV tracking and navigation}
\label{sec:tracknav}

GeantV implements basketized vectorization of geometry navigation queries following the workflow described in Section~\ref{sec:vec_workflow}. Geometry ``baskets'' are passed to a top-level navigation API, then dispatched to VecGeom to benefit from its vectorization features as described in Section~\ref{sec:navigation_performance}. The geometry queries are integrated into the stepping procedure in a special \textit{GeomQuery} stage, providing a large number of handlers, one per logical volume in the user geometry. Each query for computing the distance to the next boundary and safety within the current volume can be executed in either scalar or vector mode. The efficiency in the vectorized case depends strongly on the volume shape and number of daughters. The track position and direction data are internally gathered into SOA data structures by VecGeom and dispatched to the 3D solid algorithms, updating navigation states held by the GeantV track structure. Even in the scalar calling sequence, VecGeom vectorizes the calls to the internal navigation optimizer.

An initial attempt to basketize and call the vector \textit{DoIt} method for all volumes in a complex geometry such as CMS proved to be inefficient. The main reason was not VecGeom vectorization inefficiency, but the impact of SIMD basketizing on the GeantV workflow. CMS geometry has O(4K) volumes, out of which $\sim80\%$ of the steps are performed in only $\sim10\%$ of the volumes. In a GeantV vector flow scenario, after an initial propagation out of the central vertex volumes, most tracks become isolated in SIMD baskets belonging to many different, less important volume handlers. The workflow enters starvation mode and has to force frequent flushes of these baskets and execution of geometry queries in scalar mode. The effect worsens for complex geometry setups, typical for detectors at the LHC. In simple setups, composed of just a few geometry volumes, this scenario does not happen and basketization gains are evident in the case that the geometry code takes a sizeable fraction of the execution time.

A new \textit{dynamic basketizing} feature is implemented to alleviate this effect. Initially all volume basketizers are switched to ON, but the frequency of flushes versus vectorized executions is measured and triggers scalar mode for inefficient baskets. Depending on the tuned efficiency threshold, the prototype will end up disabling most volume basketizers and keeping only about 5\% active. Due to the fact that some shapes with intensive computation (such as polycones and polyhedra) are not vectorized in VecGeom in multi-particle mode, the overall vectorization efficiency is rather poor and is reduced further by scatter/gather overheads.

Related to the geometry, GeantV uses a different strategy per step for the boundary crossing algorithm in a magnetic field, compared to Geant4. The algorithm first estimates the deviation of the particle moving with a given step along the helix arc using a small angle approximation, compared to a straight-line propagation with the same step. Constraining this bending error to be less than an acceptable tolerance gives the maximum allowed step in magnetic field, $\delta_{\mathrm{field}}$. The geometry navigation interface is queried for the distance to the next boundary along a straight line, $\delta_{\mathrm{boundary}}$, as well as the isotropic safe distance within the current volume, $\delta_{\mathrm{safe}}$. The propagation step is first constrained by the minimum between the physics step limit, $\delta_{\mathrm{physics}}$, and the next boundary limit, $\delta_{\mathrm{boundary}}$. The second constraint is the maximum between the magnetic field limitation and the safety. This allows particles with small momenta to ignore trajectory bending in the limit of nearby volumes, and particles with large momenta to ignore nearby volumes and travel forward much farther in the case that the deflection is small. In practice, this allows larger steps to be taken near volume boundaries without the risk of crossing accidentally. Finally, the step to be taken is the minimum of either the geometry or the physics limit, within the field/safety constraint. This step is passed to the integrator algorithm to move the track to the new position. If it was the geometry that limited the step, the geometry is queried for a possible final relocation after the propagation; otherwise, the algorithm is repeated until either the physics or geometry distance limits are reached. Note that after the track kinematics are updated, tracks limited by geometry that have not completed crossing into the next volume are copied back into the geometry stage basket and considered in a subsequent execution of this stage. The tracks having reached the next boundary are forwarded to the \textit{PostPropagation} stage, as shown in Fig.~\ref{fig:simulation_stages}.

\subsection{EM Physics models and vectorization}

The ultimate goal of the GeantV R\&D project is to exploit the possible computational benefits of applying vectorization techniques to HEP detector simulation code. From the physics modeling point of view, the most intensively used and computationally demanding 
part of these simulations is the description of electromagnetic (EM) interactions of $e^-$, $e^+$, and $\gamma$ particles with matter. This is what motivated the choice of 
EM shower simulation code to demonstrate the possible computational benefits of applying track-level vectorization.

Geant4 provides a unique variety of EM physics models to describe particle interactions with matter~\cite{ref:Geant4-gen3}, from the eV to PeV energy range, with different levels of physics accuracy. 
Each application area can find a suitable set of models with the appropriate balance between the accuracy of the physics description and the 
corresponding computational complexity. Moreover, Geant4 provides a pre-defined collection of EM physics models and processes for different application areas in the form of EM 
physics constructors~\cite{Geant4:PhysListGuide}. Among these, the so-called EM \textit{standard} physics constructor (sometimes called EM Opt0) is recommended by the developers for HEP detector simulations.

A corresponding set of EM models has been provided for the GeantV transport engine,
together with the appropriate physics simulation framework.
The accuracy of each GeantV model implementation was carefully tested through individual, model-level tests by comparing the computed final states and integrated 
quantities (e.g. cross-sections, stopping power) to those produced by the corresponding Geant4 version of the given model. Moreover, several simulation applications have been developed to test and verify the GeantV EM shower simulation accuracy, including both a general, simplified sampling calorimeter and a complete CMS detector setup. In all cases, the GeantV simulation results, measured using quantities such as energy deposit distributions in a given part of the detector, number of charged and neutral particle steps, secondary particles, etc., agreed with the corresponding Geant4 simulation results to within 0.1\% (see more in Section \ref{sec:ApplicationsAndValidations}).

The final state generation, or interaction description, pieces of these models are the most computationally demanding subset of the physics simulation. At the same time, they provide the physics code that can be the most suitable for track level vectorization. 
The final state generation usually includes the generation of stochastic variables, such as energy transfer, scattering, or ejection angles, from their probability distributions, determined by the corresponding energy or angle differential cross sections (DCS) of the underlying physics interaction. The probability density functions (PDF) are proportional to the DCS, which is usually a complex function and very often only available in numerical form. This implies that the analytical inversion of the corresponding cumulative distribution function (CDF) is unknown. For this reason, to generate samples of the stochastic variables needed to determine the final state of a primary particle that underwent a physics interaction, different numerical techniques have to be used.

The composition-rejection method, for example, is one of the most extensively used method in particle transport simulation codes to generate random samples according to a given PDF. However, it is not very suitable for vectorization, being based on an unpredictable number of loop executions depending on the outcome of the random variable. This implies that if this algorithm was vectorized over primary tracks, the different lanes of the vector would reach exit conditions in a non-deterministic way, at different moments, thus reducing the number of used lanes, eventually causing a loss of any potential computational gain. 
For this reason, special care was taken to find new sampling algorithms, more suitable for vectorization, and also to implement solutions that could provide the maximum possible benefits of track-level vectorization, even when applied to existing and well known sampling techniques.

As an outcome of this R\&D phase, two methods were implemented and tested in the GeantV EM Physics library. In GeantV jargon, the first method is known as ``sampling tables'', which makes it possible to use an alias method in combination with sampling tables, thanks to the introduction of a discrete random variable as an intermediate step. The second method is known as ``lane-refilling rejection'', or simply ``rejection'', which takes advantage of vectorization even in the presence of non-deterministic sampling techniques. More details about the implementation of the EM physics models and their vectorization can be found in Ref.~\cite{EMPhysics:CHEP2018}. 

\begin{figure}
    \centering
    \includegraphics[width=\columnwidth]{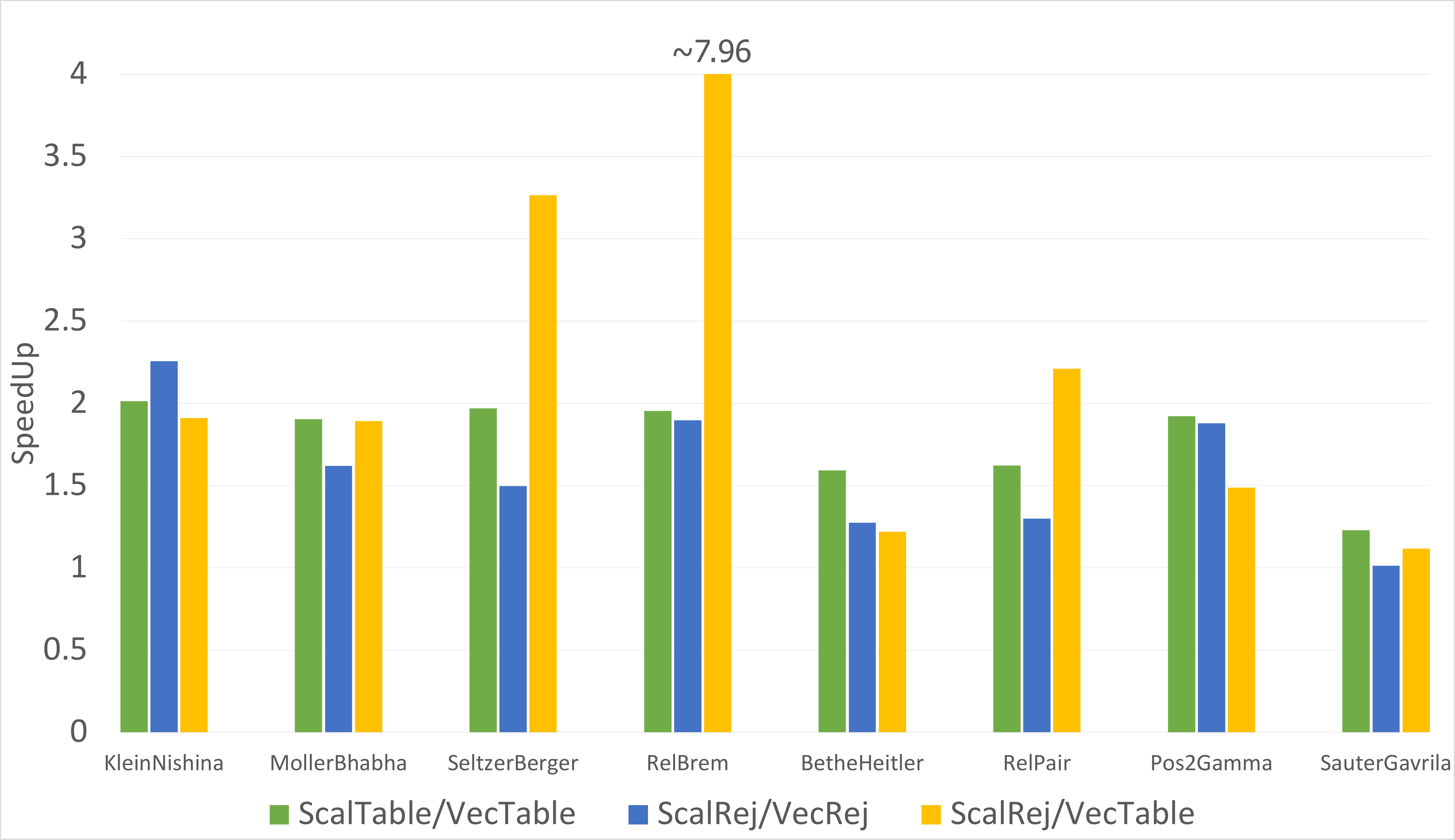}
    \caption{Speedup of the final state generation of different electromagnetic physics models obtained with SIMD vectorization in case of different sampling algorithms. The results were obtained by using Google Benchmarks~\cite{googlebenchmarks} on an Intel{\textregistered} \thinspace Haswell Core\textsuperscript{TM} i7-6700HQ, 2.6 GHz, with the Vc backend and AVX2 instruction set processing 256 tracks. ``Scal'' and ``Vec'' refer to scalar and vector implementations, while ``Table'' and ``Rej'' refer to the sampling table and rejection methods, respectively.}
    \label{fig:VecPhys:PerformanceGeneral}
\end{figure}

Using the above mentioned unit tests to analyze the  
performance of the vectorized EM models compared to their (optimized) scalar versions, excellent 1.5--3$\times$ and 2--4$\times$ vectorization gains were achieved on Haswell and Skylake architectures, respectively. The instruction set used on both architectures was AVX2, since AVX512 was not supported by the Vc backend. Figure~\ref{fig:VecPhys:PerformanceGeneral} shows the speedup of the final state generation of different electromagnetic physics models obtained with SIMD vectorization in the cases of the two different methods. 

As a result of these developments, the physics simulation part of the GeantV R\&D project provides the possibility of exploiting the benefits offered by applying track-level vectorization on a complete EM shower simulation suitable for HEP detector simulation applications.   

\begin{figure}
    \centering
    \includegraphics[width=\columnwidth]{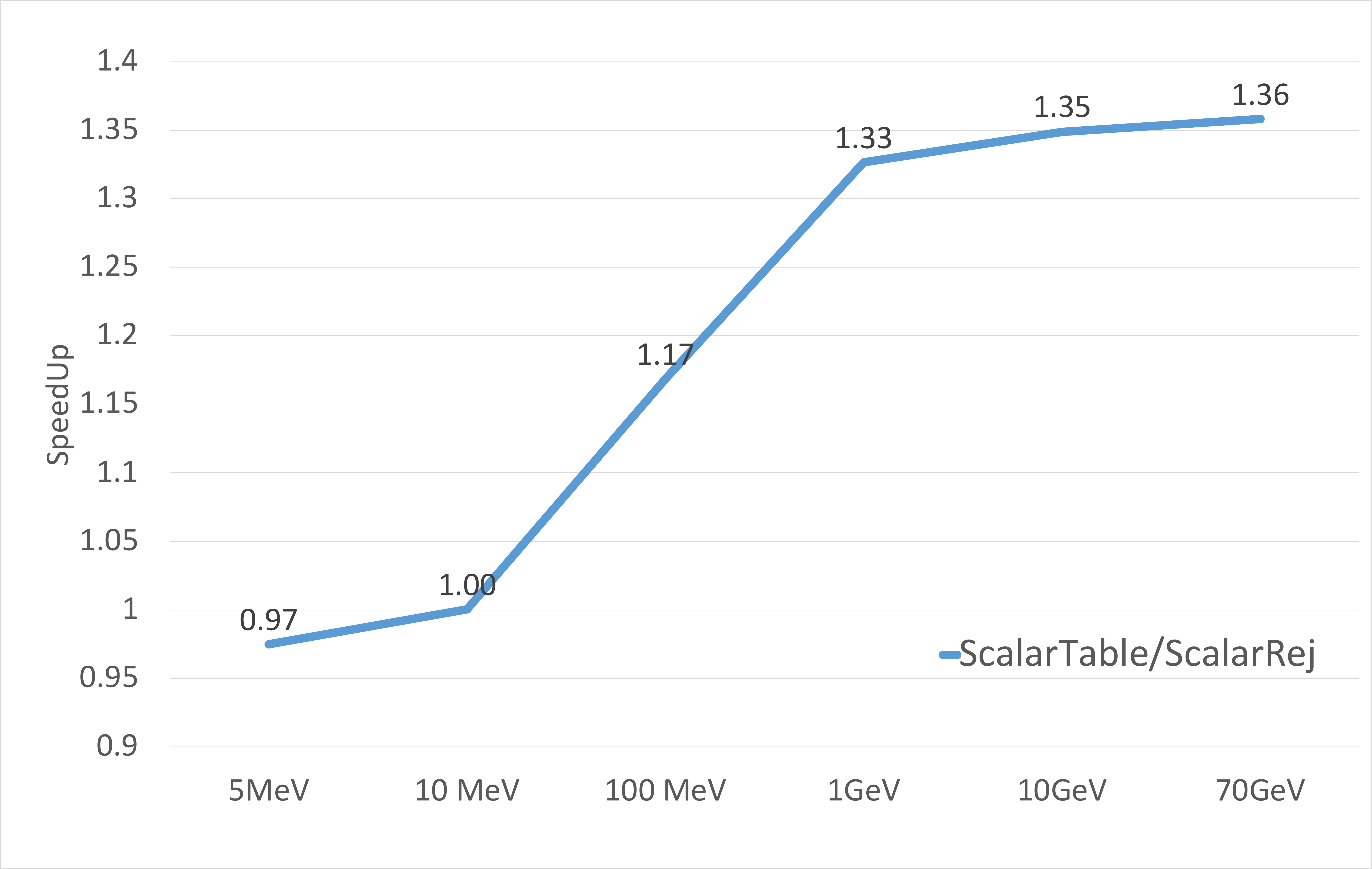}
    \caption{Speedup of the \textit{rejection}-based final state sampling compared to the \textit{sampling table}-based one in case of the Bethe-Heitler $e^-/e^+$ pair production model.}
    \label{fig:VecPhys:BetheHeitlerRejVsAlias}
\end{figure}

A relatively wide range of performance variation in the algorithms and their vectorization gains is observed. This is due to the fact that each of the EM physics models under study translates to a final state sampling algorithm with unique computational characteristics more favorable for one sampling technique or the other.
In addition to this, while the sampling table-based final state generations have a constant run time under any external conditions, the efficiency of a given rejection algorithm can change significantly with the primary particle energy. This is illustrated in Fig.~\ref{fig:VecPhys:BetheHeitlerRejVsAlias} that shows the relative speedup of these two techniques applied to the Bethe-Heitler $e^-/e^+$ pair production model, as a function of the primary $\gamma$  particle energy. The two algorithms perform similarly at lower energies, while the rejection algorithm becomes ${\sim}35\%$ faster at higher $\gamma$  energies simply because of the smaller rejection rate.

The results shown indicate that the GeantV vectorized EM physics library has to be tuned to select the most efficient algorithm for any specific physics process, depending on the specific conditions. The complexity of the underlying DCS, the target material composition, or the primary particle energy are examples of conditions that can heavily affect the physics performance. 
The GeantV physics framework has been designed to take all of these considerations into account and to allow the choice of the most efficient algorithm for final state generation, depending on the primary particle energy or detector region. This makes it possible to obtain the maximum possible performance, while keeping the memory consumption of the algorithms under control, even in the case of the most complex HEP detector simulation applications.

\subsection{Magnetic field integration}

The integration of the equations of motion for a charged particle in a non-uniform pure magnetic field (or an electromagnetic field) accounts for about 15--20\% of the CPU time of a typical HEP particle transport simulation.
This integration is typically performed using the family of Runge-Kutta methods,
which involves the generation of multiple intermediate states $(x, p)$,
and the evaluation of the field and the corresponding equation of motion.
Many floating point operations are carried out for each step of each track,
providing substantial work for each initial data point, but without
any expensive functions such as logarithms or trigonometric functions.
In GeantV, the input to the field propagation stage is a basket of tracks.
Each track has a requested step length for integration (see Section~\ref{sec:tracknav}), obtained from the
physical step size, the distance to the nearest boundary, and the curvature of
the track (to avoid missing boundaries).

The tracks' positions are typically scattered throughout the detector.
The integration of separate tracks is carried out in separate 
vector lanes in order to create the most portable code,
and to make the best potential use of vector units with different widths.
The vectorization of this part of a particle transport simulation has an important requirement.
All steps of charged particles must be integrated, so long as the step can affect the deposition of energy or other quantities measured.

The lower level parts of the integration can be fully vectorized, because the
operations proceed in lockstep, synchronously over the lanes of a vector with each lane corresponding to a different track:
\begin{itemize}
\item the evaluation of the EM field at each track's current or predicted location, either using interpolation (as in our benchmark example) or other methods such as the evaluation of a function;
\item the evaluation of the `force' part of the equation of motion using the Lorentz equation;
\item and a single step of a Runge-Kutta algorithm, which provides an estimate of the end state of a track (position, momentum) and the error in this estimate.
\end{itemize}

The top level of Runge-Kutta integration involves
checking whether the estimated error conforms to the required accuracy and 
checking if a successful step finishes the required integration interval.
If the integration goes on, it must also calculate the size of the next integration step. 
Different actions are required depending on whether a step succeeded or not.
In case a step was not successful, integration must continue for those tracks.
A lane with a finished track, or one that reached the maximum allowed number
of integration substeps, must be refilled from the potentially remaining pool of tracks in the
basket.

Since all charged particles are involved, there is a large population of particles undertaking integration steps during a simulation.  
Larger-size baskets to accumulate work in field integration were created, and can
be configured separately from the general basket size.
With larger baskets, the fraction of lanes doing useful work increases substantially, getting close to unity, as shown in Table~\ref{tab:active_lanes}.

\begin{table}[t]
\centering
\caption{Fraction of `inactive' lanes, in which the integration has already finished, for different values of the basket size. Two configurations were measured: the default, in which tracks were processed in the original order of the basket, and the other (`pre-processed'), in which selected lanes with estimated work (length $s$ over radius of curvature $R$) above a threshold value were brought to the front of the basket. Measurements are from single-threaded simulations of 100 events, each with 10 
primary electrons of 10 GeV energy in the CMS setup. The threshold used was $s/R > 3$. }
\label{tab:active_lanes}   
\begin{tabular}{rrr} \hline\noalign{\smallskip}
            &  \multicolumn{2}{c}{Percentage of idle lanes} \\
Basket size &   Default  &  Pre-processed \\ \noalign{\smallskip}\hline\noalign{\smallskip}
      16    & 18.6   & 14.0 \\
      32    & 13.0   & 6.6 \\ 
      64    & 7.3   & 2.5 \\ 
     128    & 3.9   & 0.3 \\ 
     256    & 2.3   & 0.0 \\ 
     512    & 1.5   & 0.1 \\ 
    1024    & 0.7   & 0.1 \\ \noalign{\smallskip}\hline
\end{tabular}
\end{table}

\begin{figure}
    \centering
    \includegraphics[width=\columnwidth]{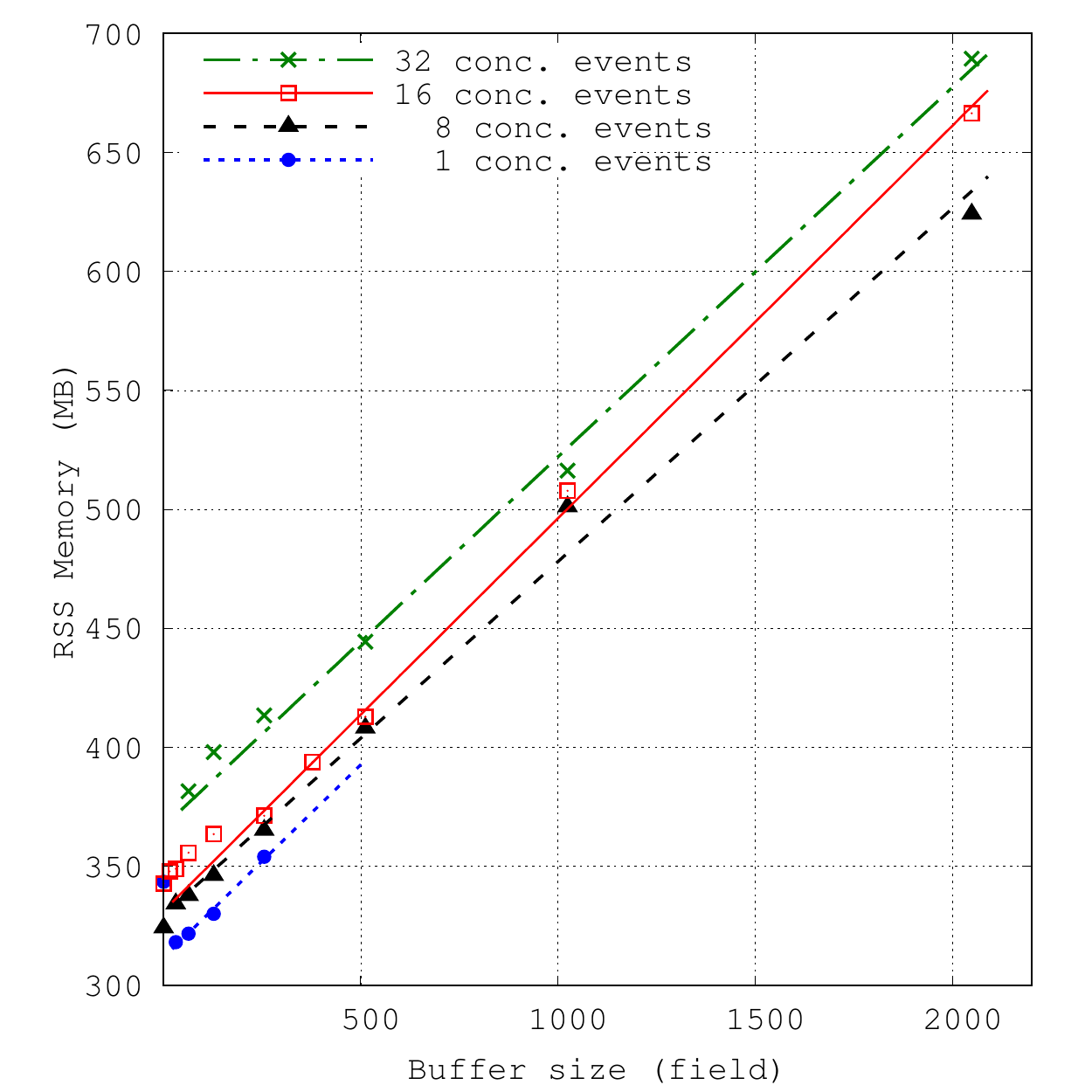}
    \caption{Memory size of a simulation versus the basket size for field propagation. 
             Simulations with different number of simultaneous events are shown. 
             The number of primaries per event is also varied, using 16 (default) and 8.
             All are single-threaded simulations of 100 events, each with 16 primary electrons with 10 GeV energy in the CMS setup.   
             The simulations were run on a MacBook Pro 2016 with 16GB RAM running MacOS 10.13.6.}
    \label{fig:memsize_field_baskets}
\end{figure}

Unfortunately, increasing the size of the buffer for field propagation has an additional effect, which counteracts the increase in efficiency from the reduction of idle lanes. 
It increases the number of simultaneous tracks in flight, as larger baskets accumulate more tracks. In turn, this increases the memory usage, which is proportional to the the number of tracks in flight.
The effects can be seen clearly in Fig.~\ref{fig:memsize_field_baskets}, where a linear relation between the buffer size for field propagation and
the memory usage is evident. It is possible, however, to reduce memory usage by starting fewer simultaneous events, as seen in the additional measurements with 1, 8, 16, or 32 simultaneous events. 

\subsection{Input and output (I/O)}

\subsubsection{Input}
\begin{figure}[ht]
    \centering
    \vspace*{-0.6cm}
    \includegraphics[width=\columnwidth]{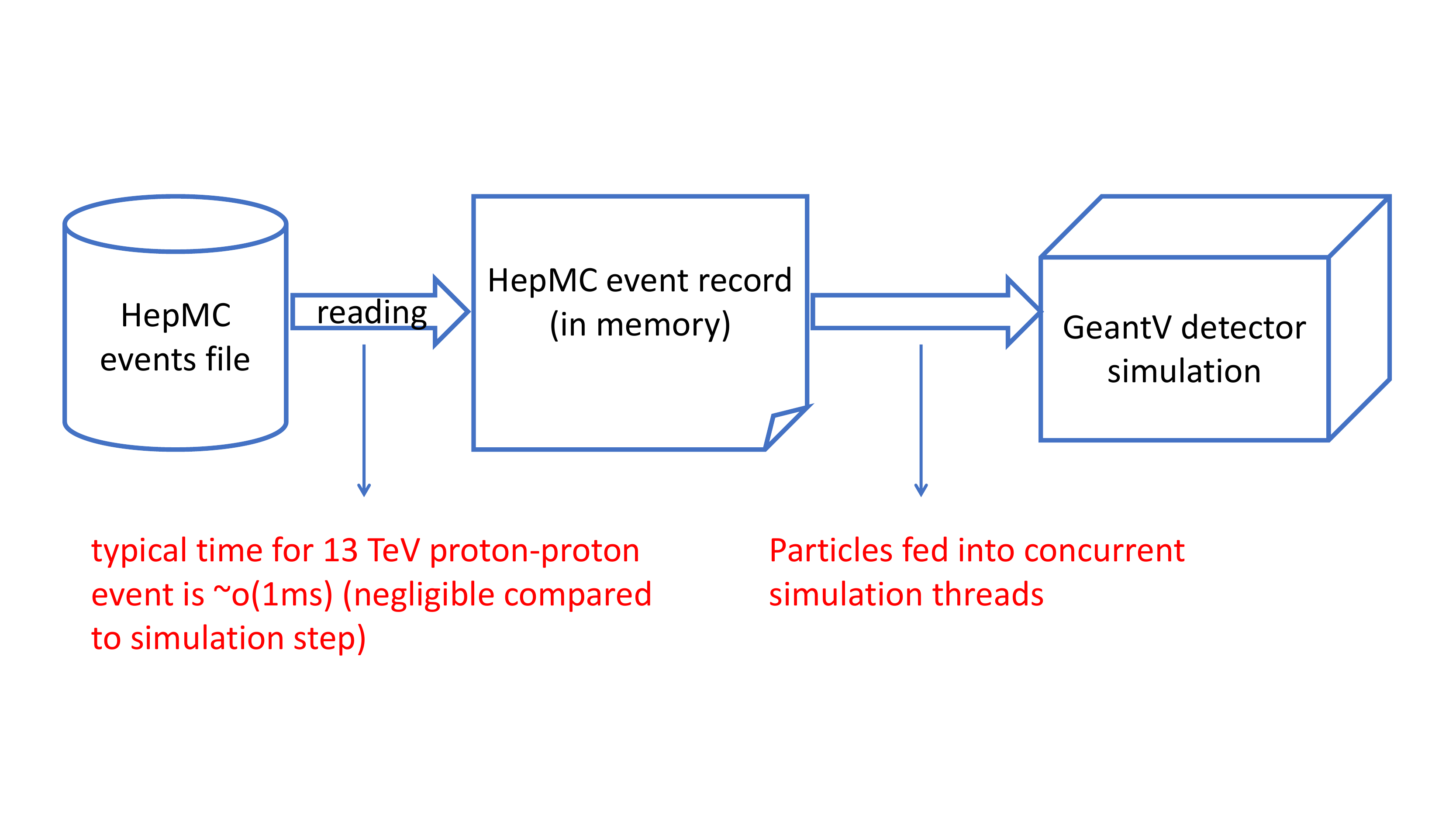}
    \caption{A diagram showing how the HepMC event record is used as the GeantV input format.}
    \label{fig:hepmctogv}
\end{figure}

Simulation input consists of particles to be transported through the detector. These can be either realistic collision events produced by Monte Carlo event generators or single particles (similar to a test beam) to study a particular response. The use of an interface (the so-called \emph{event record}) makes the generation and the simulation steps independent, as schematically shown in Fig.~\ref{fig:hepmctogv}. For the GeantV transport engine, it is irrelevant how the 'primary' (input particles) are produced. The simulation threads concurrently process particles from the input.

The interface to the HepMC3~\cite{Buckley:2019xhk} event record has been implemented (the \textit{HepMCGenerators} class). This interface can read both HepMC3 ASCII data and ROOT files containing serialized objects. The different types of input files are recognized by their file extensions. The interface selects the stable (outgoing) particles from the provided event and passes them to the transport engine. It can also apply optional cuts, for instance on $\eta$ (pseudorapidity), $\phi$ (azimuthal angle), or momentum.

\subsubsection{Output}
\begin{figure}[ht]
    \centering
    \includegraphics[width=\columnwidth]{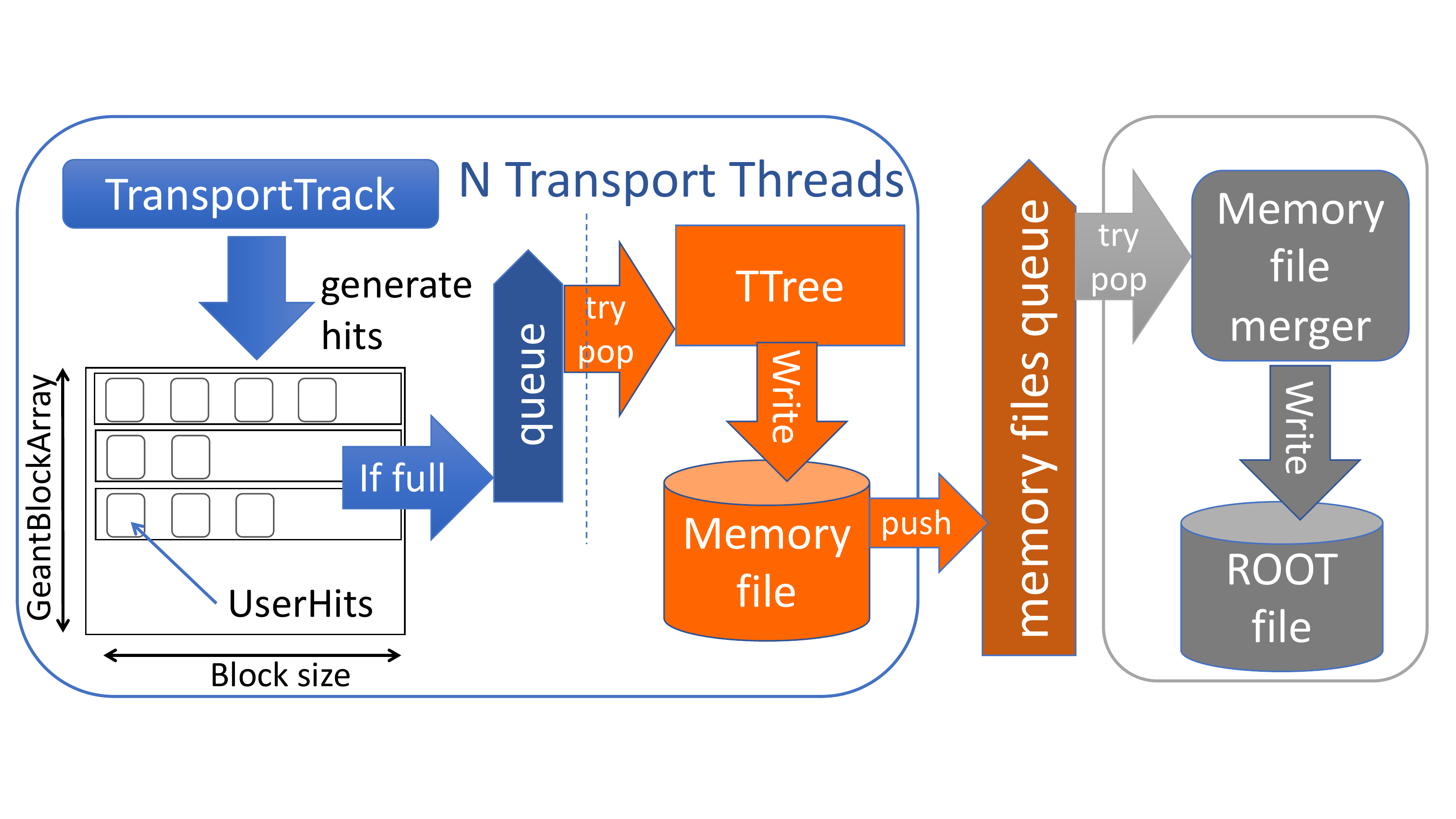}
    \caption{A diagram of the output architecture, which is based on ROOT's TBufferMerger.}
    \label{fig:gvtobuffer}
\end{figure}

The detector simulation produces \emph{hits} that contain energy deposition and timing information in the sensitive parts of the detector, which are the output of the program. In the case of GeantV, those hits are produced concurrently by all the simulation threads and need to be recorded properly. Thread-safe queues have been implemented to handle the asynchronous generation of hits from several threads simultaneously. The {\textit{GeantFactory}} machinery takes care of grouping the hits into so-called HitBlocks and putting them in the queues. Two possible approaches were tried for saving the hits into a file. In the first, all the hits produced by different threads were given to one `output thread' for serialization. This approach turned out not to scale properly and became a bottleneck. The problem was solved by the second approach, where the serialization was performed by each transport thread and the `output thread' was only responsible for the actual writing of the data to the file. This approach did not adversely affect the memory consumption in any visible way. The implementation is based on the {\textit{TBufferMerger}} class provided by ROOT~\cite{Brun:1997pa}. Each transport thread fills, in parallel, its ROOT TTree objects, and the TBufferMerger merges these TTrees and saves them into the file on disk, as shown in Fig.~\ref{fig:gvtobuffer}. 

\begin{figure}[ht]
    \centering
    \includegraphics[width=\columnwidth]{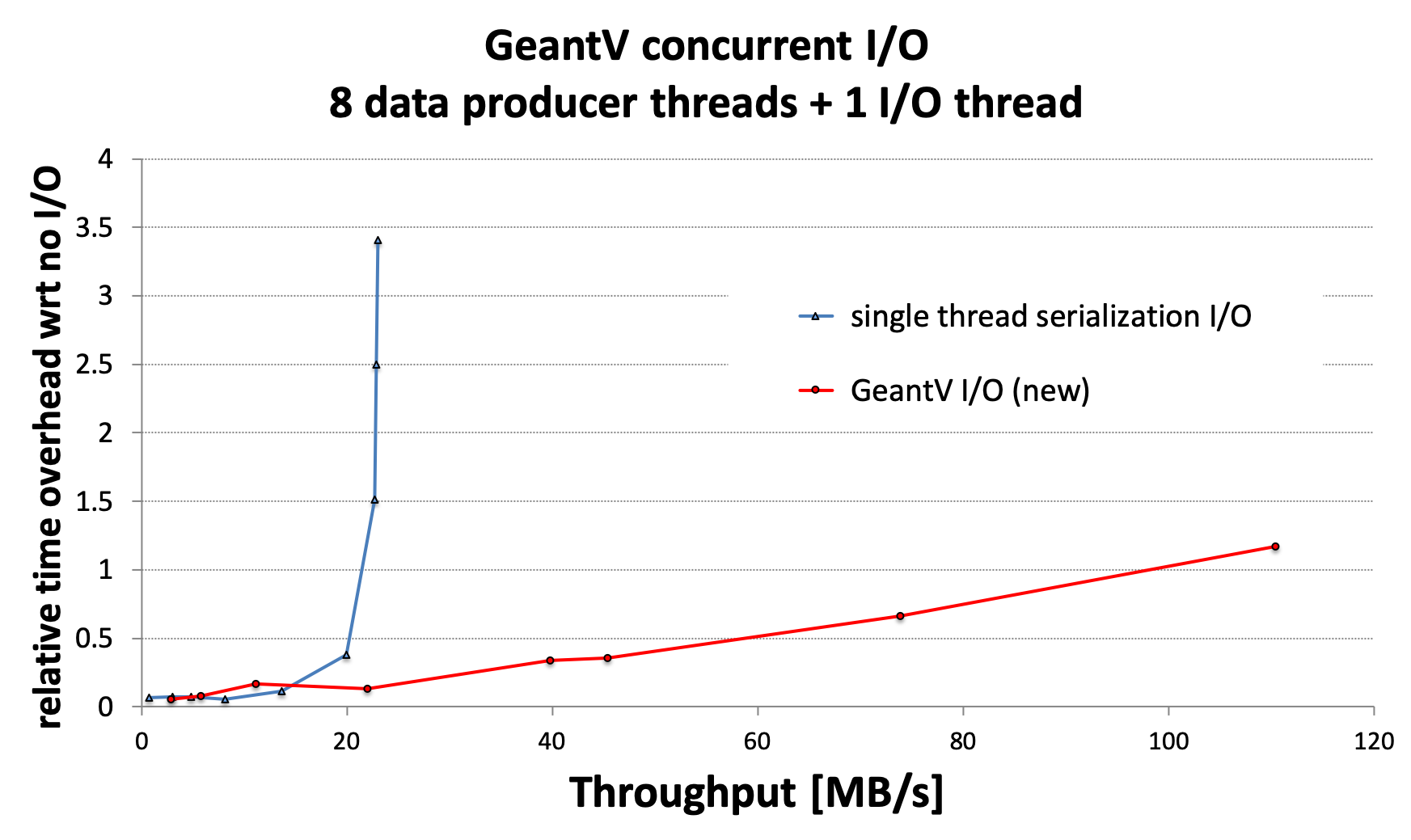}
    \caption{The I/O performance compared to the `single thread serialization' approach. These tests used an Intel{\textregistered} Xeon{\textregistered} CPU E5-2630 v3 @ 2.40GHz (Haswell), 2x8 cores, HT=2 (16 native threads, 32 in hyperthreading mode), disk: SSD ~430 MB/s non-cached write speed (measured with: dd if=/dev/zero of=/tmp/testfile bs=1G count=1 oflag=direct).}
    \label{fig:ioperf}
\end{figure}

This architecture has been profiled and shows very good scaling behavior, as seen in Fig.~\ref{fig:ioperf}. In particular, it solves the bottleneck problem of the `single thread serialization' approach. More details on the usage of the multithreaded output are provided in Sections~\ref{sec:scoreint} and~\ref{sec:exp-fwk-int}.

\subsubsection{MC truth}
In addition to the hits, users may be interested in saving the kinematic output, so-called \emph{MC truth} information. This consists of the generated particles (or at least some of them) that produced those hits. The handling of MC truth is quite detector dependent and no general solution exists. The algorithms selecting which particles to store, how to keep connections between them, and how to associate hits to them are not straightforward and, most of the time, require some trade-offs between the completeness of the stored event information and its size. In general, it is not desirable to store all the particles, as this would only waste the disk space without providing any useful information. For instance, typically delta rays (low-energy secondary electrons), low-energy gamma showers, etc. are not stored. It is best to store all the particles needed to understand the given event and to associate to the output hits. In all cases, it is necessary to set the particles' connections in order to form consistent event trees. 

In addition to all of the above points, multi-threading and concurrency only increase the complexity, because the order of processing the particles is non-deterministic. There are situations where, depending on the load of the processor, processing of the `daughter' particle may be completed before the `mother' particle's propagation ends. Once finished, the events need to be reassembled from the products generated by the different threads after parallel processing. 

\begin{figure*}[h]
    \centering
    \includegraphics[width=\textwidth]{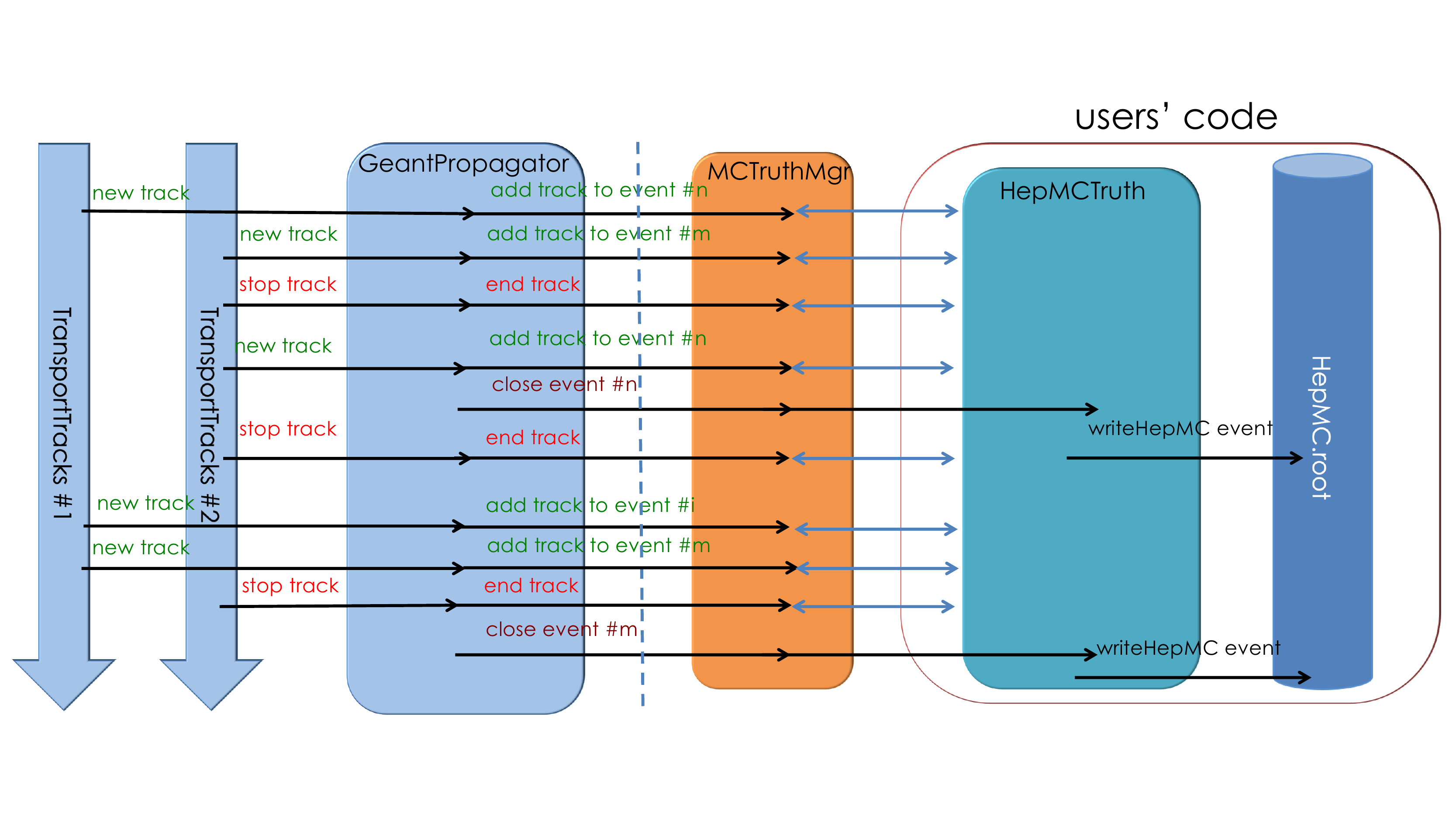}
    \caption{An example MC truth-handling architecture based on the GeantV MCTruthMgr class, with the HepMC event record as the user's output format.}
    \label{fig:mctruth}
\end{figure*}

Following the idea that there is no perfect or complete strategy for handling MC truth, users must be able to decide which particles to store. The GeantV particle transport, therefore, must provide the possibility of flagging particles as `to be stored' according to some user-defined rules and, at the same time, to ensure that the stored event has consistent mother-daughter links, as well the correct hit associations. Taking all these requirements into account, MC truth handling was implemented using an architecture that has a light coupling to the transport engine, with minimal interaction with the transport threads, and at the same time provides the flexibility to implement custom particle history handlers. In this design, shown in Fig.~\ref{fig:mctruth}, the interface provided by the MC truth manager ({\textit{MCTruthMgr}} class) receives concurrent notifications from transport threads about adding new primary or secondary particles, ending particles, or finishing events. It then delegates the processing of particles' history to a concrete MC truth implementation. In other words, the implementation is composed of the interface from the {\textit{MCTruthMgr}} class and the underlying infrastructure for the particles' history (with a light-weight, transient, intermediate event record) and the user code that implements the decision-making (filtering) algorithm, as well as the conversion to the users' event format. As a proof of principle, an example using HepMC3 as the MC truth output format is provided. This demonstrates how to implement a simple filtering algorithm based on particles' energy, allowing a consistent particle history to be serialized into an output file. 

\subsection{User interface}

GeantV provides `user actions' (similar to those in the Geant4 toolkit) that allow users to control the program flow at the level of run, event, track and step. Concurrent containers allow users to accumulate different kinds of information and merge the information from the different threads at the end of the run. Scoring is done using dedicated stepping actions in which information from the sensitive volumes is accessible.

\subsubsection{Scoring interfaces}
\label{sec:scoreint}
The GeantV prototype implements specific scoring interfaces that aim to facilitate handling user-defined data structures for mixed concurrent events. The concurrency aspect is handled by having multiple copies of the scoring data structures attached to GeantV task data objects. Each running worker thread picks up a different task data object, percolating it as an argument to the user scoring interfaces. Since the maximum number of events transported concurrently, $N_{\mathrm{max}}$, is limited, the user scoring data structures have to be indexed in arrays having the same limit: $N_{\mathrm{max}}$. The user class must have a trivial default constructor and copy constructor, and has to implement methods to merge and clear the data for a given event slot. The users are provided the interface to attach custom data types to each task, usable subsequently in their application for scoring in a thread-safe way. A detailed example using this pattern can be seen in the \href{https://gitlab.cern.ch/GeantV/geant/blob/master/examples/physics/FullCMS/GeantV/src/CMSFullApp.cxx}{CMSFullApp} example from the GeantV Git~\cite{geantv:git} repository. Another example is presented in Section~\ref{sec:exp-fwk-int}.

\section{GeantV applications and physics validations}
\label{sec:ApplicationsAndValidations}

The complexity of detector simulation software requires rigorous testing and 
continuous monitoring during development in order to ensure 
code correctness and to keep simulation precision and computing performance under control. Several tests and applications, with different levels of complexity, have been developed in order to meet these needs.

Particle transport simulation is composed of several individual components, including the geometry modeler, material description,
physics models and processes, held together by the simulation framework. The 
framework is used to set up a flexible modeling environment, including a 
generic computation workflow controlled by high-level manager objects. As a consequence, 
the individual building blocks are accessed through their interfaces and 
provide their functionalities through the framework. Checking the correctness of individual components is a pre-requisite for ensuring the above-mentioned quality criteria. Subsequently, executing complete simulation applications that exploit and exercise the whole framework is an essential final step in this testing procedure.

Model-level tests allow the verification of the responses 
of individual physics models by directly calling their interface methods. This makes it possible to test in an isolated way the correctness of 
the integrated quantities (e.g. atomic cross section, stopping power, etc.) and differential quantities
(e.g. energy or angular distribution of the final-state particles) that the 
physics models provide during the simulation. 
The production of such model-level tests was enforced as part of the physics model development procedure. The results were verified by comparing with the theoretical expectations and the corresponding Geant4 tests.
To test and validate the overall simulation 
framework, including its building blocks, complete simulation applications were developed, along with the corresponding 
Geant4 applications, if these were not already available.

An application with a simple setup (\textit{TestEm5} \footnote{whenever possible, identical names are used for the GeantV and Geant4 applications}), with a configurable particle gun and a configurable target, was developed as a first-level test 
application. In spite of its relative simplicity, this application can produce
several integrated quantities (e.g. mean energy deposit, track length, number of steps, 
backscattering and transmission coefficients, etc.) and differential quantities (e.g. transmitted/backscattered particle angular/energy distributions). 
The primary particle and target properties can be modified easily. 
This application was the perfect tool for primary testing, 
validation, and debugging during the development of the physics framework.

\begin{figure}[ht]
    \centering
    \includegraphics[width=\columnwidth]{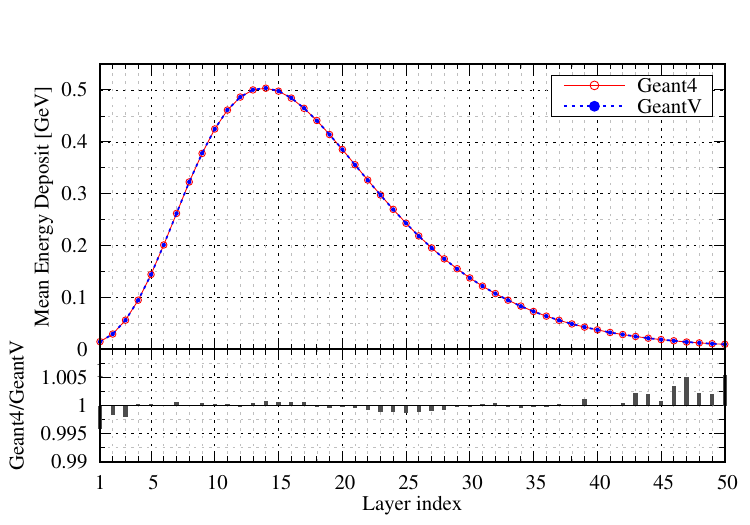}
    \caption{Mean energy deposit by E$_0$=10 GeV e$^-$ in a simplified sampling calorimeter as a function of the layer number simulated by GeantV and Geant4 (\textit{10.4.patch03}). The calorimeter is 50 layers of 2.3 mm lead and 5.7 mm liquid argon. A 4 T transverse magnetic field was applied and 0.7 mm secondary production threshold was used.}
    \label{fig:phys1}
\end{figure}

The second application developed was a generic, simplified sampling calorimeter 
simulation (\textit{TestEm3}), similar to that used for monthly validation by the 
Geant4 electromagnetic (EM) physics developers. With its intermediate complexity, this application 
was used for verification of the simulation by comparing several differential 
and integrated quantities to those provided by the corresponding Geant4 
simulation. As an example of such a differential quantity, the mean energy 
deposit in the calorimeter by electrons generated with 10 GeV energy as a function of the layer number 
(proportional to the depth) is shown in Fig.~\ref{fig:phys1} and 
compared to the corresponding Geant4 (version 10.4.patch03) simulation results.
Integrated results, such as the mean energy deposit, track lengths in the absorber 
and gap materials, or the mean number of secondary particles and simulation steps 
obtained from the same simulation setup, are summarised in Tables~\ref{tab:phys1} and~5.
All measured quantities demonstrate agreement within the per mil level compared to the corresponding values obtained from Geant4. 

Finally, a simulation application using the complete CMS detector (\textit{FullCMS}) was 
developed in order to validate and verify the correctness and robustness of the 
overall framework when reaching the complexity of an 
LHC experiment. While a similar level of agreement as mentioned above was found 
between the GeantV and the corresponding Geant4 simulation results, this application was mainly used for performance analysis and profiling.

\begin{table*}[ht]
	\label{tab:phys1}
	\begin{center}
		\caption{Detailed results of the simulation from irradiating a simplified sampling calorimeter (50 layers of 2.3 mm Pb + 5.7 mm LAr; cut = 0.7 mm) with $3\times10^5$ electrons generated with energy 10 GeV in a 4 T magnetic field: mean and RMS values for the energy deposit ($E_{\mathrm{dep}}$) and the charged particle track length ($L_{\mathrm{tr}}$). 
	Note: $^{*}$more events would be required to see the agreement between the RMS values.}
	\begin{tabular}{c | c c c c | c c c c }
        \hline\noalign{\smallskip}
        \multicolumn{1}{c}{} &  \multicolumn{4}{|c|}{Geant4}  &  \multicolumn{4}{c}{GeantV}  \\
    	material  & $E_{\mathrm{dep}}$ [GeV] & rms$^*$  [MeV ] & $L_{\mathrm{tr}}$ [m] & rms  [cm] &    $E_{\mathrm{dep}}$ [GeV] & rms$^*$  [MeV ]& $L_{\mathrm{tr}}$ [m] & rms  [cm]\\
	\noalign{\smallskip}\hline\noalign{\smallskip}
	Pb        & 7.6220   & 68.787 &    5.4071  & 5.0523    &   7.6383      &   68.857       &   5.4187      &   5.0566   \\
	LAr        & 2.2367   & 53.0346 &  11.1017  & 27.2538   &   2.2207     &   52.5708      &   11.0255    &   27.0225  \\
	\noalign{\smallskip}\hline
	\end{tabular}
	\end{center}
\end{table*}

\begin{table}[ht]
	\label{tab:phys2}
	\begin{center}
	\caption{Detailed results of the simulation from irradiating a simplified sampling calorimeter (50 layers of 2.3 mm Pb + 5.7 mm LAr; cut = 0.7 mm) with $3\times10^5$ electrons generated with energy 10 GeV in a 4 T magnetic field: mean number of secondary $e^-$, $e^+$ and $\gamma$ particles, as well as the mean number of steps made by charged and neutral particles. 
	Note: $^{*}$the geometry is always called before the physics step limit (at the pre-step point), which, in the case of GeantV, results in a slightly different step limit for e$^-$/e$^+$.}
	\begin{tabular}{ l r r r }
	    \hline\noalign{\smallskip}
	    Mean number of: & Geant4 & GeantV & \%--diff.\\
	    \noalign{\smallskip}\hline\noalign{\smallskip}
        gamma                & 5181  & 5179  & -0.04 \\
        electron             & 8891  & 8899  &  0.09 \\
        positron             & 534.5 & 534.5 &  0.00 \\
        charged steps$^{*}$  & 36572 & 35887 & -1.87 \\
        neutral steps        & 35030 & 35063 &  0.09 \\
        \noalign{\smallskip}\hline
	\end{tabular}
	\end{center}
\end{table}

\section{Usability aspects}

\subsection{Reproducibility}

Due to the stochastic nature of particle physics processes, detector simulation results are influenced by the generated random number sequence. Different sequences will generally produce slightly different, but statistically compatible, results. Such sequences are pseudo-random, controlled and reproducible based on an initial `seed' (the next generated number is fully determined by the current generator state). Reproducibility is an important requirement in the case of HEP detector simulation: simulations with the same initial configuration (primary particles and pRNG choice and seed) must give the same results. Even
in case of non-sequential processing, the simulation must be repeatable when starting from the same initial configuration of the pRNG engine. This must hold true even if different choices are 
made during a run, e.g. using vector kernels for a set of 
physics processes of selected tracks.
A key practical reason is the need to reproduce and debug problems that
occur during the simulation of a particular event or initial particle. 
In general, the reliability of a simulation that cannot be exactly repeated 
is more difficult to assess.

In GeantV, baskets of tracks undergoing the same interaction are accumulated to enable computations based on vectors of track properties, 
with the goal to perform the bulk of the computational work using these vectors.
The remaining tail of tracks is treated with sequential (non-vectorized) 
code, but using the same algorithms as the vector code.
Multi-threading is used to gather larger populations of
tracks having similar properties and enable wider vectors, targeting more efficient
use of the vector code and higher performance.
Due to the indeterminate order of execution in multi-threading, 
the track content of baskets is not preserved in each run.
In addition, a different set of remaining `unbasketized' tracks is processed in 
scalar mode in each run, particularly during the final ramping down phase of the simulation.
To be reproducible, a particular algorithm must obtain the same pRNG output value (variate)
for a track, whether it is processed as part of a vector in a
basket of tracks (`vector' mode)
or as a single track using the non-vectorized code (`scalar' mode).

To obtain the same results for a track's physics interactions (or other 
operations), the same sequence of output values of a pRNG is needed.
This is achieved by associating a single pRNG state with each track.
Whenever a new track is created, either as a primary particle or in a process,
a new state of the pRNG must be generated deterministically and 
associated with the track.  This idea, called `pseudo-random' trees, was first proposed
in the 1980s in a particle transport application~\cite{ref:halton1989}.
A first implementation was also created using linear congruential generators.
Applications in other parallel and branching computations have been proposed 
since then. The recent review in Ref.~\cite{ref:Scaathun} has an overview 
and an evaluation of the proposed methods.  
One such method for constructing seeds, called the {\it pedigree} method, was developed by Leiserson {\it et al.}~\cite{ref:pedigreeLSS} to exploit deterministic parallel 
computations written in Cilk. This method was implemented in particle 
transport simulation~\cite{ref:G4pedigree} using Geant4~\cite{ref:Geant4-gen3}
as a test-bed.

The approach adopted for GeantV depends on two pieces: first, an initial seed 
for the scalar mode or a set of seeds for the vector mode is assigned; second, a unique sub-stream index is 
determined for each track. The stream index for the primary track consists of 
high precision bits set by the event number and low precision bits by the 
track index. For the secondary (or daughter) tracks, the stream index is 
generated in a collision-resistant way using the current state of the pRNG 
carried by the (mother) track that undergoes an interaction.  
 
\begin{sloppypar}To enable reproducibility of the vector code for physics processes, a vector pRNG must be created in 
order to generate the output in each vector lane of the pRNG corresponding to each track index in the basket.  In GeantV, this 
role is played by an instance of a proxy class, acting
as a vector pRNG. The proxy provides all the expected outputs
in each vector lane (as individual pRNGs would behave for each track) and advances the
state of each track's pRNG accordingly. The first implementation of a proxy 
class gathers the contents of the scalar pRNGs into an instance of the 
corresponding {\textit VecRng} class (e.g. gathering {\textit{MRG32k3a$<$double$>$}} 
into {\textit{MRG32k3a$<$Double\_v$>$}}). The proxy instance is reusable, 
by explicitly attaching and detaching the set of track pRNG states.
\end{sloppypar}

Reproducibility was tested using a limited set of GeantV physics
processes, including bremsstrahlung, ionisation, and Compton 
scattering, which undergo a self-contained $e^{-}-\gamma$ cascade process.
The pRNG used in the tests is ThreeFry from the Random123 package~\cite{ref:random123}.
This counter-based generator was chosen because the stream is easily split 
in separate sequences, the state size is moderate (128 bits), and the method of 
initialization from a seed is trivial.

\begin{figure}
\centering
\includegraphics[width=8.0cm]{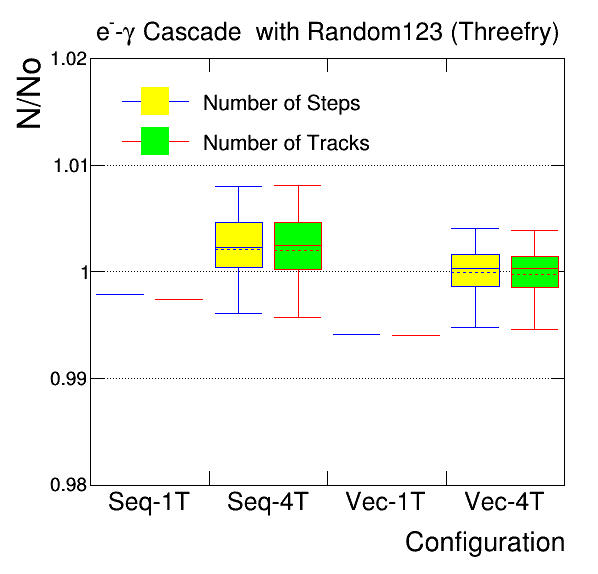}
\caption{The ratio of the total number of tracks (steps) of the default 
(non-reproducible) mode normalized to the reproducible configuration, from which 
the total number of tracks (steps) is $N_0$. 10 GeV $e^{-}$ tracks are tested with scalar (Seq) and vector (Vec) configurations with 1 thread (1T) and 4 threads (4T).}\label{fig:nsteps}
\end{figure}

The numbers of tracks and steps in a simulated dataset were measured using several different settings. 1000 events were simulated, each
comprised of ten 10 GeV electrons impinging on a 50-layer lead and liquid argon
calorimeter. The simulation was run using either 1 or 4 threads and in one of two modes: the default mode, in which a per-thread
state of one serial pRNG and one vector pRNG are used in each thread; and the
`reproducible' mode in which the method described above is used. Using the value ($N_o$) for the 
`reproducible' mode run with 1 thread as a baseline, the ratios of the 
number ($N$) of tracks and steps are shown in Fig.~\ref{fig:nsteps}. 

It is verified that the reproducible mode maintains the constant number of 
tracks and steps for all tested configurations, as required.  In addition, 
those numbers are different from the single-threaded mode for each of the 
scalar (Seq-1T) and vector (Vec-1T) modes by $0.2-0.6\%$ - the single-threaded mode is reproducible by definition if a pRNG with a fixed seed (for scalar) or seeds (for vector) is repeatable on the same hardware, but its sequence is different from that of the reproducible mode which maintains reproducibility for all configurations with the set up described earlier.
In non-reproducible multi-threaded modes, the number of tracks and steps fluctuates for each trial as 
expected, with averages (and variances) within one sigma of the values of the reproducible mode.

\begin{figure}
\centering
\includegraphics[width=8.0cm]{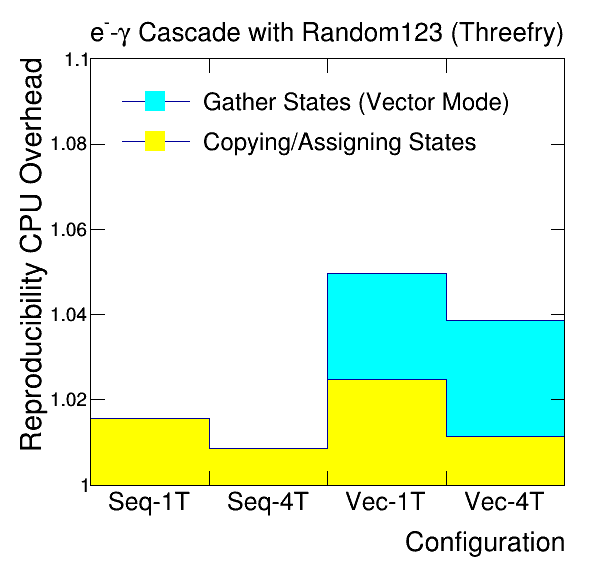}
\caption{The overhead of the reproducibility in simulation (CPU) time for the 
strategy using gathering scalar states to a vector state for different 
configurations and splitting states in reverse. 10 GeV $e^{-}$ tracks are simulated with scalar (Seq) and vector (Vec) configurations with 1 thread (1T) and 4 threads (4T).}\label{Fig:overhead}
\end{figure}

Reproducibility introduces an overhead in the simulation from
copying and assigning pRNG states during the simulation workflow,
gathering scalar states to a SIMD vector state or joining and splitting 
states in the proxy approach, and synchronizing the index of states in output
(Random123 specific). Figure~\ref{Fig:overhead} shows an example of the CPU 
overhead as the ratio of the time to execute in reproducible mode to the time to execute in default mode. The method tested is gathering scalar pRNG states to a vector state and then splitting states in reverse.  Another approach, using the join-split method, shows a similar (2-5\%) 
performance degradation for the reproducible mode.

Alternative proxy-based implementations are under development, including one that
avoids the cost of copying the data.  This is most interesting for the cases 
in which the average number of variates required is small and/or the pRNG 
state is large.

\subsection{Experiment framework integration}\label{sec:exp-fwk-int}

The Compact Muon Solenoid (CMS) experiment uses a custom, fully-featured, multi-threaded software framework called CMSSW~\cite{CMSSW,Elvira:2007xby,Lange:2015sba,Hildreth:2017vpw}. This software framework is used to produce billions of simulated events every year, employing the Geant4 simulation toolkit. In addition, CMSSW handles various other components including event generation, detector geometry, magnetic field, and scoring. The last component includes the creation of simulated hits that are used as input for custom electronics simulations.

The most important test of GeantV with CMSSW was to demonstrate the compatibility of the threading models. In production, CMSSW uses event-level parallelism with Geant4. This approach isolates each event in its thread. By avoiding communications between threads, the thread-safety of the application is easier to guarantee. In contrast, GeantV may process tracks from multiple events together in multiple threads. The GeantV approach was first tested in a simplified multi-threaded framework that uses the same Intel{\textregistered} TBB task-based processing as CMSSW. This test was successful and led to the development of the external loop mode for GeantV (Section~\ref{sec:concmod}), in order to allow the experiment's software framework to control the distribution of tasks to threads.

\begin{figure}
\centering
\includegraphics[width=0.95\linewidth]{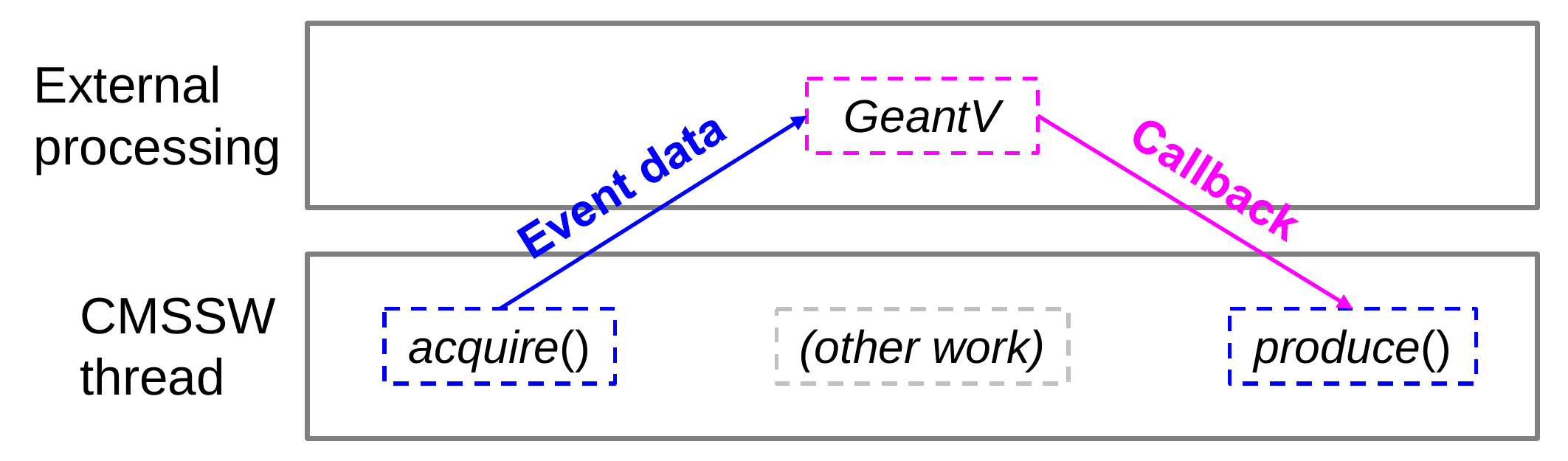}
\caption{The ExternalWork feature in CMSSW, showing the communication between the experiment software framework and GeantV.}
\label{fig:externalwork}
\end{figure}

Subsequently, the GeantV engine was fully integrated into CMSSW. To allow a more efficient trading of tasks between CMSSW and GeantV, a new CMSSW framework feature called ExternalWork is employed~\cite{makortelCHEP2019}. With ExternalWork, the actions of the CMSSW module that runs GeantV are split into two steps: acquire and produce. In the acquire step, the input event data is obtained and sent to GeantV. The acquire step is non-blocking, so it returns control of the thread after spawning a task for GeantV to process the event. Once GeantV has finished processing the event, it executes a callback function, which adds the produce step to the TBB task queue. In the produce step, the CMSSW output products are created and placed in memory. This is depicted in Fig.~\ref{fig:externalwork}. Without the use of asynchronous callbacks, the framework could be blocked if an event is loaded in one thread but then finishes processing in a different thread, after GeantV basketizes the event's tracks together with other events. In the future, it may be possible to decouple the loading of event data from the spawning of tasks by making the external loop mode more sophisticated, which could further increase the efficiency of this kind of parallel processing.

To demonstrate the full compatibility of GeantV with the experiment software framework and the steps necessary to reuse Geant4-based applications with the new transport engine, all of the additional components mentioned above are important. It is straightforward to convert generated events, stored in the HepMC format, into native GeantV input. The CMS detector geometry can be converted into a TGeo representation, which is automatically recognized by GeantV and can be navigated by VecGeom. For simplicity, a constant magnetic field of $3.8\mathrm{T}$ is used.

The scoring code required significantly more effort to adapt. There are two approaches to scoring in Geant4: sensitive detectors or watchers. Sensitive detectors are classes assigned to sensitive volumes, whose methods are automatically called when tracks traverse those volumes. This approach is the most efficient, because the volume name does not have to be checked. In contrast, watchers check every volume before deciding if they should execute their methods and record hit data. The second approach was chosen for the compatibility demonstration because the first approach is not available in GeantV and, in addition, as currently implemented in CMSSW, the first approach has more dependencies on Geant4 classes.

The full suite of scoring classes for the CMS detector comprises roughly 10,000 lines of code. A simplified scoring class that handles the electromagnetic and hadronic calorimeters was used as a demonstrator. These detectors were chosen because their scoring algorithms are relatively complex, relying on many Geant4 objects and interfaces, and because they are sensitive to the electromagnetic physics processes that have been vectorized in GeantV. The goal was to be able to use the exact same scoring code with both Geant4 and GeantV, to avoid regressions or increased maintenance burdens. Both the objects and interfaces differ between Geant4 and GeantV, so the demonstrator class is turned into a class template, where the template parameter is a traits class that collects all relevant objects and aliases them to common names. To address the differences in interfaces, specialized wrapper classes, with consistent methods, are provided for both Geant4 and GeantV. These wrapper classes handle the event, step, and geometry volume objects. This approach, using template wrappers and traits classes, has several benefits. It allows complete reuse of the scoring code with virtually no changes in the implementation, and it has no impact on performance, because the templates are evaluated at compile time.

\begin{figure}
\centering
\includegraphics[width=0.95\linewidth]{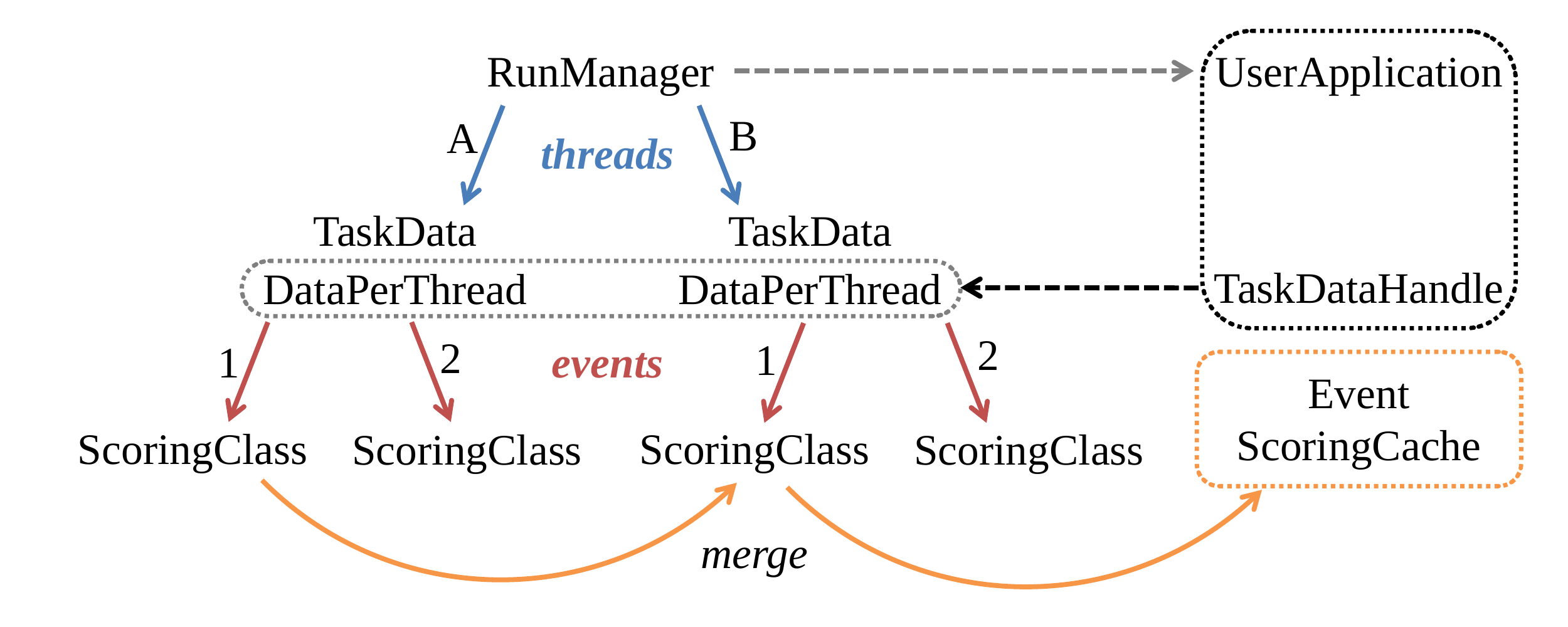}
\caption{The process of aggregating scoring information from events being processed in multiple threads.}
\label{fig:scoringaggr}
\end{figure}

However, there is another element to scoring in GeantV: thread-safety. In Geant4, as mentioned, each event is isolated in its thread, so having one instance of each scoring class per thread suffices. In GeantV, because tracks from multiple events are processed in multiple threads, steps for a given event may occur in different threads simultaneously. Rather than upsetting the complex scoring code by trying to make the existing classes accept input from multiple threads without causing data races, the chosen approach creates one instance of the scoring class per thread, per event. When a given event finishes processing in GeantV, the per-thread scoring classes dedicated to that event merge their output into a cache associated with the event, which is also accessible to CMSSW. This aggregation process is supported by the \textit{TaskData} construct in GeantV, as depicted in Fig.~\ref{fig:scoringaggr} and also discussed in Sections~\ref{sec:tracknav} and~\ref{sec:scoreint}. The duplication of scoring class instances can increase the memory usage; however, this is mitigated by sharing read-only members, such as maps of detector volumes, between instances of the class.

With all of these elements in place, equivalent simulations can be run in CMSSW using Geant4 and GeantV. This allows testing of both physics and computing performance, which are reported in Section~\ref{sec:perf-user}. The CMSSW module and supporting code that demonstrates the integration of GeantV can be found at Ref.~\cite{SimGVCore}.

\section{Performance results}
\label{sec:perfResults}

In this section, an investigation of various performance aspects of the GeantV prototype is presented. Both a simplified example of a sampling calorimeter and a complex application that uses the CMS geometry and complete EM physics were used as benchmarks.

Several configurations of the core GeantV engine were tested to highlight the contribution of different components to the observed overall performance. Two key aspects of performance were examined: the intrinsic performance of the GeantV applications as measured by various performance counters, and comparisons between different configurations of the GeantV applications with the equivalent applications running in Geant4. Finally, the performance from the user's perspective, by modifying the CMS simulation application to accommodate the GeantV engine with scoring, and comparing with the existing, similarly configured Geant4-based application was examined as well.

Simulation performance has multiple dependencies on different parameters. The most important is the complexity of the application itself. Changing production cuts, tracking cuts, or tracking precision can result in orders of magnitude differences in the number of simulated particles and steps. The benefits of running an application in the GeantV framework can vary hugely when different cuts are applied. The geometry complexity and the magnetic field settings are other application-dependent parameters that can greatly affect the CPU profile.

Another dimension to explore is the performance dependence on the hardware architecture (CPU, vector architecture, cache layout), and on the compiler and optimization flags. Measurements on several different systems were performed, although the coverage is far from complete. This analysis gives insight into how the application manages scarcity or abundance of various resources and consequently highlights areas where performance is good, as well as areas to improve.

Finally, different configurations of GeantV were explored: varying the basket size and the size of the event cache, and switching basketization off to emulate single track transport. This provides insight into the performance of individual components or scheduling features.

\subsection{Global performance}
A set of global performance metrics to compare the GeantV examples with the equivalent Geant4 ones were selected: total execution wall clock time, instructions per cycle (IPC) and FLOPS per cycle (FPC), computational intensity (FLOPS per memory operation - FMO), and the fraction of vector instructions (single and double precision). Cache misses at different levels, as well as TLB (translation lookaside buffer) misses were also evaluated. These global counters were measured using the default GeantV settings, varying only the complexity of the application. Different platforms with different vector architectures and CPU cache configurations were tested.

For the performance benchmark results and comparisons described in this section, equivalent standalone CMS applications of GeantV and Geant4 were used, unless otherwise stated.
These utilised the 2018 CMS GDML 
detector description and a magnetic field map interpolating a grid of field values extracted from CMSSW. The default scheduling mode used for GeantV is an optimized, vectorized mode in which basketization and vectorization are turned on for all sub-modules except for geometry; other modes will be specifically mentioned whenever appropriate. The 8.3.0 version of GCC was used with the default optimization level (\textit{-O3}) and build type (\textit{Release}). 
An input of 1000 events of sixteen 10 GeV electrons was simulated on dedicated machines, with no other running processes, where the measurement uncertainty
was less than 0.5\%. 

\begin{sloppypar}Table~\ref{tab:hardware-platforms} shows the characteristics of the hardware platforms used for the tests. The results of the CPU benchmark of GeantV (version beta) compared to Geant4 (version 10.5) with the CMS detector are given in Table~\ref{tab:CPU-per-event-cms}. The hardware platforms considered are Intel{\textregistered} E2620 (Sandy Bridge), Intel{\textregistered} E2680 (Broadwell), and AMD{\textregistered} 6128 (Opteron). As shown in the ``Speedup'' column in Table~\ref{tab:CPU-per-event-cms}, the relative CPU performance ratio of Geant4 to GeantV widely varies on different platforms. The impact of different configurations of the magnetic field on the relative speedup is marginal on the same hardware platform; the results from Intel{\textregistered} E2620 are shown in Table~\ref{tab:perf-field-config} as an example.\end{sloppypar}

\begin{table}[h]
\caption{Properties of the hardware platforms used for performance tests: CPU [GHz], Memory [GB] and L3 Cache size [MB].}
\label{tab:hardware-platforms}
\begin{center}
\begin{tabular}{lcrc}
\hline\noalign{\smallskip}
Processor &  CPU & Memory &  L3 Cache  \\
\noalign{\smallskip}\hline\noalign{\smallskip}
Intel E2620 (Sandy Br.) &   2.0  & 32   &  15\\
Intel E2680 (Broadwell) &   2.4  & 128   & 35 \\
AMD 6128 (Opteron) &   2.3  & 64   & 12 \\
\noalign{\smallskip}\hline
\end{tabular}
\end{center}
\end{table}

\begin{table}[h]
\caption{Performance comparison between GeantV and Geant4: the average CPU 
time in seconds per event for simulating sixteen 10 GeV electrons 
propagating through the CMS detector and the magnetic field.}
\label{tab:CPU-per-event-cms}
\begin{center}
\begin{tabular}{lcccc}
\hline\noalign{\smallskip}
Processor & SIMD  &  Geant4   & GeantV  &  Speedup  \\
\noalign{\smallskip}\hline\noalign{\smallskip}
Intel E2620 & AVX &   4.94  & 2.33   & 2.12 \\
Intel E2680 & AVX2 &   2.18  & 1.63   & 1.43 \\
AMD 6128 & SSE4     &   6.63  & 4.33   & 1.53 \\
\noalign{\smallskip}\hline
\end{tabular}
\end{center}
\end{table}

\begin{table}[h]
\caption{Performance comparison between GeantV and Geant4: the impact of 
different configurations of the magnetic field on Intel E2620 (Sandy Bridge).
In the CPU performance ratios G4/GV and G4/GV(vect), the denominators refer to GeantV in scalar mode and vector mode, respectively.}
\label{tab:perf-field-config}
\begin{center}
\begin{tabular}{lccc}
\hline\noalign{\smallskip}
Configuration & GeantV [s]  &  G4/GV   & G4/GV(vect) \\
\noalign{\smallskip}\hline\noalign{\smallskip}
Zero field & 1794  & 1.86   & 1.95 \\
Uniform (3.8T) & 2412  & 1.97   & 2.19 \\
CMS field map & 2621  & 1.88   & 2.12 \\
\noalign{\smallskip}\hline
\end{tabular}
\end{center}
\end{table}

It turns out that the Geant4 performance is more sensitive to the size of cache memory and fluctuates more
compared to GeantV. An extended performance benchmark study on various hardware platforms
and different compilers is also available in Appendix~\ref{app:perf_benchmark}.  

There are a variety of performance metrics (hardware counters) provided by PAPI (performance application programming interface)~\cite{perf:PAPI}. A combination of
PAPI counters provides useful information about code performance, such as floating-point operations, instruction per cycle, cache behaviors, memory access patterns, and so on. For example, floating-point operations per cycle is a good measure for the CPU utilization, while instruction per cycle quantifies good balance with minimal stalls. To collect
profiling information along with hardware counters,
Open$\mid$Speedshop~\cite{perf:openspeedshop} is employed as the primary profiler. This program provides an integrated toolkit and analysis framework for various performance experiments and measurements. 
Table~\ref{tab:perf-ipc} shows
the IPC of GeantV compared to Geant4, which indicates that GeantV executes relatively more instructions per cycle. Since the number of instructions completed is approximately
proportional to the total floating-point operations, the FPC of GeantV with
respect to Geant4 follows the same pattern.
\begin{table}[h]
\caption{The IPC, instructions (INS) per cycle (CYC), of GeantV compared to Geant4.}
\label{tab:perf-ipc}
\begin{center}
\begin{tabular}{l|cc|cc}
\hline\noalign{\smallskip}
 & \multicolumn{2}{c|}{GeantV} & \multicolumn{2}{c}{Geant4} \\
Processor &  INS/CYC  & IPC &  INS/CYC  &  IPC  \\
\noalign{\smallskip}\hline\noalign{\smallskip}
Intel E2620  &  7038/6610    & 1.06       &  8388/10788  &  0.78 \\ 
Intel E2680 &  6474/5521    & 1.17       &  8914/5514   &  1.62 \\
AMD 6128 &  7813/8839    & 0.88       &  8459/11228  &  0.75 \\
\noalign{\smallskip}\hline
\end{tabular}
\end{center}
\end{table}

Another important performance metric is FMO, which quantifies data locality or computational intensity.  Table~\ref{tab:perf-fmo} shows the FMO of GeantV compared to Geant4, 
which implies that GeantV has better data locality than Geant4 in the platforms tested, even with widely varying cache sizes and policies.  The reason that FMO on Intel{\textregistered} E2680 (Broadwell) is much larger (better) than other architectures may be due its relatively large memory and L3 cache shown in Table~\ref{tab:hardware-platforms}, which leads the smaller number of load and store instructions.
Nevertheless, the FMO is relatively small for both Geant4 and GeantV, which indicates that
the typical HEP detector simulation is a memory-bound application.

\begin{table}[h]
\caption{Floating-point instructions per memory operation (FMO) in terms of 
floating point operations (FO) over the sum of load instructions (LD) and store
instructions (SR).}
\label{tab:perf-fmo}
\begin{center}
\begin{tabular}{lcc}
\hline\noalign{\smallskip}
 & \multicolumn{2}{c}{FLOP/(LD+SR)} \\
Processor &  GeantV  &  Geant4    \\
\noalign{\smallskip}\hline\noalign{\smallskip}
Intel E2620 &  1718/3402 (0.50) & 2181/5509 (0.40) \\
Intel E2680 &  2347/1758 (1.34) & 3824/3100 (1.23) \\
AMD 6128    &  3191/3704 (0.86) & 1620/5515 (0.29) \\
\noalign{\smallskip}\hline
\end{tabular}
\end{center}
\end{table}

Nonetheless, the resulting speedup and the platform dependency are not driven by 
a specific set of functions or libraries. For example, the percentage of the total CPU time by 
Geant4 library, shown in Table~\ref{tab:geant4-CPU-by-libs}, is very comparable on 
different hardware platforms. This is also true for GeantV, as shown in Table~\ref{tab:geantv-CPU-by-libs}, which indicates that
the relative speedup seems to be a global effect spread over all the code.
\begin{table}[h]
\caption{The percentage of CPU time spent in each Geant4 library for simulating sixteen 10 GeV electrons propagating the CMS detector.}
\label{tab:geant4-CPU-by-libs}
\begin{center}
\begin{tabular}{lrrr}
\hline\noalign{\smallskip}
 & Intel & Intel & AMD \\
Library (\%) &  E2620 &  E2680 & 6128 \\
\noalign{\smallskip}\hline\noalign{\smallskip}
libG4geometry.so    & 41.8 &  43.6 &  42.3  \\
libG4processes.so   & 22.0 &  20.8 &  21.0  \\
libG4global.so      &  7.3 &   8.0 &   7.5  \\
libG4tracking.so    &  7.3 &   6.5 &   7.2  \\
libG4track.so       &  6.0 &   4.7 &   5.8  \\
full\_cms            &  5.2 &   6.1 &   6.6  \\
libG4clhep.so       &  3.3 &   3.0 &   3.0  \\
libm-2.12.so        &  2.7 &   3.5 &   2.9  \\
libG4particles.so   &  1.2 &   0.7 &   1.0  \\
libG4digits\_hits.so &  1.1 &   1.3 &   1.0 \\
\noalign{\smallskip}\hline
\end{tabular}
\end{center}
\end{table}

\begin{table}[h]
\caption{The percentage of CPU time spent in each GeantV library for simulating sixteen 10 GeV electrons propagating the CMS detector.}
\label{tab:geantv-CPU-by-libs}
\begin{center}
\begin{tabular}{lrrr}
\hline\noalign{\smallskip}
 & Intel & Intel & AMD \\
GeantV Library (\%) &  E2620 &  E2680 & 6128 \\
\noalign{\smallskip}\hline\noalign{\smallskip}
libGeant\_v.so        &   42.1 & 46.3  & 43.2 \\
libRealPhysics.so     &   36.0 & 34.2  & 37.3 \\
libGeantExamplesRP.so &   14.1 & 14.1  & 14.5 \\
libc-2.12.so          &    3.8 &  1.8  &  1.1 \\
libVmagfield.so       &    3.1 &  2.8  &  3.1 \\
libm-2.12.so          &    0.6 &  0.6  &  0.6 \\
\noalign{\smallskip}\hline
\end{tabular}
\end{center}
\end{table}

To understand the underlying cause of the overall performance difference between Geant4 and 
GeantV, instruction and data cache misses at different levels were also studied.
Tables~\ref{tab:L1-cache-miss} and~\ref{tab:L2-cache-miss} show instruction 
and data cache misses in L1 and L2, respectively. The GeantV application shows far 
fewer instruction cache misses in L1, which is attributed to the fact that GeantV has much
simpler code structure and consists of smaller libraries.

\begin{table}[h]
\caption{L1 cache misses in 1 billion hardware counters between Geant4 (G4) and GeantV (GV).
ICM and DCM are Instruction and Data Cache Misses, respectively. The Level 1 latency is typically 3 cycles.}
\label{tab:L1-cache-miss}
\begin{center}
\begin{tabular}{l|cc|cc}
\hline\noalign{\smallskip}
 & \multicolumn{2}{c|}{ICM} & \multicolumn{2}{c}{DCM} \\
Processor   &  GV & G4 & GV & G4 \\
\noalign{\smallskip}\hline\noalign{\smallskip}
Intel E2620   &  54       & 429    &  218    & 269    \\ 
Intel E2680  &  39       & 511    &  188    & 272    \\
AMD 6128  &  49       & 309    &  141    & 144    \\
\noalign{\smallskip}\hline
\end{tabular}
\end{center}
\end{table}

\begin{table}[h]
\caption{L2 cache misses in 1 billion hardware counters between Geant4 (G4) and GeantV (GV).
ICM and DCM represent instruction and data cache misses, respectively. The Level 2 latency is typically 12 cycles.}
\label{tab:L2-cache-miss}
\begin{center}
\begin{tabular}{l|cc|cc}
\hline\noalign{\smallskip}
 & \multicolumn{2}{c|}{ICM} & \multicolumn{2}{c}{DCM} \\
Processor   &  GV & G4 & GV & G4 \\
\noalign{\smallskip}\hline\noalign{\smallskip}
Intel E2620 &  19   & 36  &  86      & 46  \\ 
Intel E2680 &  23   & 29  &  101     & 51  \\
AMD 6128    &  17   & 3.6 &  55      & 10  \\
\noalign{\smallskip}\hline
\end{tabular}
\end{center}
\end{table}

Most modern hardware systems have a TLB that severs as the cache
for page tables that map addresses between virtual memory and physical memory.
Table~\ref{tab:TLB-cache-miss} shows both instruction and data TLB cache misses for
GeantV compared to Geant4.
In general, GeantV leads substantially fewer TLB misses compared to Geant4.
However, it turns out that the total cost for TLB misses is a relatively small fraction of 
the total elapsed time. For example, the 330 million TLB misses on the Intel{\textregistered} E2620 cost about one second.

\begin{table}[h]
\caption{TLB misses in 1 million hardware counters between Geant4 (G4) and GeantV (GV).
IM and DM represent instruction and data TLB misses, respectively. The TLB miss latency is typically 6 cycles.}
\label{tab:TLB-cache-miss}
\begin{center}
\begin{tabular}{l|cc|cc}
\hline\noalign{\smallskip}
 & \multicolumn{2}{c|}{IM} & \multicolumn{2}{c}{DM} \\
Processor   &  GV & G4 & GV & G4 \\
\noalign{\smallskip}\hline\noalign{\smallskip}
Intel E2620 &  53   & 4256  &  3168   & 4626  \\ 
Intel E2680 &  N/A  & N/A   &  24    & 82  \\
AMD 6128    &  55   & 149   &  88    & 1628  \\
\noalign{\smallskip}\hline
\end{tabular}
\end{center}
\end{table}

\subsection{Scheduler performance}
The performance of the GeantV workload scheduler was evaluated to quantify the impact of different parameters: number of events in flight, basket size, scalar emulated mode against basketized mode. The observed vectorization gains per component were measured. The performance impact of the cool-down phase when basketization is less efficient was also evaluated.

The main task of the GeantV scheduler is to maximize the amount of work executed via SIMD baskets compared to scalar. Handling too many baskets being filled concurrently during stepping becomes challenging when all the available tracks to be transported are exhausted and the limit of the number of events in flight is reached. To avoid the stall, the scheduler has to fire some of the partially filled baskets in the scalar mode, allowing work to continue at the cost of reducing the basket (and subsequently vectorization) efficiency. The efficiency drop increases with the simulation stage complexity, so for example basketizing all geometry volumes in a complex setup becomes prohibitively expensive. For this reason, the scheduler needs to keep active only a limited set of baskets, ideally those that are most ``popular'' in terms of fraction of tracks processed by the associated algorithms. For example, the basket associated with vectorized field propagation is most popular, since it handles all charged particles. The same applies to multiple scattering and to some physics processes. 

For the geometry case, it is difficult to predict which are the volumes handling most steps.
Switched from scalar to vector execution mode for a given basket based on a ``track popularity'' score,
a dynamic basetizing strategy was adopted.
The simulation starts with no active geometry basket, executing navigation in scalar mode, and activating a given basket only when the number of handled tracks reaches a high watermark. This strategy gradually activates basketization for the volumes getting the highest particle multiplicity (i.e. the central barrel for collider experiment setups). As transported tracks are exhausted and baskets have to be processed in the scalar mode, their popularity is demoted until reaching a low watermark, which triggers their deactivation. This strategy allows to maintain a rather constant active (i.e. not stalled in SIMD baskets) track population while maximizing the SIMD flow via popular baskets.

\begin{figure}[ht]
    \centering
    \includegraphics[width=\columnwidth]{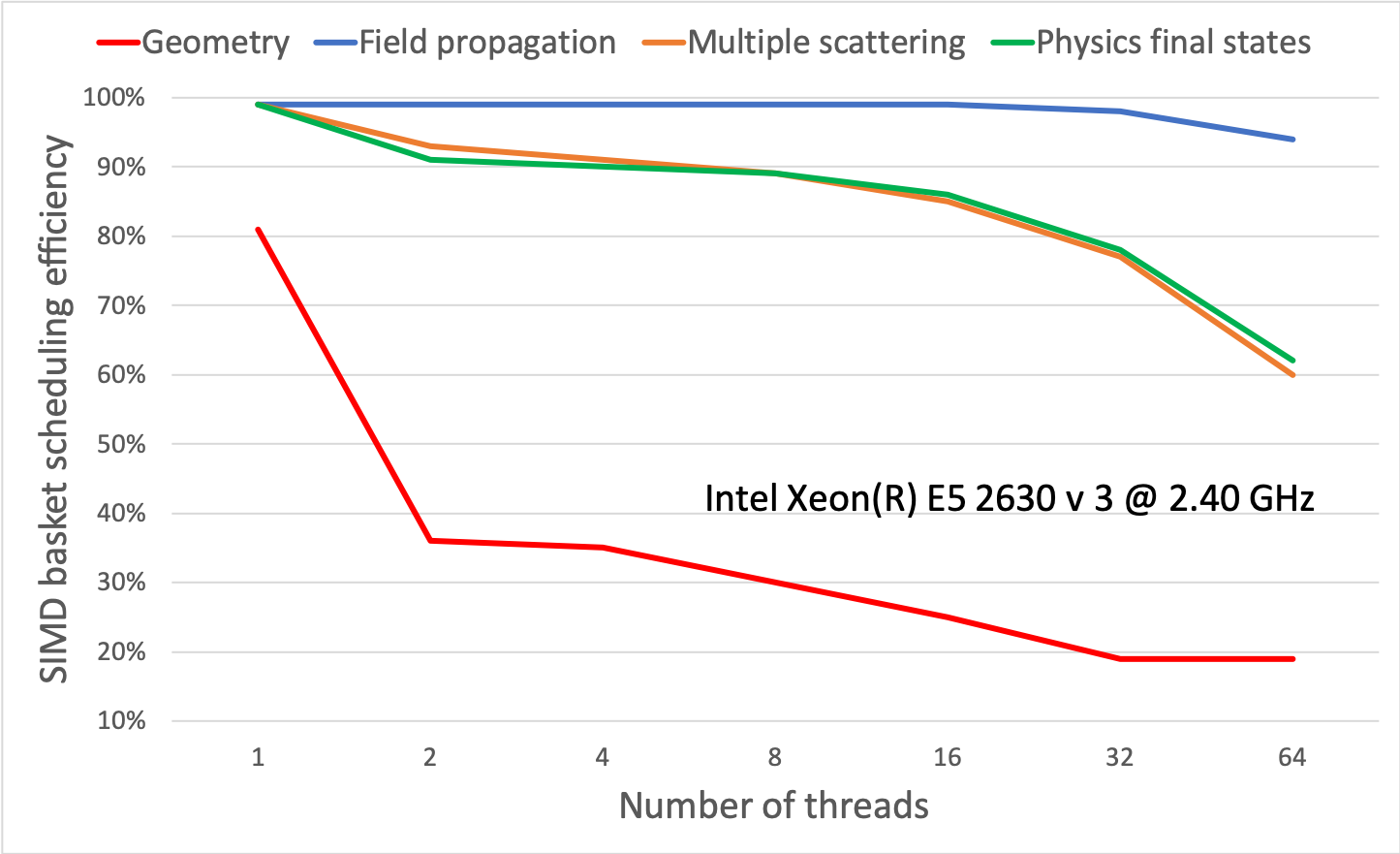}
    \caption{Scheduling efficiency for SIMD baskets of different categories depending on the number of threads. CMS benchmark shooting 50 events with 100 GeV electrons, using 16 events in flight.}
    \label{fig:results_simd_sched_eff}
\end{figure}

In the multi-threaded mode, SIMD baskets are filled concurrently in order to maximize the available population per category. The drawback is that multi-threading consumes much faster the reserve of buffered tracks, which forces more frequent scalar executions if the number of events slots is kept constant. Figure~\ref{fig:results_simd_sched_eff} shows the fraction of total tracks processed in SIMD mode for the main simulation stages, depending on the number of threads. While the very popular stages such as field propagation have a rather constant high SIMD dispatch efficiency, unpopular baskets suffer depletion much faster with increasing number of threads. For example, geometry basketization becomes very inefficient for more than few tens of active geometry baskets. This behavior calls for replicating SIMD baskets per thread rather than sharing them, which gives very good results for the propagation and multiple scattering stages. For geometry this strategy does not bring unfortunately any improvement.
Besides the scheduler efficiency to dispatch baskets, the global performance is highly impacted by the intrinsic efficiency of the basketizing procedure, involving gather and scatter actions as well as concurrent access. As presented in detail in Ref.~\cite{Top:bottom:vectorization:CHEP2018}, the main conclusion is that the basketizing dynamics strongly depends on the complexity of the workflow and on state parameters, such as number of tracks in flight, particle production budget, or percent to completion for a given event.

The GeantV scheduler has an option to run in single track mode, which emulates Geant4-style sequential tracking. 
Table~\ref{tab:perf-single-track} compares the performance of GeantV single track mode
to the default (basketized) mode, showing, but for  marginal variations on different platforms, that the impact of the GeantV scheduler or data locality from basketization is not the primary source of the performance difference between Geant4 and GeantV in scalar mode. Note that computing performance 
depends on the basket sizes for the magnetic field, for physics, and for the multiple scattering process, and may need to be optimized for each hardware platform separately. The default number of tracks per basket
used for these comparisons was 16.

\begin{table}[h]
\caption{The relative CPU performance of the GeantV single track mode, \emph{GV-strk}, which 
emulates Geant4-style tracking, compared with the default GeantV basketized mode, \emph{GV-bskt}.}
\label{tab:perf-single-track}
\begin{center}
\begin{tabular}{lccc}
\hline\noalign{\smallskip}
Processor    &  GV-bskt  & GV-strk  &  GV-strk/GV-bskt \\
\noalign{\smallskip}\hline\noalign{\smallskip}
Intel E2620 & 2621 sec  &  2960 sec       &  1.13  \\ 
Intel E2680 & 1628 sec  &  1533 sec       &  0.94  \\
AMD 6128    & 4457 sec  &  4817 sec       &  1.08  \\
\noalign{\smallskip}\hline
\end{tabular}
\end{center}
\end{table}

\subsection{Profiling analysis}
This section presents detailed profiling information that illustrates the relative performance of different components (geometry, physics and magnetic field propagation) along with the major hotspots. Results are compared with Geant4 to understand which components exhibit different performance features. These profiles are also presented for different configurations of the CMS application with varied cuts and for the simplified calorimeter example. Tables~\ref{tab:perf-profiling-func-scalar}, \ref{tab:perf-profiling-func-vector}, \ref{tab:perf-profiling-func-vector-nogeom} are lists of the top 10
functions of GeantV for different configurations, ranked by the exclusive CPU time, while 
Table~\ref{tab:perf-profiling-func-geant4} is the list of top functions from 
the Geant4 application. In general, there are no unexpected hotspots or bottlenecks, which indicates that both GeantV and Geant4 applications are reasonably modular and granular already. On the other hand, there exist noticeable differences between the scalar and the vector mode of GeantV. For example, the CPU fraction of CMSmagField::EstimateFieldValues is significantly 
reduced in the vector mode as it is efficiently vectorized. The overhead from the geometry basketization is largely due to the extra track handling, such as Handler::AddTrack and Handler::Flush, which are shown in Table~\ref{tab:perf-profiling-func-vector} but not in 
Table~\ref{tab:perf-profiling-func-vector-nogeom}.  It is also worthwhile to note that Spline::GetValueAt of GeantV takes significantly less time than its equivalent function in Geant4, G4PhysicsVector::Value, which is 
the top CPU function in recent versions of Geant4. 
\begin{table}[h]
\caption{Top 10 functions in GeantV scalar mode.}
\label{tab:perf-profiling-func-scalar}
\begin{center}
\begin{tabular}{ll}
\hline\noalign{\smallskip}
\% time &  Function name \\
\noalign{\smallskip}\hline\noalign{\smallskip}
8.22 &  CMSmagField::EstimateFieldValues                             \\
5.44 &  ScalarNavInterfaceVGM::NavIsSameLocation         \\
5.36 &  DormandPrinceRK45::StepWithErrorEstimate             \\
3.32 &  SimpleABBoxLevelLocator::LevelLocate          \\
2.99 &  \_\_GI\_memcpy                                    \\
2.98 &  SimulationStage::Process                         \\
2.87 &  PhysicsProcess::PostStepLimitationLength                     \\
2.75 &  Spline::GetValueAt                                           \\
2.25 &  GSMSCModel::ComputeParameters                                \\
2.19 &  HybridNavigator::GetHitCandidates\_v            \\
\noalign{\smallskip}\hline
\end{tabular}
\end{center}
\end{table}

\begin{table}[h]
\caption{Top 10 functions in GeantV vector mode.}
\label{tab:perf-profiling-func-vector}
\begin{center}
\begin{tabular}{ll}
\hline\noalign{\smallskip}
\% time &  Function name \\
\noalign{\smallskip}\hline\noalign{\smallskip}
5.45 & ScalarNavInterfaceVGM::NavIsSameLocation \\
4.95 & Handler::AddTrack \\
4.66 & Handler::Flush \\
4.32 & CMSmagField::EstimateFieldValues \\
3.44 & SimulationStage::CopyToFollowUps \\
3.13 & SimpleABBoxLevelLocator::LevelLocate \\
2.78 & SimulationStage::Process \\
2.78 & PhysicsProcess::PostStepLimitationLength \\
2.55 & memcpy \\
2.48 & GeomQueryHandler::DoIt \\
\noalign{\smallskip}\hline
\end{tabular}
\end{center}
\end{table}

\begin{table}[h]
\caption{Top 10 functions in GeantV vector mode, except Geometry (i.e, using the scalar mode for Geometry).}
\label{tab:perf-profiling-func-vector-nogeom}
\begin{center}
\begin{tabular}{ll}
\hline\noalign{\smallskip}
\% time &  Function name \\
\noalign{\smallskip}\hline\noalign{\smallskip}
6.68 & ScalarNavInterfaceVGM::NavIsSameLocation \\
4.84 & CMSmagField::EstimateFieldValues         \\
4.17 & SimpleABBoxLevelLocator::LevelLocate     \\
3.56 & SimulationStage::Process                 \\
3.11 & PhysicsProcess::PostStepLimitationLength \\
2.86 & memcpy (libc-2.12.so)                    \\
2.74 & HybridNavigator::GetHitCandidates\_v      \\
2.68 & Spline::GetValueAt                       \\
2.56 & SimulationStage::CopyToFollowUps         \\
2.47 & DormandPrince5RK::StepWithErrorEstimate  \\
\noalign{\smallskip}\hline
\end{tabular}
\end{center}
\end{table}

\begin{table}[h]
\caption{Top 10 functions in Geant4.}
\label{tab:perf-profiling-func-geant4}
\begin{center}
\begin{tabular}{ll}
\hline\noalign{\smallskip}
\% time &  Function name \\
\noalign{\smallskip}\hline\noalign{\smallskip}
6.52 &  G4PhysicsVector::Value	      	      	 \\
5.20 &  G4ScalarRZMagFieldFromMap::GetFieldValue \\
3.36 &  G4Navigator::LocateGlobalPointAndSetup \\
2.52 &  G4DormandPrince745::Stepper \\
2.38 &  G4Navigator::ComputeStep \\
2.36 &  G4VEmProcess::PostStepGPIL \\
1.92 &  G4PropagatorInField::ComputeStep \\
1.77 &  G4Transportation::AlongStepGPIL \\
1.67 &  G4VoxelNavigation::ComputeStep \\
1.64 &  G4Mag\_UsualEqRhs::EvaluateRhsGivenB \\
\noalign{\smallskip}\hline
\end{tabular}
\end{center}
\end{table}

\subsection{Vectorization performance}
The performance of basketization is compared by switching it on or off for each component. This requires a mode that allows the grouping of tracks into baskets, but dispatching tracks in scalar mode. With this mode, the overheads of basketization can be evaluated.

Table~\ref{tab:perf-vec-mode} shows the fraction of vector instructions in
each module of GeantV and the relative CPU gain introduced by vectorization with respect
to the scalar mode for each of the enabled vectorization options.  The CPU gain is relatively small, even though
the fraction of vector instructions is significant.  This is due to
several factors, including the basketization overheads (10--25\%, as shown in in Appendix~\ref{app:perf_benchmark}) and inefficiency from gather/scatter and mask operations in vectorization.  
In addition, the poor vector performance of the geometry is not due to a lack of vectorization, but to the execution of sequential algorithms used in navigation. Note that the sizable amount of vector instructions in scalar mode (15.67\%) comes from both compiler auto-vectorization and VecGeom internal vectorization.

\begin{table}[h]
\caption{Vector instructions, the fraction of vector instructions (PAPI\_DP\_VEC)/(PAPI\_DP\_OPS), and the relative gain in CPU usage by vectorization of a specific module with respect to the scalar mode.
Here, PAPI\_DP\_OPS and PAPI\_DP\_VEC are floating point (double precision) 
operations and double precision vector/SIMD instructions in 1-billion counters, respectively. MSC-vec is the case when vectorization in the GeantV multiple scattering (MSC) model is turned on. The Opt-vec mode is the same as All-vec, but Geom-vec is turned off.}
\label{tab:perf-vec-mode}
\begin{center}
\begin{tabular}{lcccc}
\hline\noalign{\smallskip}
 & PAPI\_ & PAPI\_ & & \\
Mode    &  DP\_OPS & DP\_VEC   & \%  &  Gain \\
\noalign{\smallskip}\hline\noalign{\smallskip}
Scalar    &  1770  &   277 &  15.67 &   -    \\
Geom-vec  &  1771  &   333 &  18.82 &  0.96  \\
Field-vec &  1858  &   814 &  43.83 &  1.08  \\
MSC-vec   &  1789  &   397 &  22.24 &  1.02  \\
Phys-vec  &  1785  &   343 &  19.25 &  1.00  \\
All-vec   &  1868  &  1051 &  56.26 &  1.00  \\
Opt-vec   &  1868  &   996 &  53.35 &  1.12  \\
\noalign{\smallskip}\hline
\end{tabular}
\end{center}
\end{table}

\subsection{Concurrency performance}
\begin{sloppypar}This section presents the multi-threaded performance of the GeantV applications compared to the Geant4 equivalent. This includes a discussion of the scalability features and pros and cons for track-level parallelism versus event-level parallelism.\end{sloppypar}

\begin{figure}[ht]
    \centering
    \includegraphics[width=\columnwidth]{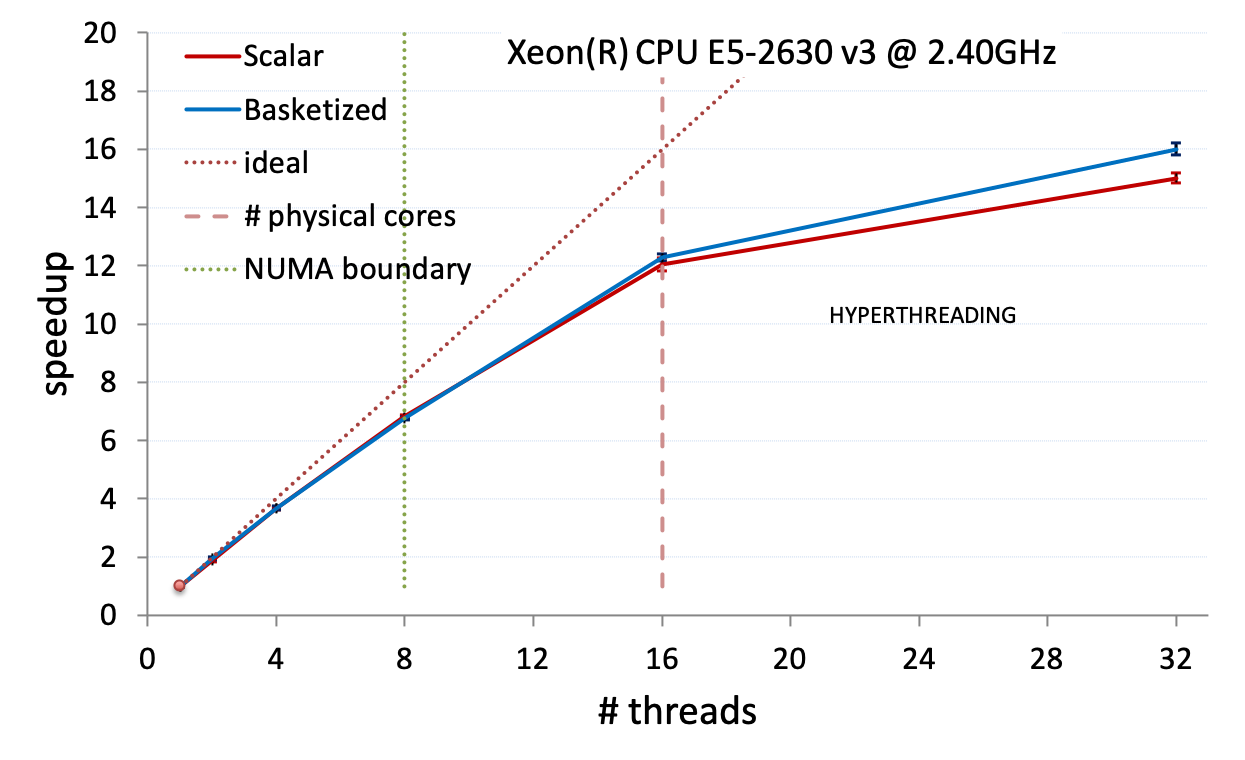}
    \caption{Strong scaling versus threads for the CMS example benchmark on a dual-socket Xeon{\textregistered} CPU E5-260 v3 @ 2.40 GHz with 8 cores per socket. The simulation was performed separately for scalar and basketized workflows.}
    \label{fig:strong_scaling}
\end{figure}

The strong scaling behavior of the GeantV prototype is shown in Fig.~\ref{fig:strong_scaling}. The efficiency loss of about 25\% when filling 16 physical cores is not ideal. It is caused both by extra memory contention for shared track basketizers, and by a decrease in basket efficiency as the number of threads increases.

\begin{figure}[ht]
    \centering
    \includegraphics[width=\columnwidth]{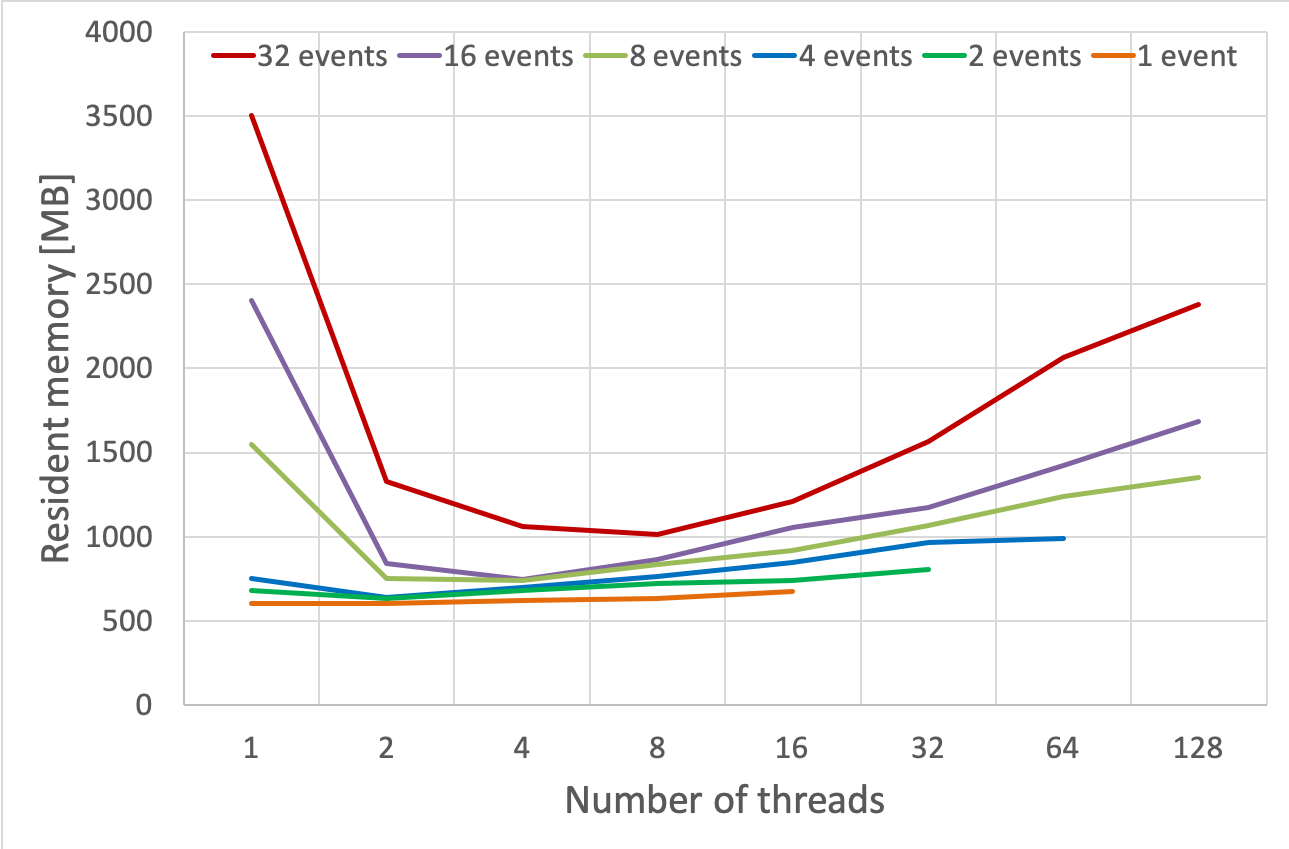}
    \caption{Maximum resident memory dependence on the number of threads for different size of the event buffer in the full basketized mode. Measurements are done with a fixed number of 100 events in the CMS example benchmark on a dual-socket Xeon{\textregistered} CPU E5-260 v3 @ 2.40 GHz having 8 cores per socket.}
    \label{fig:memory_scaling}
\end{figure}

Figure~\ref{fig:memory_scaling} shows the memory usage versus the number of threads, in a configuration keeping all basketizers active. As a general remark, the memory footprint is largely dominated by the number of tracks in flight. Increasing the number of threads requires more tracks for load balancing, but the memory can be kept under control at the price of lowering the basket efficiency. A particular inverse slope effect is observed for small number of threads, more accentuated for large event buffers. In this domain more track data needs to be allocated for the same amount of tracks when fewer threads are used. The main reason for this ``abnormal'' behavior is related to the track reuse policy, in connection with the fact that basketization ``steals'' tracks from the workflow. The single-thread mode tends to release tracks from baskets later compared to the multi-thread one, which increases the fragmentation for track memory blocks. This effect degrades also the computing efficiency, and reflects the complexity of the scheduling mechanism in basketized mode, which is subject to further optimization procedures.

\subsection{Performance in an experiment framework}\label{sec:perf-user}
The performance of GeantV is analyzed after integration into the CMS simulation framework as discussed in Section~\ref{sec:exp-fwk-int}. In order to compare with Geant4, it is necessary to configure the application as similarly as possible to the options available in GeantV. These settings include an EM-only physics list that supports the same models that have been vectorized in GeantV, as well as the same production cuts and other cuts. The same magnetic field integrator and stepper are used. The CMS detector geometry corresponds to the version operated in 2018. CMS has also introduced several optimizations to improve the CPU performance of the Geant4-based simulation, including Russian roulette and shower libraries~\cite{Apostolakis:2018ieg,PedroCHEP2018}; the optimizations that are not compatible with GeantV are disabled. There is good agreement in the physical output quantities from equivalent GeantV and Geant4 runs in the CMS software, validating the performance comparisons~\cite{PedroCHEP2019}.

The tests are conducted using 500 generated events with two electrons, each at $E = 50\,\mathrm{GeV}$, with random directions in $\eta$ and $\phi$. A constant magnetic field of $B = 3.8\,\mathrm{T}$ is used. The CMSSW tests compare single-threaded to multi-threaded performance, as multi-threaded jobs are necessary for efficient use of the WLCG resources. To ensure a constant workload, the number of events per thread is kept constant in each test by reusing the initial 500 generated events. Unused threads are kept busy to simulate production conditions with all cores in use. The CPU and memory usage of the main program are estimated with ROOT file output disabled, as the overhead from output is the same for Geant4 and GeantV, and therefore irrelevant. In the GeantV tests, vectorized algorithms are enabled for multiple scattering and magnetic field propagation. Both the basketized and single track modes of GeantV operation are tested.

\begin{figure}
\centering
\includegraphics[width=0.95\linewidth]{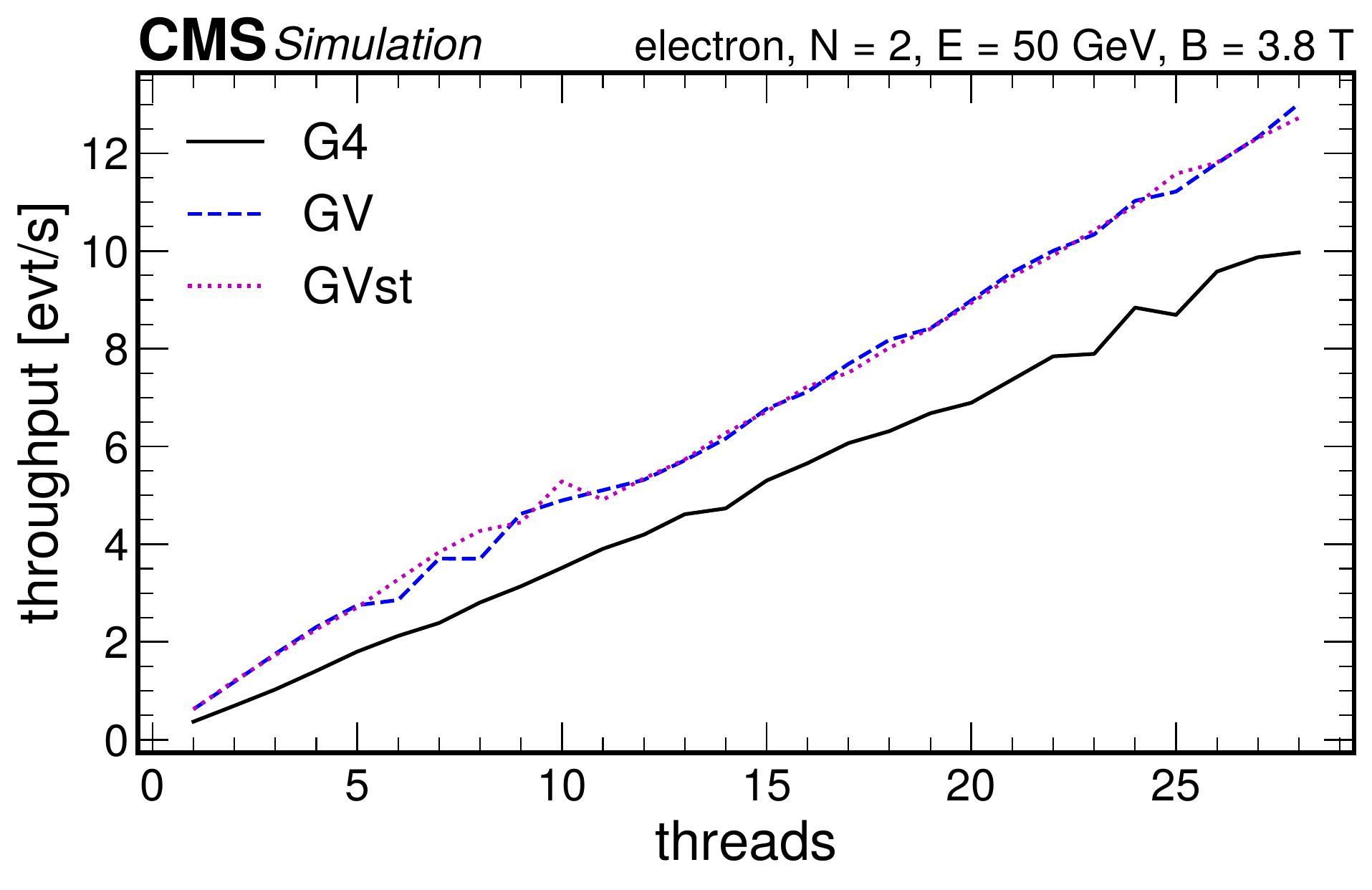}
\includegraphics[width=0.95\linewidth]{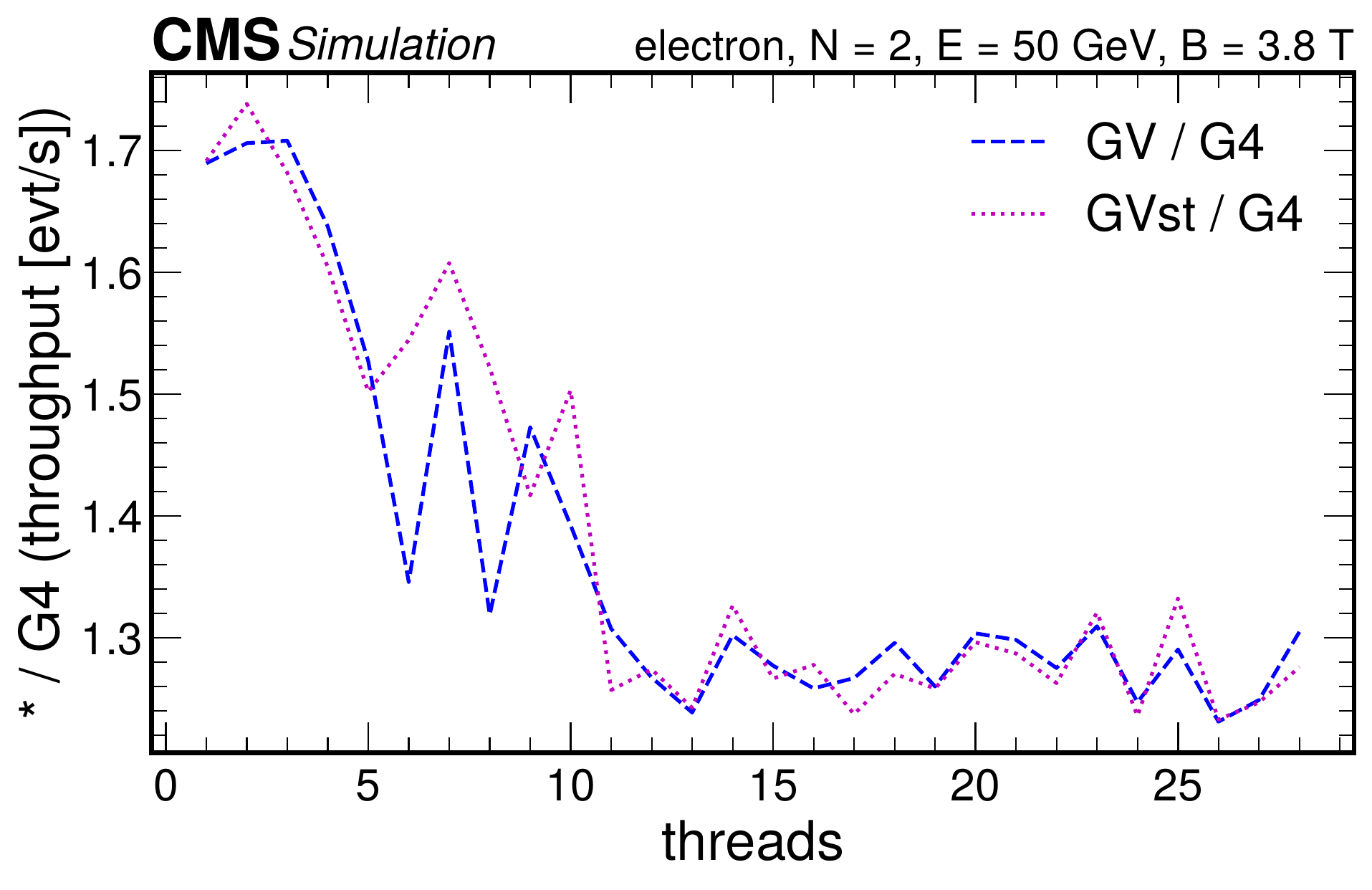}
\includegraphics[width=0.95\linewidth]{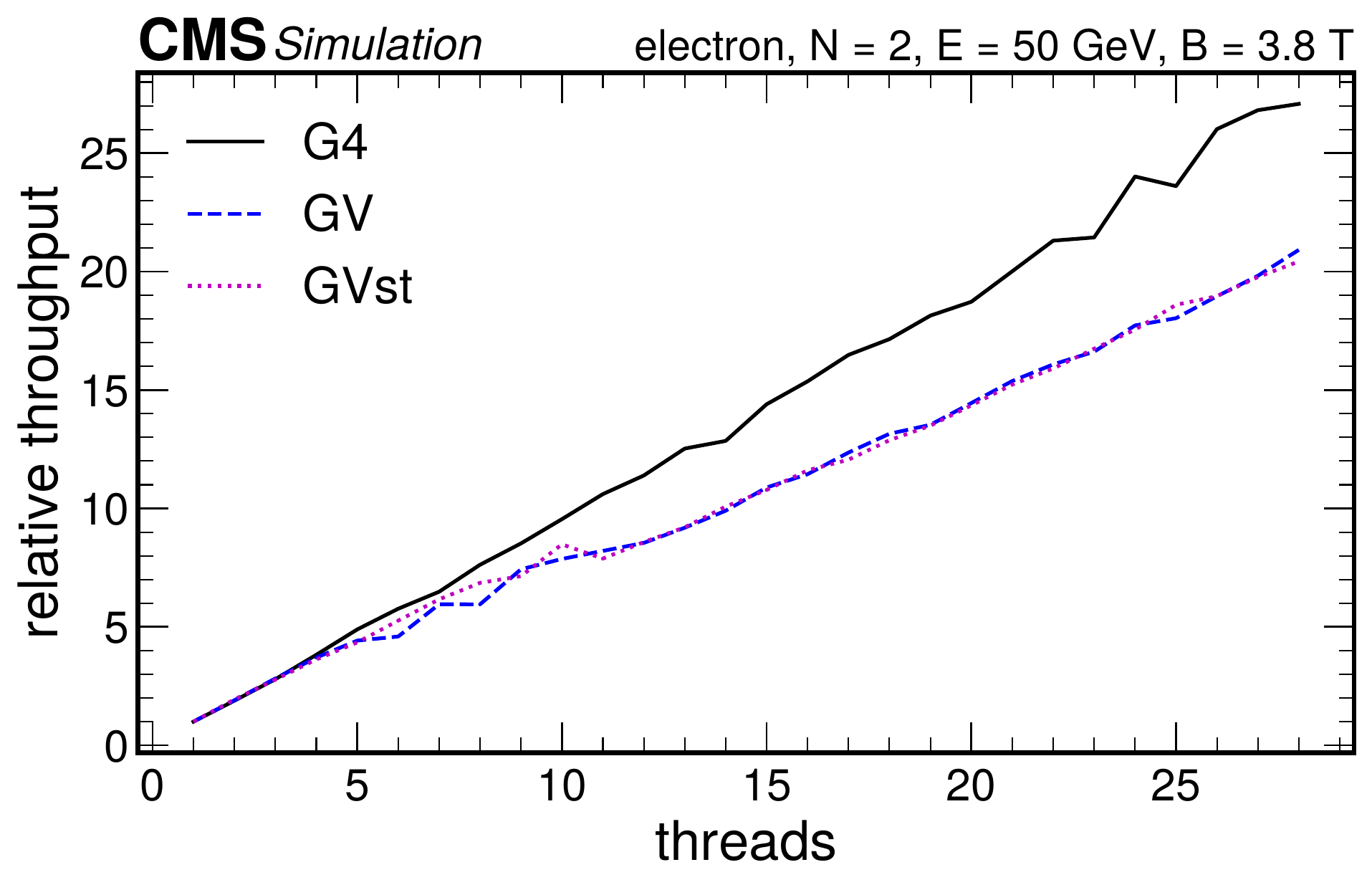}
\caption{Top: throughput in events per second for Geant4 (G4, black), GeantV (GV, blue), and GeantV single track mode (GVst, purple). Middle: throughput ratio for Geant4/GeantV (blue) and Geant4/GeantV single track mode (purple). Bottom: speedup calculated as throughput(N threads) / throughput(1 thread). The E5-2683 v3 CPU was used for these tests.}
\label{fig:cms-throughput}
\end{figure}

\begin{figure}
\centering
\includegraphics[width=0.95\linewidth]{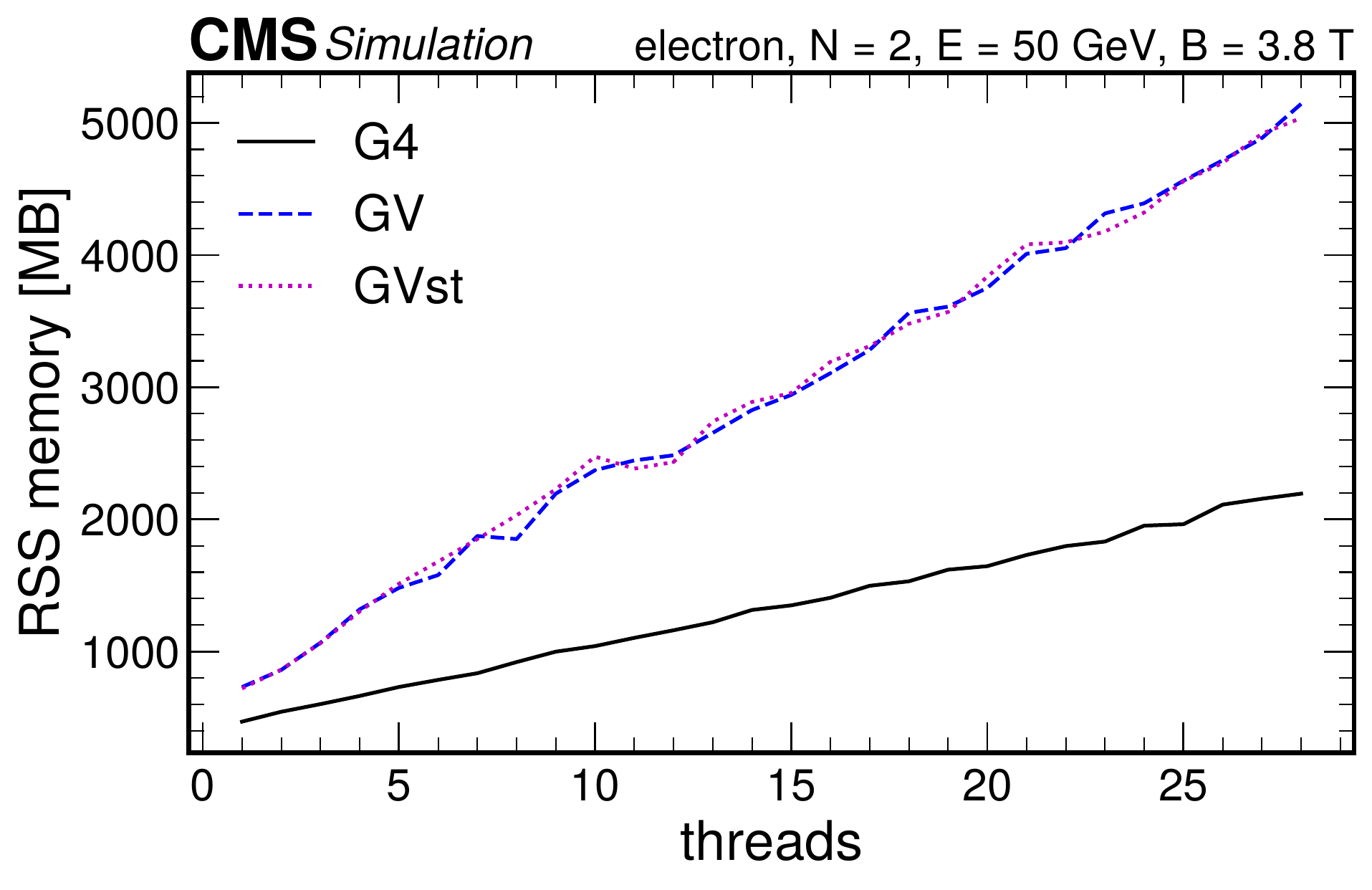}
\includegraphics[width=0.95\linewidth]{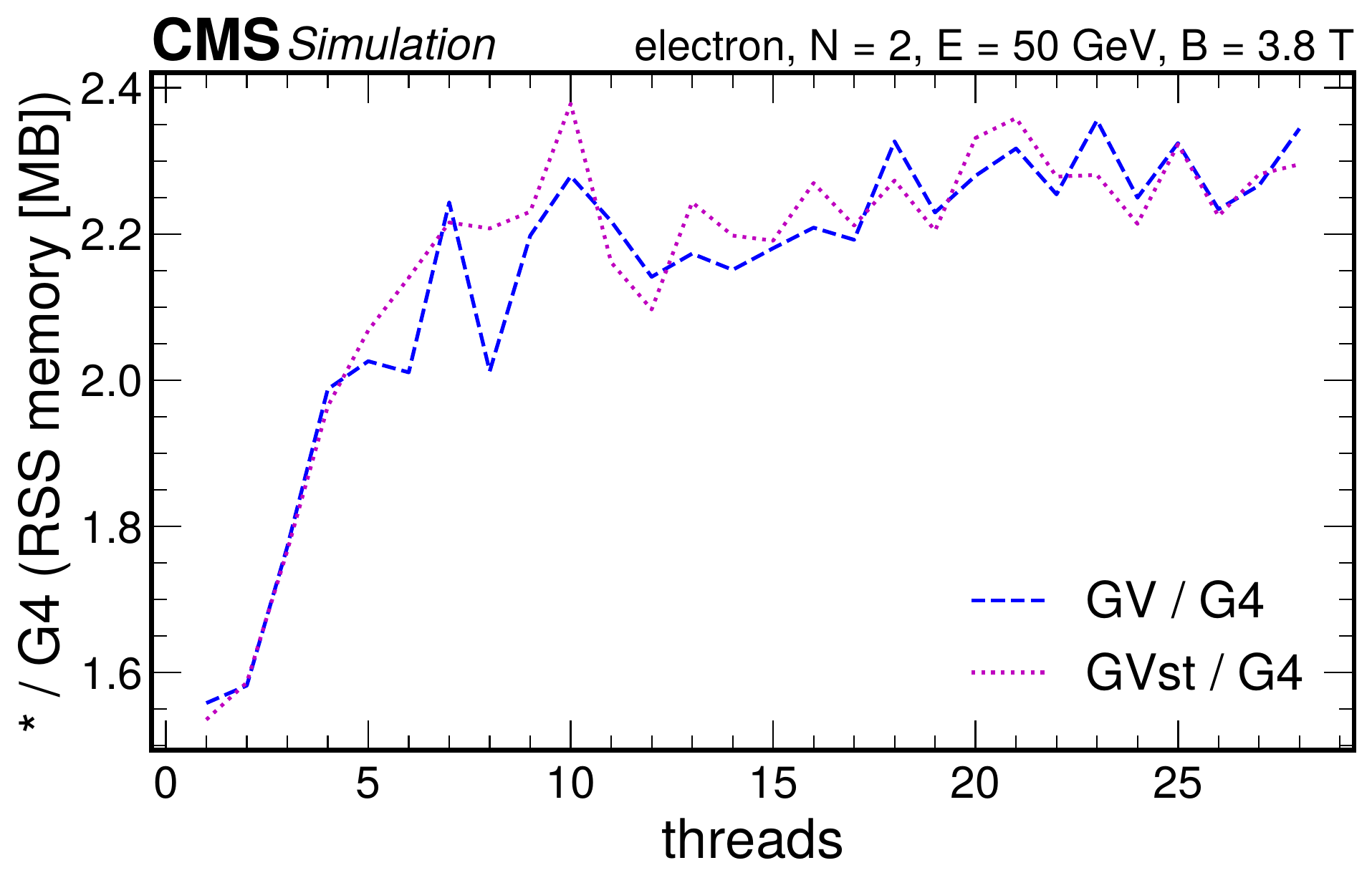}
\caption{Top: RSS memory in megabytes for Geant4 (G4, black), GeantV (GV, blue), and GeantV single track mode GVst, purple). Bottom: RSS memory ratio for Geant4/GeantV (blue) and Geant4/GeantV single track mode (purple). The E5-2683 v3 CPU was used for these tests.}
\label{fig:cms-memory}
\end{figure}

Several different Intel{\textregistered} machines were used for the tests, with different cache sizes and other parameters. Table~\ref{table:cms-speedups} summarizes the results. This table also includes results from the GeantV built-in standalone CMS test with similar settings, in order to characterize the performance observed in the full CMSSW framework. There is virtually no difference in performance between basketized mode and single track mode. In all cases, the single thread speedup in CMSSW exceeds the single thread speedup from the standalone test. This is likely due to the additional instructions included when running in the CMSSW framework. The smaller instruction size from GeantV plays an even more important role in this case, as it allows more CMSSW instructions to be cached by the CPU. More pronounced speedups, along with more pronounced differences between CMSSW and the standalone, are seen in machines with smaller caches, supporting this explanation. Unfortunately, the speedup declines as the number of threads is increased, because GeantV does not scale as well as Geant4 with multiple threads. As expected, GeantV uses more memory than Geant4. For both programs, the memory usage increases linearly with the number of threads. Figures~\ref{fig:cms-throughput} and~\ref{fig:cms-memory} depict the scaling behavior of throughput and memory for Geant4 and GeantV as the number of threads increases, using the E5-2683 v3 machine.

\begin{table*}
 \caption{CMSSW test results with different machines, both single-threaded and multi-threaded. Standalone results are included as a comparison. Ratios of throughput (\#events/s) and memory usage are both shown. The ``N threads'' column shows the result for the maximum number of threads (physical cores) for each machine. The cache value corresponds to the largest cache for each processor: L3 for the E5-2683 v3 and Gold 6248, and L2 for the E5-2660 v2.}
  \label{table:cms-speedups}
  \centering
  \begin{tabular}{crrr|rrr|rr}
\hline\noalign{\smallskip}
          & & & & \multicolumn{3}{c|}{Throughput [GV/G4]} & \multicolumn{2}{c}{RSS memory [GV/G4]} \\
          & & & & \multicolumn{1}{c}{Standalone} & \multicolumn{2}{c|}{CMSSW} & \multicolumn{2}{c}{CMSSW} \\
  Machine & Clock [GHz] & Cache [kB] & Cores & 1 thread & 1 thread & N threads & 1 thread & N threads \\
  \noalign{\smallskip}\hline\noalign{\smallskip}
  E5-2683 v3 & 2.00 & 35840 & 28 & 1.60 & 1.69 & 1.30 & 1.56 & 2.34 \\
  Gold 6248  & 2.50 & 28160 & 20 & 1.49 & 1.66 & 1.18 & 1.54 & 2.31 \\
  E5-2660 v2 & 2.20 &  4096 &  8 & 2.14 & 2.63 & 2.18 & 1.42 & 2.36 \\
  \noalign{\smallskip}\hline
  \end{tabular}
\end{table*}

\section{Lessons learned}
 
The GeantV R\&D project performed an in-depth investigation of alternative particle transport scheduling models for simulation. While the main objective was to achieve important speedups from vectorization and extra locality, there were several other direct or derived studies producing important  results and conclusions. These are briefly discussed in the following subsections.

\subsubsection{Vectorization model, basketization and parallelism}

In order to take advantage of code that executes in using vector instructions, tracks that are similar need to be gathered together, and this introduces many significant challenges.

One challenge is the trade-off between memory usage and efficiency.  For example, it was necessary to restrict the number of shapes for which tracks were collected into shape-specific baskets, in order to avoid memory explosions from both the number of baskets and the number of tracks in flight needed to fill those baskets. However, this means that only the shapes that are selected can benefit from vectorization.  

Similarly, in order to fit within the user memory budget, the number of events in flight must be limited. In practice, this means that when the number of tracks in flight for a given event starts to ramp down significantly, the event should be closed out so that a new one can start, avoiding starvation. To close out
an event is an expensive operation whose cost grows with the total number of baskets: it requires finding all outstanding baskets that still have at least one track belonging to the event and processing them to completion in scalar mode, since they have not reached the threshold to run in vector mode. All of these factors, plus the lack of vectorized navigation, means that the gains from vectorization of the geometry stage are marginal at best, even though the VecGeom primitives are among the best vectorized code.

When introducing multiple threads, load balancing becomes yet another challenge where synchronization points (however small they may be) are needed to determine how much of the work can be and should be shared between threads, and to do the actual sharing. In order to increase scaling, it has been necessary to reduce the amount of sharing between threads several times, at the expense of vector efficiency. An early version copied track data from one thread to another thread's input stack essentially every time a track left a volume to go into a volume of a different type. In a later version, this happened only for tracks in overflowing baskets when a different thread was idle (ran out of local tracks to process). Even in this limited case, the cost was noticeable, in particular during event tails. In order to reduce contention in the shared resources, a notion was introduced of a group of threads all pinned to a NUMA domain. Each group of threads is essentially independent of the others, with the tracks always staying within a single group. This arrangement also helps to reduce the amount of memory transfer across NUMA domains.

At first, it was assumed that there would be a benefit from gathering the memory fetches as closely as possible.  Even though improved CPU data cache usage was observed in the implementation that passed the tracks from baskets to basket by copying the data, the real bottleneck, and major user of CPU time, was \textit{memcpy} itself. The gain from avoiding \textit{memcpy} was measured to be very large, even though it resulted in a much less efficient access pattern for filling the vector register, due to a more fragmented data layout in memory. On the other hand, the cost of filling the vector register from this fragmented memory layout is noticeable, and in fact it is one of the major problems blocking any efficiency gain from vectorization.

\subsubsection{Geometry}

The VecGeom library was one of the first software components put in place for GeantV. It helped to pave the way for GeantV as a whole, as it led to the development of VecCore, a framework for abstracting vector operations for different processor architectures, which demonstrated that SIMD acceleration can be achieved in a portable manner. The programming model developed for VecGeom was generalized by introducing VecCore, which has been used in GeantV to write code that, with a single implementation, can be instantiated to support either scalar or vector inputs. This model was not only used to improve the run-time performance of several algorithms with vector inputs, but also to improve the run-time performance of some algorithms even when the input is scalar. Which type of gain is more relevant depends strongly on the detector being simulated. It was also shown that automatic code generation and specialization of the algorithms, tailored to the target geometry, can lead to significant further performance improvements.

\subsubsection{Physics}

Since the modeling of electromagnetic (EM) interactions of  $e^-,e^+$ and $\gamma$ particles with matter is the most intensively used and most computationally demanding part of most high-energy physics detector simulations, EM physics processes were selected to be vectorized. Beyond vectorization, the existing code was first reviewed, overhauled and improved, based on an exhaustive review of the relevant literature. This resulted in optimized versions of the code that brought significant performance improvements. Using model-level tests to analyze the performance of the vectorized EM models compared to their (optimized) scalar versions, excellent vectorization gains were achieved: 1.5--3$\times$ and 2--4$\times$ on Haswell and Skylake (AVX2) architectures, respectively. Unfortunately, these synthetic model test gains are not visible when integrated into a full application that has to deal with the entire range of models, particle types, and energies in a stochastic manner.

\subsubsection{Magnetic field}

The integration of the equations of motion of a charged particle in a non-uniform pure magnetic field (or an electromagnetic field) accounts for about 15--20\% of the CPU time of a HEP application. After reengineering and vectorizing the implementation, improvement in the ratio of the Geant4 run-time over the GeantV run-time was measured in the benchmark example from a factor 1.88 in fully scalar mode to a factor 2.12 with the integration of the equations of motion implementation executed in vector mode.

\subsubsection{Interfacing with user task-parallel frameworks}

Concerning the ability to integrate the GeantV prototype in the experiments' simulation frameworks, two essential questions are whether the run-time performance gains are reproduced when the simulation is executed within the experiment framework and how much effort is needed to replace the simulation engine with the new implementation. GeantV and CMS software developers worked closely together to explore this integration. During this co-development effort, there were several iterations on some of the fundamental features of the internal scheduler and its interfaces in order to ease the integration effort and improve run-time efficiency. Thanks to this collaboration, one of the results is the realization that the integration of the prototype of the GeantV toolkit within an experiment framework is relatively straightforward. The other major result is that the run-time performance gain seen in the standalone example is also seen in the integrated example, and is even slightly better.

\section{Summary and conclusion}

The GeantV R\&D project has reached its conclusions after several years of development and study undertaken in the context of an international collaboration with the participation of the LHC experiments and under the umbrella of  the HEP Software Foundation (HSF). Its main objective of demonstrating an achievable speedup of a novel approach based on parallel particle transport has been realized with the delivery of a prototype that simulates full electromagnetic showers in a realistic and complex calorimeter. It has been shown that the performance gain from the vectorization of the individual software components is largely lost in the process of reshuffling the particles for the vector operations. On the other hand, it has also been observed that significant improvements in the performance of the simulation software can be obtained by better exploitation of data and code locality, as well as through more compact code based on modern programming idioms. These findings are informing the direction of future improvements of the Geant4 toolkit, including the investigation of architectural revisions.

Furthermore, the GeantV project has delivered the modular software packages VecGeom, VecCore, and VecMath, which are having a significant impact in different software areas within high energy physics. Those packages have already gone through all the phases of development, validation, and integration. They are now used in production by toolkits like Geant4 and ROOT and are delivering noticeable gains in performance. 

In summary, the GeantV R\&D project has contributed a set of useful libraries to the HEP software community, as well as valuable knowledge which has been used to inform further development of detector simulation toolkits. Future lines of work include modernization of the Geant4 simulation toolkit code, R\&D for efficient utilization of accelerators in modern hardware platforms, and investigation of fast simulation techniques that promise to provide the necessary speed and physics fidelity needed for a larger fraction of use cases in future HEP experiments. 

\begin{acknowledgements}

\begin{sloppypar}This work was supported by Fermi Research Alliance, LLC under Contract No. DE-AC02-07CH11359 with the U.S. Department of Energy, Office of Science, Office of High Energy Physics. Part of this work was supported by funds provided by Intel Corporation to the Center for Scientific Computing at the S\~{a}o Paulo State University under FUNDUNESP Grant No. 2323/2014. The authors would like to thank Graeme A Stewart (CERN) for his careful review of this article. We also owe a lot to many colleagues who supported the GeantV project and gave valuable feedback during the development period, including Krzysztof Genser (Fermilab) and Chris Jones (Fermilab).\end{sloppypar}

\end{acknowledgements}

%
\section*{Conflict of interest}
The authors declare that they have no conflict of interest.

%

\appendix

%
%

\section{Performance benchmark}
\label{app:perf_benchmark}

\renewcommand{\arraystretch}{2.5}
\newcommand{\pms}{\hspace{-4pt}\pm\hspace{-4pt}}
\setlength\tabcolsep{3pt}

\begin{table*}[ht]
\caption{Performance analysis by architecture. The {\bf GV} column shows the absolute execution time in seconds for a scalar, single thread GeantV configuration, while all other columns provide speedups or ratios, as indicated, relative to this column. The GeantV single track mode is indicated by {\bf strk}. See the text for more details.}
\label{tab-ap-perf}
{\scriptsize
\begin{center}
\begin{tabular}{cccccccrclrclrclrcl}
\hline\noalign{\smallskip}
\bf{CPU specs} & \bf{OS} & \bf{gcc} & \bf{SIMD} & \parbox{1.cm}{\centering\bf{L1\\cache}} & \parbox{1.cm}{\centering\bf{L2\\cache}} & \parbox{1.cm}{\centering\bf{L3\\cache}} & 
\multicolumn{3}{c}{\bf{GV [sec]}} & 
\multicolumn{3}{c}{\bf{G4/GV}} & 
\multicolumn{3}{c}{\bf{strk/GV0}} & \multicolumn{3}{c}{\parbox{1.5cm}{\centering\bf{Vector\\ gain}}} \\
\noalign{\smallskip}\hline\noalign{\smallskip}
Intel i7 2.5GHz & Ubuntu 16.04 & 5.4.0 & AVX2 & 126KB & 1MB & 8MB & 941 & $\pms$ & 5 & 1.41 & $\pms$ & 0.04 & 1.02 & $\pms$ & 0.02 & 1.09 & $\pms$ & 0.01 \\
\parbox{2.1cm}{Intel Core i7-4510U 2GHz} & Ubuntu 16.04 & 5.4.0 & AVX & 128KB & 512KB & 4MB & 1303 & $\pms$ & 3 & 1.09 & $\pms$ & 0.01 & 0.95 & $\pms$ & 0.07 & 1.09 & $\pms$ & 0.08 \\
AMD A10-7700k & \parbox{1.9cm}{Fedora Work\-station 29} & 8.2.1 & AVX & \parbox{1.6cm}{2x96 KB I,\\ 4x16 KB D} & 2x2MB & - & 1828 & $\pms$ & 5 & 1.80 & $\pms$ & 0.04 & 1.25 & $\pms$ & 0.03 & 1.01 & $\pms$ & 0.01 \\
\parbox{2.1cm}{Intel Celeron 1000M 1.8GHz} & \parbox{1.9cm}{Fedora Work\-station 29} & 8.3.1 & SSE4 & 64KB & 512KB & 2MB & 2769 & $\pms$ & 10 & 1.03 & $\pms$ & 0.01 & 1.11 & $\pms$ & 0.01 & 0.84 & $\pms$ & 0.01 \\
Intel Centrino 2 & \parbox{1.9cm}{Fedora Work\-station 29} & 8.2.1 & AVX & - & 2x2MB & - & 2592 & $\pms$ & 2 & 1.92 & $\pms$ & 0.01 & 1.24 & $\pms$ & 0.01 & 1.01 & $\pms$ & 0.01 \\
\noalign{\smallskip}\hline
\end{tabular}
\end{center}
}
\end{table*}

\noindent
\begin{table*}[h]
\caption{Basketization overhead study, by architecture. $\mathcal{B}_{oh}$ is the basketization overhead of GeantV, which is measured separately for different stages: propagation in magnetic field, physics, geometry, multiple scattering (MSC), and combined field+physics+MSC (FPM). The latter configuration was shown in Section~\ref{sec:perfResults} to be the best performing configuration in the high-end, multi-threading machines. See the text for more details.}
\label{tab-ap-baskOvh}
{\scriptsize
\begin{center}
\begin{tabular}{ccccccrclrclrclrclrcl}
\hline\noalign{\smallskip}
\bf{CPU specs} & \bf{gcc} & \bf{SIMD} & 
\parbox{1.cm}{\centering\bf{L1\\cache}} & 
\parbox{1.cm}{\centering\bf{L2\\cache}} & 
\parbox{1.cm}{\centering\bf{L3\\cache}} & 
\multicolumn{3}{c}{{$\mathcal{B}_{oh}$} (B-field)} & 
\multicolumn{3}{c}{{$\mathcal{B}_{oh}$} (phys)} & 
\multicolumn{3}{c}{{$\mathcal{B}_{oh}$} (geom)} & 
\multicolumn{3}{c}{{$\mathcal{B}_{oh}$} (MSC)} & 
\multicolumn{3}{c}{{$\mathcal{B}_{oh}$} (FPM)} \\
\noalign{\smallskip}\hline\noalign{\smallskip}
Intel i7 2.5GHz & 5.4.0 & AVX2 & 128KB & 1MB & 8MB & 2\% & $\pms$ & 1\% & 2\% & $\pms$ & 1\% & 6\% & $\pms$ & 1\% & 0\% & $\pms$ & 1\% & 3\% & $\pms$ & 1\% \\
\parbox{2.cm}{Intel Core i7-4510U 2GHz} & 5.4.0 & AVX & 128KB & 512KB & 4MB & -1\% & $\pms$ & 7\% & -3\% & $\pms$ & 7\% & 12\% & $\pms$ & 9\% & -4\% & $\pms$ & 8\% & 2\% & $\pms$ & 8\% \\
AMD A10-7700k & 8.2.1 & AVX & \parbox{1.6cm}{2x96 KB I,\\ 4x16 KB D} & 2x2MB & - & 15\% & $\pms$ & 1\% & 4\% & $\pms$ & 1\% & 15\% & $\pms$ & 1\% & 1\% & $\pms$ & 1\% & 13\% & $\pms$ & 1\% \\
\parbox{2.cm}{Intel Celeron 1000M 1.8GHz} & 8.3.1 & SSE4 & 64KB & 512KB & 2MB & 9\% & $\pms$ & 1\% & 5\% & $\pms$ & 1\% & 9\% & $\pms$ & 1\% & -1\% & $\pms$ & 1\% & 9\%& $\pms$ & 1\% \\
Intel Centrino 2 & 8.2.1 & AVX & - & 2x2MB & - & 6\% & $\pms$ & 1\% & 3\% & $\pms$ & 1\% & 13\% & $\pms$ & 1\% & -1\% & $\pms$ & 1\% & 7\% & $\pms$ & 1\% \\
AMD e-300 & 8.2.0 & SSE2 & 64KB & 1MB & - & 1\% & $\pms$ & 1\% & 3\%& $\pms$ & 1\% & -3\% & $\pms$ & 1\% & -2\% & $\pms$ & 1\% & & - &\\
\noalign{\smallskip}\hline
\end{tabular}
\end{center}
}
\end{table*}

Most of the multi-thread scaling and vectorization performance analysis was based on data collected using high-end machines, and the most relevant results have been described on Section~\ref{sec:perfResults}. Another approach to performance analysis has also been carried out: a comparative study using several different, typical end-user machines, in order to collect performance data for a wide spectrum of machine specifications. The goal of using a more heterogeneous set of machines was to assess how lower-grade processors and low-memory conditions would affect the performance of the prototype.

The study was based on 1000 events per job with a single 10 GeV electron per event, using the 2018 CMS geometry available from the GeantV repository. For Geant4, release {\tt 10.4.p03} in single thread mode was the baseline configuration. To minimize external interference, jobs were submitted to machines with no other running processes. Each job configuration was run ten times, and the results are based on simple averages of CPU times.

Some of the performance numbers from different architectures are provided in the following tables. The first few columns describe details of each machine's configuration, including processor, brand, operating system, memory and compiler version used. The last columns represent performance results, providing timing averages and uncertainty estimates.

Table~\ref{tab-ap-perf} shows global performance measurements. The absolute timing measurements in the GeantV column roughly agree with processor power and clock speeds. The ratio of Geant4 to GeantV performance ranges from 1.03 to 1.92 for different hardware platforms.
GeantV in single track mode (strk) shows some expected correlation with the Geant4 to GeantV ratio. The best vector gains are achieved for hardware with the best SIMD capabilities, but a low level of vectorization density is also observed, probably due to the difficulty of vectorizing HEP simulations.

Table~\ref{tab-ap-baskOvh} shows the basketization overheads, in an attempt to assess the performance costs attributable to packing the data for vectorization efficiency, but without the corresponding vectorization gains, for jobs where the vectorized algorithms had been explicitly disabled. The basketization process collects tracks with similar characteristics, in order to maximize the SIMD synchronization (e.g. vectorization efficiency), as described earlier.

The track basketization was done separately per simulation stages (magnetic-field propagation, physics, geometry, multiple scattering), as the requirements for each stage are different. The geometry basketization requirements are the most strict,
since tracks need to be in physical volumes which are instances of the same logical volume.
This produces a very large number of baskets, as compared to the basketization requirements for other stages, corresponding to a very significant performance degradation. Ultimately, the best performing GeantV configuration had geometry basketization (and vectorization) disabled at the job initialization level, which is shown in the {\bf FPM} and {\bf Vector gain} table columns.

Different machines present different constraints for the GeantV run-time environment. The observed performance can reflect some of those aspects, and some trends can be observed. However, a more precise interpretation of the effects of different parameters is hard to derive unambiguously from the final numbers.


\end{document}